\documentclass[PAPER, atlasdraft=false, texlive=2023, USenglish, coverpage=false,cernpreprint=true, texmf, orcidlogo]{atlasdoc}

\usepackage[subfigure]{atlaspackage}
\usepackage{atlasbiblatex}

\usepackage{atlasphysics}
\usepackage{atlasjetetmiss}

\graphicspath{{logos/}{figures/}}

\usepackage{ANA-STDM-2020-04-PAPER-defs}

\AtlasTitle{Measurements of jet cross-section ratios in 13 TeV proton--proton collisions with ATLAS}

\AtlasAbstract{
Measurements of jet cross-section ratios between inclusive bins of jet multiplicity are performed in 140~fb$^{-1}$\ of proton--proton collisions with $\sqrt{s}=13$~TeV center-of-mass energy, recorded with the ATLAS detector at CERN's Large Hadron Collider.
These ratios are constructed from double-differential cross-section measurements that are made in bins of jet multiplicity and other observables that are sensitive the energy-scale and angular distribution of radiation due to the strong interaction in the final state.
Additionally, the scalar sum of the two leading jets' transverse momenta is measured triple-differentially, in bins of the third jet's transverse momentum and of jet multiplicity.
These measurements are unfolded to account for acceptance and detector-related effects.
The measured distributions are used to construct ratios of the inclusive jet-multiplicity bins, which have been shown to be sensitive to the strong coupling $\alpha_{\textrm{S}}$ while being less sensitive than other observables to systematic uncertainties and parton distribution functions.
The measured distributions are compared with state-of-the-art QCD calculations, including next-to-next-to-leading-order predictions for two- and three-jet events.
These predictions are generally found to model the data well, and perform best in bins with a modest requirement on the third jet's transverse momentum.
Significant differences between data and Monte Carlo predictions are observed in events with large rapidity gaps and invariant masses of the leading jet pair.
Studies leading to reduced jet energy scale uncertainties significantly improve the precision of this work, and are documented herein.
}

\author{The ATLAS Collaboration}

\AtlasRefCode{STDM-2020-04}

\PreprintIdNumber{CERN-EP-2024-119}

\arXivId{2405.20206}

\HepDataRecord{https://www.hepdata.net/record/ins2791854}

\AtlasJournalRef{Phys. Rev. D 110, 072019 (2024)}
\AtlasDOI{10.1103/PhysRevD.110.072019}

\AtlasJournal{PRD}

\hypersetup{pdftitle={ATLAS document},pdfauthor={The ATLAS Collaboration}}

\begin{document}

\maketitle

\setcounter{tocdepth}{4}
\tableofcontents

\clearpage
\section{Introduction}
%

Quantum chromodynamics (QCD) is the theory of the strong interaction, which describes the interactions of quarks and gluons.
It is therefore fundamental to understanding both initial-state and final-state physics at hadron colliders such as CERN's Large Hadron Collider (LHC).
Recent theoretical advances have led to fixed-order predictions of three-jet cross-sections in proton--proton ($pp$) collisions at next-to-next-to-leading order (NNLO)~\cite{Czakon:2021mjy}.
Comparisons between experimental results and these state-of-the-art predictions provide powerful tests of perturbative QCD.
Discrepant models of QCD processes impact the accuracy of physics simulations and Monte Carlo event generators, which in turn often limits experimental precision.
Better understanding of the modeling of QCD processes is therefore needed to achieve the highest possible levels of precision in physics analysis, both at the LHC and at future experimental facilities~\cite{Begel:2022kwp,Nachman:2022emq}.

In this analysis, multiple facets of QCD are studied by measuring differential cross-sections of multijet events, and their ratios, in $\sqrt{s}=13~\TeV$ $pp$ collisions at the LHC.
One set of measured observables is sensitive to the energy scale of the hard-scattering process in the event, and can be used to test the accuracy of fixed-order matrix element predictions.
A complementary set of observables is sensitive to the angular distributions of hadronic energy flow in the final state, and hence can probe other aspects of QCD modeling.
The differential cross-sections for each observable are used to construct cross-section ratios between different inclusive jet-multiplicity bins, reducing the sensitivity of the measurements to systematic uncertainties and parton distribution functions.

This procedure was used previously by the ATLAS~\cite{STDM-2010-12} and CMS~\cite{CMS-QCD-10-012,CMS-QCD-11-003} collaborations at the LHC to measure the three-jet to two-jet cross-section ratio, \RThreeTwo, in $pp$ collisions at $\sqrt{s}=7~\TeV$.
Prior to those results, this quantity was measured at other hadron colliders by the UA1~\cite{UA1:1985abw}, UA2~\cite{UA2:1986bnn}, CDF~\cite{CDF:1995odg} and D0~\cite{D0:1995bph,D0:2000tld} collaborations.
Here, \RThreeTwo is presented for the first time in $\sqrt{s}=13~\TeV$ $pp$ collisions, and is compared with fixed-order QCD predictions at NNLO accuracy.
Although such high-precision predictions do not yet exist for events with larger jet multiplicities, the higher multiplicity ratios \RFourThree, \RFourTwo and \RFiveFour are experimentally accessible with the complete \RunTwo dataset.
These ratios were also measured in this analysis, to serve as a reference for future theoretical developments.

The structure of this paper is as follows. A description of the ATLAS detector, the \RunTwo dataset and the multijet simulations used in this analysis are given in Section~\ref{sec:atlas}.
In Section~\ref{sec:methodology}, an overview of the event selection and observables used in this analysis is provided, along with a description of the unfolding procedure used to correct the measured distributions for effects related to detector resolution.
The estimated systematic uncertainties from various sources are described in Section~\ref{sec:systematics}, and modeling improvements leading to smaller jet energy scale uncertainties are highlighted.
Fixed-order QCD calculations that are compared with the measured data are then described in Section~\ref{sec:theory}.
The main results of the analysis are presented in Section~\ref{sec:results} and compared with fixed-order QCD predictions and Monte Carlo simulated event samples.
Concluding remarks follow, in Section~\ref{sec:conclusions}.


%

%
\section{The ATLAS detector, Run 2 dataset and simulation}
\label{sec:atlas}

\subsection{The ATLAS detector}

The ATLAS detector~\cite{PERF-2007-01} at the LHC covers nearly the entire solid angle around the
collision point.\footnote{ATLAS uses a right-handed coordinate system with its origin at the nominal interaction point (IP)
in the center of the detector and the \(z\)-axis along the beam pipe.
The \(x\)-axis points from the IP to the center of the LHC ring,
and the \(y\)-axis points upwards.
Cylindrical coordinates \((r,\phi)\) are used in the transverse plane,
\(\phi\) being the azimuthal angle around the \(z\)-axis.
The pseudorapidity is defined in terms of the polar angle \(\theta\) as \(\eta = -\ln \tan(\theta/2)\).
Angular distance is measured in units of \(\Delta R \equiv \sqrt{(\Delta y)^{2} + (\Delta\phi)^{2}}\),
where $ y = (1/2) \ln[ (E + p_{z})/(E-p_{z})]$ is the object's rapidity defined by its energy and longitudinal momentum. }
It consists of an inner tracking detector surrounded by a thin superconducting solenoid, electromagnetic and hadron calorimeters,
and a muon spectrometer incorporating three large superconducting air-core toroidal magnets.

The inner-detector system is immersed in a \qty{2}{\tesla} axial magnetic field
and provides charged-particle tracking in the range \(|\eta| < 2.5\).
The high-granularity silicon pixel detector covers the vertex region and typically provides four measurements per track,
the first hit normally being in the insertable B-layer installed before \RunTwo~\cite{ATLAS-TDR-19,PIX-2018-001}.
It is followed by the silicon microstrip tracker, which usually provides eight measurements per track.
These silicon detectors are complemented by the transition radiation tracker (TRT),
which enables radially extended track reconstruction up to \(|\eta| = 2.0\).
The TRT also provides electron identification information
based on the fraction of hits (out of a typical total of 30) above a higher energy-deposit threshold corresponding to transition radiation.

The calorimeter system covers the pseudorapidity range \(|\eta| < 4.9\).
Within the region \(|\eta|< 3.2\), electromagnetic calorimetry is provided by barrel and
endcap high-granularity lead/liquid-argon (LAr) calorimeters,
with an additional thin LAr presampler covering \(|\eta| < 1.8\)
to correct for energy loss in material upstream of the calorimeters.
Hadron calorimetry is provided by a steel/scintillator-tile calorimeter (`tile calorimeter'),
segmented into three barrel structures within \(|\eta| < 1.7\), and two copper/LAr hadron endcap calorimeters.
The solid angle coverage is extended by forward copper/LAr and tungsten/LAr calorimeter modules
optimized for electromagnetic and hadronic energy measurements respectively.

The muon spectrometer comprises separate trigger and
high-precision tracking chambers measuring the deflection of muons in a magnetic field generated by
the superconducting air-core toroidal magnets.
The field integral of the toroids ranges between \num{2.0} and \qty{6.0}{\tesla\metre}
across most of the detector.
Three layers of precision chambers, each consisting of layers of monitored drift tubes, cover the region \(|\eta| < 2.7\),
complemented by cathode-strip chambers in the forward region, where the background is highest.
The muon trigger system covers the range \(|\eta| < 2.4\) with resistive-plate chambers in the barrel, and thin-gap chambers in the endcap regions.

Interesting events are selected by the first-level trigger system implemented in custom hardware,
followed by selections made by algorithms implemented in software in the high-level trigger~\cite{TRIG-2016-01}.
The first-level trigger accepts events from the \qty{40}{\MHz} bunch crossings at a rate below \qty{100}{\kHz},
which the high-level trigger reduces in order to record events to disk at about \qty{1}{\kHz}.

An extensive software suite~\cite{ATL-SOFT-PUB-2021-001} is used in data simulation, in the reconstruction
and analysis of real and simulated data, in detector operations, and in the trigger and data acquisition
systems of the experiment.

\subsection{Data} %

This analysis is performed using data from LHC $pp$ collisions with a center-of-mass energy of $\sqrt{s}~=~13$~\TeV, collected during 2015--2018 with the ATLAS detector.
The total integrated luminosity of this dataset is $140~\ifb$.
The uncertainty in the combined 2015--2018 integrated luminosity is 0.83\% \cite{DAPR-2021-01}, obtained using the LUCID-2 detector \cite{LUCID2} for the primary luminosity measurements, complemented by measurements using the inner detector and calorimeters.
Due to the high instantaneous luminosity and the large total inelastic $pp$ cross-section, there are, on average, 33.7 simultaneous (\enquote{\pileup{}}) collisions in each bunch crossing.
Data events must satisfy quality requirements to be included in the analysis~\cite{DAPR-2018-01}.

\subsection{Simulation}\label{sec:atlas:mc}

Samples of Monte Carlo (MC) simulated multijet events are used in this analysis for comparison with the data, to unfold the detector-level measurement to particle-level, and to achieve the reductions in jet energy scale uncertainties described in Section~\ref{sec:systematics}.
Since the jet production cross-section is much larger than the cross-sections for other processes, these multijet samples are sufficient to describe data in the fiducial region of the measurement.

\PYTHIA[8.230]~\cite{Sjostrand:2014zea} is used as the nominal MC generator in this analysis, and is also referred to here as the \enquote{nominal} simulation.
Samples of $\TwoToTwo$ dijet events were generated using the A14 set of tuned parameters (\enquote{tune})~\cite{ATL-PHYS-PUB-2014-021}, the Lund string hadronization model~\cite{Andersson:1983ia,} and the \NNPDF[2.3lo] leading-order (LO) parton distribution function (PDF) set~\cite{Ball:2012cx}.
The \PYTHIA parton shower (PS) algorithm uses a transverse-momentum-ordered evolution~\cite{Sjostrand:2004ef}, and its renormalization and factorization scales were set to the geometric mean of the squared transverse masses of the outgoing particles.
\EVTGEN~\cite{Lange:2001uf} was used to model decays of heavy-flavor hadrons.

Three sets of \SHERPA[2.2.5]~\cite{Sherpa:2019gpd} multijet events were used with the default AHADIC cluster hadronization model~\cite{Webber:1983if} or with the \SHERPA interface to the Lund string hadronization model as implemented in \PYTHIA[6.4], and its decay tables. %
These samples include LO matrix element calculations for \TwoToTwo processes, and use the \SHERPA parton shower algorithm based on Catani--Seymour dipole subtraction~\cite{Schumann:2007mg}.
The \CT[14nnlo] next-to-next-to-leading-order (NNLO) PDF set~\cite{Dulat:2015mca} was used for matrix element calculations and the \CT[10] PDF set~\cite{Lai:2010vv} was used for multiparton interactions (MPI).
One additional \SHERPA sample was generated with \SHERPA[2.2.11], using the same settings as the sample with cluster-based hadronization described above, except that the parameters of the hadronization model were retuned to achieve better agreement with LEP data~\cite{Chahal:2022rid}.
This retuning changes the description of the baryon production rate inside jets, and was found to make the description of the ATLAS detector's jet energy response more consistent between \SHERPA and \PYTHIA jets. Further details can be found in Ref.~\cite{ATL-PHYS-PUB-2022-021}.

Another set of multijet events was generated using \HERWIG[7.1.6]~\cite{Bahr:2008pv,Bellm:2015jjp,Bellm:2017bvx} with the default cluster hadronization model and either the default angle-ordered PS or the alternative dipole PS~\cite{Webber:1983if}.
In these samples the \TwoToTwo~matrix elements are modeled with LO accuracy and interfaced with the \NNPDF[2.3lo] PDF set.
The angle-ordered sample is compared with measurements from this analysis, and both samples are used in the studies of jet energy scale systematic uncertainties presented in Section~\ref{sec:systematics}.

Two additional samples of multijet events with next-to-leading-order (NLO) matrix element accuracy were produced with \POWHEG[v2]~\cite{Nason:2004rx,Frixione:2007vw,Alioli:2010xa} using the multijet process implemented in \POWHEGBOX[v2]~\cite{Alioli:2010xd}, matched to either the \PYTHIA[8] or angle-ordered \HERWIG[7] parton shower configured as for the corresponding samples described above.
The renormalization and factorization scales in these samples were set to the transverse momentum (\pt) of the underlying Born-level configuration.
For the \PYTHIA PS, the default Lund string hadronization model was used with the \NNPDF[3.0nlo] PDF set~\cite{Ball:2014uwa} and A14 tune.
The \NNPDF[3.0nlo] PDF set was also used for the \HERWIG sample, but with the default \HERWIG cluster-based hadronization model.
These samples are referred to as the \enquote{\POWPY{}} and \enquote{\POWHER{}} samples.

All generated events were passed through a full detector simulation~\cite{SOFT-2010-01} based on \GEANT~\cite{Agostinelli:2002hh} and overlaid with simulated minimum-bias interactions generated using \PYTHIA[8]~with the A3 tune~\cite{ATL-PHYS-PUB-2016-017} and \NNPDF[2.3lo] PDF set to represent \pileup interactions.
The distribution of the average number of \pileup interactions in simulation is reweighted during data analysis to match that observed in \RunTwo data.

Additional details of the MC samples used in this measurement may be found in Ref.~\cite{ATL-PHYS-PUB-2019-017}.


%

%
\section{Methodology}
\label{sec:methodology}

\subsection{Object and event selection}\label{sec:evsel}

All jets in this analysis are reconstructed using the \akt algorithm~\cite{Cacciari:2008gp} as implemented in \Fastjet~\cite{Fastjet}, with a jet radius parameter $R=0.4$.

\enquote{Particle-level} jets are reconstructed in simulated events without detector simulation. All detector-stable particles, with a lifetime $\tau$ in the laboratory frame such that $c\tau > 10$~mm, are used, except those that are expected to deposit little or no energy in the calorimeters (\emph{i.e.}, muons and neutrinos).
Particle-level jets are required to have $\pt > 60$~\GeV and absolute rapidity $|y|<4.5$ to enter this analysis.

Detector-level jets are reconstructed from particle flow (PFlow) objects~\cite{PERF-2015-09}, which combine measurements from the ATLAS inner detector and calorimeters~\cite{PERF-2014-07} to improve the jet energy resolution and increase the jet reconstruction efficiency, especially at low jet \pt.
Jets are calibrated such that the average detector-level jet energy scale (JES) matches that of the corresponding particle-level jets, using a combination of simulation-based and \emph{in situ} techniques~\cite{JETM-2018-05}.
Signals originating from detector noise, cosmic rays and beam-induced backgrounds can be reconstructed as spurious jets, but these are efficiently rejected by following the methodology described in Ref.~\cite{ATLAS-CONF-2015-029}, updated for particle flow jets but utilizing the same observables.
For this study, detector-level jets are required to have $\pt > 60$~\GeV and absolute rapidity $|y|<4.5$.
After applying these kinematic selections, the likelihood that a particle flow jet originates from a \pileup interaction is sufficiently low that no additional \pileup-jet rejection is applied~\cite{PERF-2014-03,PERF-2016-06}.

To be included in the analysis, both the particle-level and detector-level events are required to have at least two selected jets ($\njets\geq2$), and the scalar \pt sum of the leading jet pair (ordered in \pt), $\HTTwo = p_{\text{T},1}+p_{\text{T},2}$, must satisfy $\HTTwo \geq 250~\GeV$.

All detector-level events are required to have at least one vertex reconstructed from two or more inner-detector tracks with $\pt > 500$~\MeV, and to pass the data quality requirements described in Ref.~\cite{DAPR-2018-01}.
The data were collected using a set of single-jet triggers~\cite{TRIG-2012-01}, whose thresholds depended on the data-taking year during \RunTwo.
By design, the minimum \HTTwo requirement ensures that the measurement is performed in a region where the single-jet triggers are fully efficient for the analysis selection.
It also ensures that the measurement's fiducial region does not include phase-space regions that are divergent at fixed order~\cite{Salam:2021tbm,Frixione:1997ks}.

Combinations of central and forward single-jet triggers are used to select events in ranges of \HTTwo where the combination is fully efficient.
For triggers that were prescaled during data-taking, events in data are reweighted by the appropriate prescale factor to recover a smoothly falling jet-\pt spectrum.
The prescale factors applied to central-jet and forward-jet triggers differ, so they are logically combined using the \enquote{inclusion method} described in Ref.~\cite{Lendermann:2009ah}.

\subsection{Measured observables}

The two observables chosen for their sensitivity to fixed-order effects are \HTTwo and \ptnincl.
As introduced in Section~\ref{sec:evsel}, \HTTwo~is the scalar sum of the transverse momenta of the leading two jets in the event: $H_{\text{T2}} = p_{\text{T},1}+p_{\text{T},2}$.
It is a proxy for the energy scale of the hard-scattering interaction.
For events with more than two jets, the \pt of the third-leading jet, \ptthree, determines the sensitivity to resummation effects, and varying the \ptthree threshold leads to better understanding of these effects.
The cross-section is measured triple-differentially, as a function of \HTTwo and in bins of jet multiplicity \njets and bins of \ptthree.

The \ptnincl~distribution is the inclusive jet \pt spectrum, measured in bins of inclusive jet multiplicity.
For example, the \ptTwoIncl~distribution is the \pt spectrum of the two leading jets in any selected event, and \ptThreeIncl would include the leading and subleading jet contributions in addition to the third jet, if one passes the selection.

Configurations with large logarithmic corrections can be preferentially selected by measuring jet cross-section ratios as a function of either the absolute value of the leading jet pair's rapidity difference, or the absolute value of the maximum rapidity difference between selected jets in the event (\dyjj and \dymax, respectively).
Similarly, the invariant mass of the two leading jets and the maximum dijet invariant mass found among all selected jets in the event (forming the \mjj and \mjjmax distributions, respectively) also contain a region at large invariant mass where the cross-section receives large logarithmic contributions.
These four observables are measured to probe the resummation in different ways: for example, logarithmic corrections will be larger for \mjjmax than for \mjj.
Large invariant masses for \mjj tend to be dominated by the contributions from large \pt, which are well-described by fixed-order predictions.
Conversely, \mjjmax includes greater contributions from the large angular separations that directly probe these effects.
This set of measurements provides a novel way to indirectly test analytic descriptions of vector-boson scattering/fusion (VBS/VBF) interactions and MC calculations.
Observables such as \mjj and \dyjj are also sensitive to PDFs, and were used in prior PDF fitting studies by the CMS Collaboration~\cite{CMS-SMP-16-011}.
The analysis selection imposes a single-jet \pt requirement on all jets, which allows the logarithmic structure of VBS/VBF events to be probed without introducing additional complications from hierarchies in the jet selection.

\subsection{Unfolding}

All data presented in Section~\ref{sec:results} are unfolded using an iterative D'Agostini unfolding procedure~\cite{DAgostini:1994fjx} to account for effects arising from the limited efficiency, acceptance and resolution of the ATLAS detector.
For each observable, the binning used in this measurement is varied in accord with the detector's resolution.
The unfolding algorithm was implemented using the \textsc{RooUnfold} toolkit~\cite{Adye:2011gm}.
Four iterations of the unfolding procedure are used, because this ensures the unfolding converges well and either minimizes or minimally affects the total uncertainty from all sources for all observables.
For most observables the unfolding is performed double-differentially, in bins of the observable and in exclusive bins of jet multiplicity \njets, to allow the unfolding procedure to account for migrations between relevant bins.
The \HTTwo measurement is unfolded triple-differentially in bins of \HTTwo, \njets and \ptthree.
For the double-differential measurements and triple-differential measurement, the purity of the response matrices (\emph{i.e.}, the size of the diagonal elements) is typically above 50\% or 30\% respectively.

These exclusive bins of \njets are used to construct inclusive bins of \njets.
The unfolded absolute differential cross-sections are presented in Section~\ref{sec:results:xs}.
These unfolded cross-sections are used to construct the cross-section ratios \RThreeTwo, \RFourTwo, \RFourThree and \RFiveFour, which are presented in Section~\ref{sec:results:ratios}.


%

%
\section{Systematic uncertainties}
\label{sec:systematics}

\subsection{Jet energy scale} %

Systematic uncertainties in the $R=0.4$ jet energy scale (JES) and resolution (JER) are evaluated using a series of simulation-based techniques and \emph{in situ} measurements, documented in Ref.~\cite{JETM-2018-05}.
These uncertainties are propagated by building a response matrix from each variation representing a systematic uncertainty component, then unfolding the nominal prior distribution using the varied response matrix.
The difference between the unfolded nominal and systematically varied cross-sections is taken as the systematic uncertainty.
They are the dominant sources of experimental uncertainty in the analysis.

The impact of certain components of the JES uncertainty has been reduced by updating the prescriptions given in Ref.~\cite{JETM-2018-05}.
Uncertainties can also arise because the relative jet energy response for simulated quark- and gluon-initiated jets varies between different MC generators; they are called \enquote{jet-flavor response / composition} uncertainties in Ref.~\cite{JETM-2018-05}.
These uncertainties were significantly reduced relative to their prior treatment, and the improvements are explained below in Section~\ref{sec:systematics:flavor}.
Following this updated treatment, the component of the JES uncertainty due to the jet flavor is reduced from a leading source of uncertainty in the measurement to a completely subdominant effect.

Improvements were also made to the component of the jet energy scale uncertainty related to the extrapolation of single-hadron response measurements~\cite{PERF-2012-01,PERF-2015-05,JETM-2018-08} to jets, discussed below in Section~\ref{sec:systematics:eoverp}.
This component of the JES uncertainty is reduced by roughly a factor of three compared to that reported in Ref.~\cite{JETM-2018-05}.

The JES/JER uncertainties can result in asymmetric variations, and they are left unsymmetrized in the presentation of the measured cross-sections and their ratios.

\subsubsection{Updated jet flavor response uncertainty}\label{sec:systematics:flavor}

The internal dynamics of a jet are determined in part by the flavor of the parton that initiated it.\footnote{The notion of
a \enquote{quark-initiated} or \enquote{gluon-initiated} jet is not well-defined beyond leading order in QCD~\cite{Gras:2017jty,Komiske:2018vkc}:
quark-initiated jets are narrower and have fewer constituents and a harder particle spectrum, on average, than gluon-initiated jets with the same \pt.
For simplicity, these studies use labels based on the identity of the highest-energy ghost-associated~\cite{Cacciari:2007fd} parton that the MC generator's \enquote{truth} record matches to the reconstructed jet.
While this definition can be MC-generator-dependent, this label nevertheless reflects the expected differences between quark- and gluon-initiated jet fragmentation in experimental settings~\cite{PERF-2013-02,STDM-2017-16}, and is therefore well suited to characterize differences in how these jets interact with the detector material.
}
The response of the ATLAS detector to jets depends on the underlying particle spectra, which can vary significantly between different MC generator setups.
An uncertainty related to our limited knowledge of the actual spectra in data and its relationship with the JES is therefore necessary: this component of the ATLAS JES uncertainties is referred to as the \enquote{flavor response uncertainty}. Historically, it has been defined by the difference between the responses to gluon-initiated jets from different MC generator setups multiplied by the fraction of gluon-initiated jets in the measured phase space (e.g.\ in Ref.~\cite{JETM-2018-05}, as the difference between \PYTHIA[8] and \HERWIGpp).

The jet energy response is defined as the ratio of the jet transverse momentum at detector-level ($\pT^{\text{reco}}$) to that at particle-level ($\pT^{\text{true}}$).
The ratio of the average jet response, $\mu(\pT^{\text{reco}}/\pT^{\text{true}})$, between several MC generator setups and the nominal \PYTHIA sample is shown in Figures~\ref{fig:systematics:jes:flavorresp:a} and~\ref{fig:systematics:jes:flavorresp:b} for $R=0.4$ particle flow jets that are initiated by either gluons or quarks and antiquarks, respectively.
For gluon-initiated jets, differences between \PYTHIA and alternative models are as large as 2.5\% at $\pt=60~\GeV$.
For quark-initiated jets, the differences are smaller, with a spread below ${\sim}1\%$ above $\pt=60~\GeV$.
\begin{figure}[t!]
\centering
\subfigure[]{\includegraphics[width=0.49\textwidth]{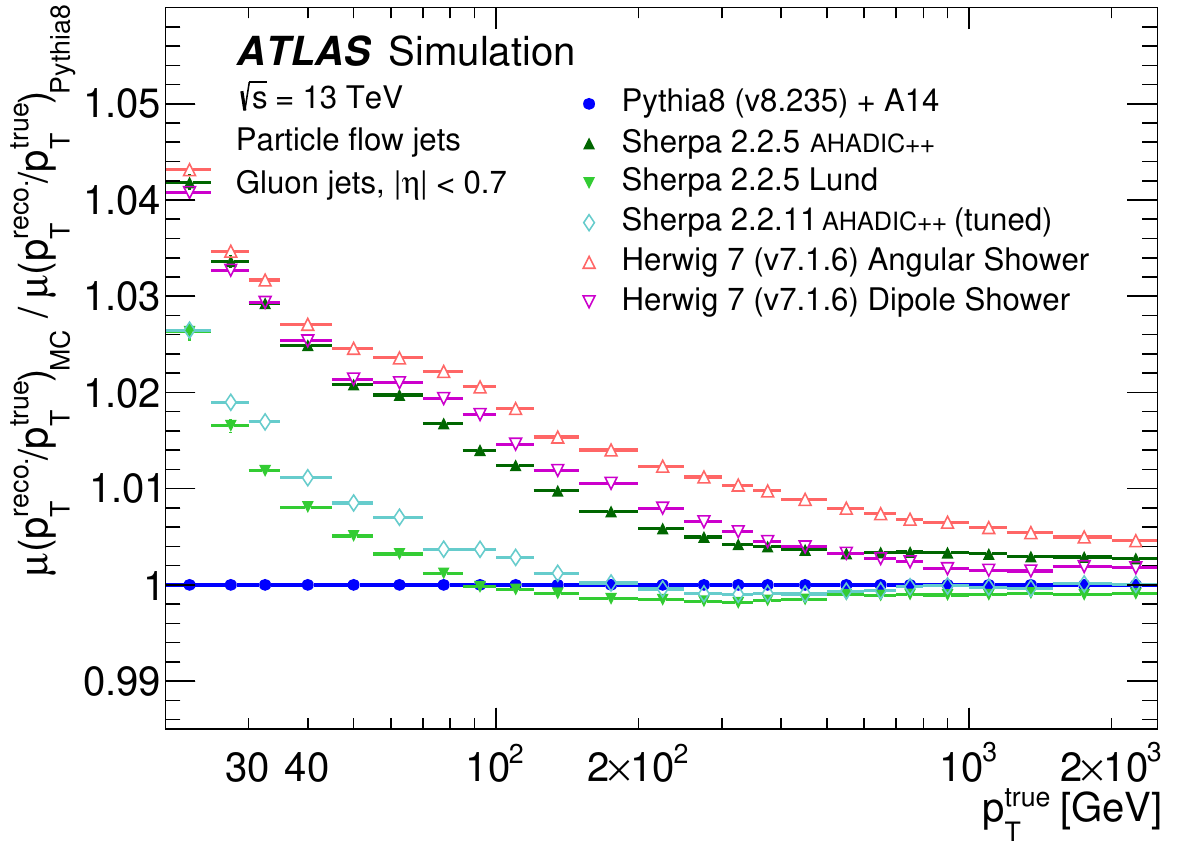}\label{fig:systematics:jes:flavorresp:a}}
\subfigure[]{\includegraphics[width=0.49\textwidth]{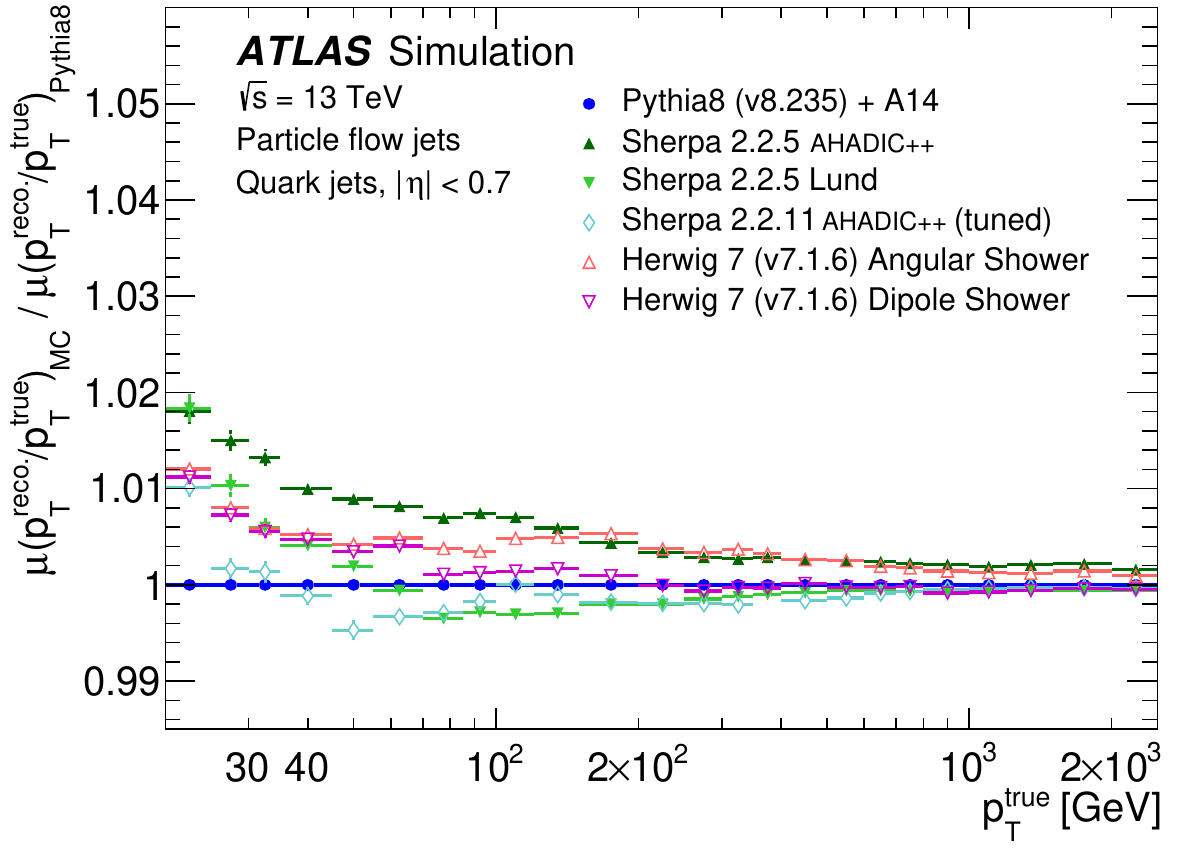}\label{fig:systematics:jes:flavorresp:b}}
\caption{The ratio of the average jet energy response for (a) gluon-initiated  and (b) quark-initiated jets in various MC simulation samples to the jet energy response of the nominal \PYTHIA sample, as a function of the true jet \pt for jets within the pseudorapidity range $|\eta|<0.7$.
}
\end{figure}

Several fragmentation- and hadronization-related effects can change the jet response.
The calorimeter energy response to hadrons rises with energy~\cite{JETM-2018-08,PERF-2015-05}, so the momentum spectrum of the particles associated with the jet is expected to play an important role in jet response modeling.
The particle composition also plays a role because the calorimeter response to neutral pions, which decay via $\pi^0\rightarrow\gamma\gamma$, is significantly higher than for hadronic showers, and for charged hadrons the particle flow algorithm is able to use track measurements.
Additionally, the ATLAS detector's response to hadrons has also been found to vary slightly depending on the species of particle~\cite{PERF-2015-05}, which is consistent with analysis of test-beam data for pions, protons and charged kaons~\cite{Adragna:2010zz,Abdallah:2021gai}.
While the particle composition is partly determined by isospin symmetries, the production of baryons and kaons occurs via different mechanisms in hadronization models that are parameterized and tuned to experimental data~\cite{Christiansen:2015yqa,Gieseke:2017clv,Gieseke:2018gff}.

Since both the particle spectra from jet fragmentation and the particle content of a jet can affect the detector response, their modeling must be tuned to experimental measurements.
Many measurements of jet fragmentation functions and other pertinent substructure observables have been performed at the SPS~\cite{Arnison:167874,BAGNAIA1984291,ua22}, LEP~\cite{Alexander:1991ce,Acton:1993jm,xyz,Buskulic1995389,Alexander1996659,Buskulic1996353,delphintrack,refId0,Abreu1999383,refId1,Akrawy1990617,opallep2}, the LHC~\cite{STDM-2017-16,STDM-2010-10,STDM-2011-38,STDM-2017-04,STDM-2018-57,STDM-2023-07,CMS-SMP-16-010,CMS-SMP-20-010,CMS-SMP-22-007}, and other colliders~\cite{Affolder:2001jx,PhysRevD.68.012003,Aaltonen:2008ae,PhysRevLett.102.232002,Bartel:1982ub,zphys,Derrick:163229,PhysRevLett.55.1954,PhysRevLett.72.3145,Abe:1996zi,PhysRevLett.63.1772,PhysRevD.56.17,Aid:1995up,Derrick:1995ca,H1:2023fzk,}, typically without explicit particle identification.
However, the ALICE experiment has performed some measurements~\cite{ALICE:2019cbr,ALICE:2021aqk,ALICE:2022phr} that do probe the particle content.

Figure~\ref{fig:systematics:jes:baryonfrac} shows the mean baryon energy fractions for central ($|\eta|<0.7$), particle-level, gluon-initiated jets in MC samples generated by \PYTHIA and either \HERWIG[7.1.6] with the angle-ordered or dipole PS algorithms (Figure~\ref{fig:systematics:jes:baryonfrac:a}) or \SHERPA[2.2.5] with cluster-based or string-based hadronization models and \SHERPA[2.2.11] with a cluster hadronization model that was retuned to LEP data~\cite{Chahal:2022rid} (Figure~\ref{fig:systematics:jes:baryonfrac:b}).
The mean baryon energy fraction for quark-initiated jets in the \PYTHIA MC sample is also shown, to indicate the size of possible differences arising due to the jet flavor.
Significant differences between the different MC generator setups are observed.
The mean baryon energy fraction varies between 12\% and 20\% for the different generators, and slowly decreases as the jet \pT increases.
Variation of the PS model in \HERWIG samples does not change the distribution much.
In both \HERWIG samples and the \SHERPA[2.2.5] sample with cluster-based hadronization the fraction of energy carried by baryons is lower than for the nominal \PYTHIA sample, while for the \SHERPA[2.2.5] sample with string-based hadronization it is higher.
The \SHERPA[2.2.11] sample with the retuned cluster-based hadronization is in better agreement with the \PYTHIA sample.

\begin{figure}[t!]
\centering
\subfigure[]{\includegraphics[width=0.49\textwidth]{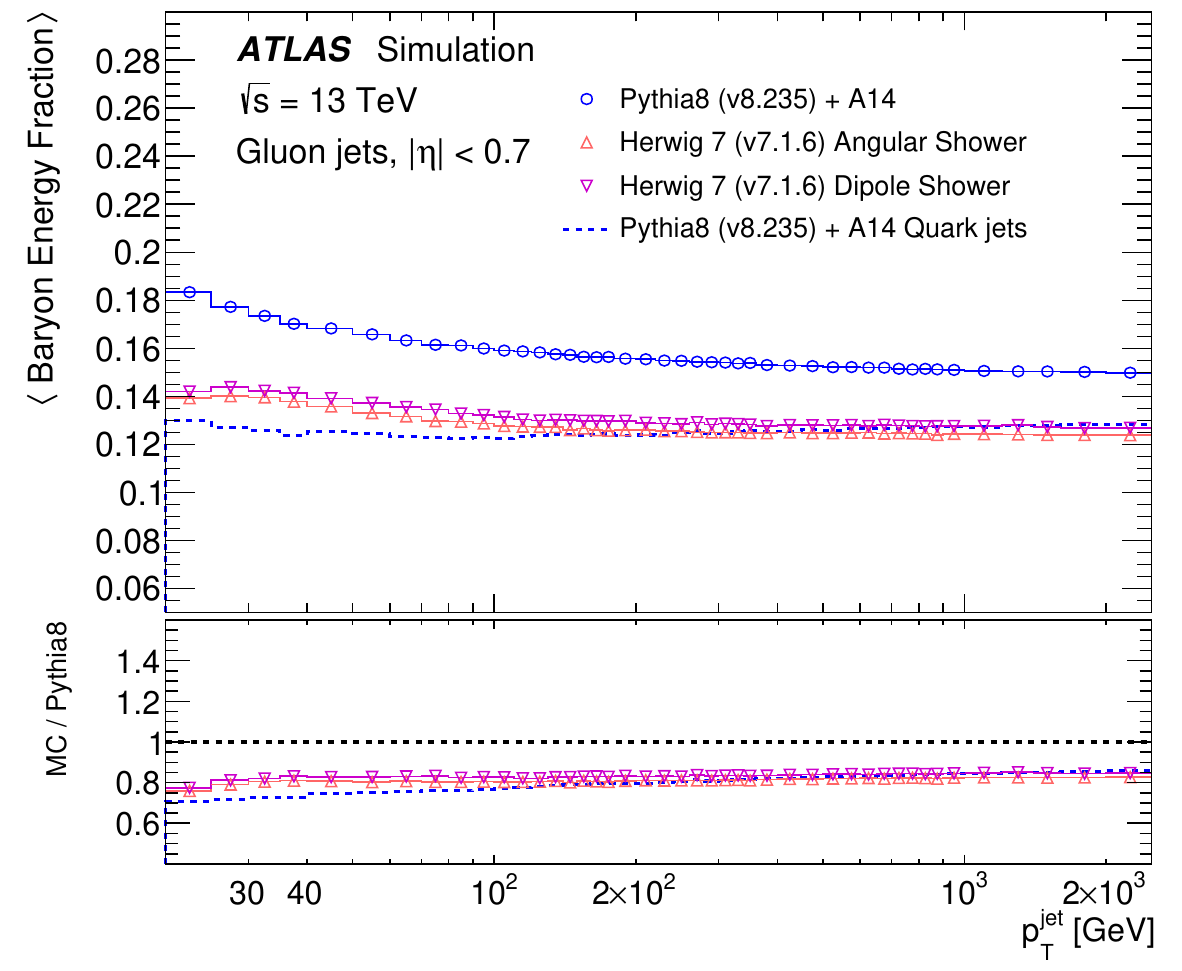}\label{fig:systematics:jes:baryonfrac:a}}
\subfigure[]{\includegraphics[width=0.49\textwidth]{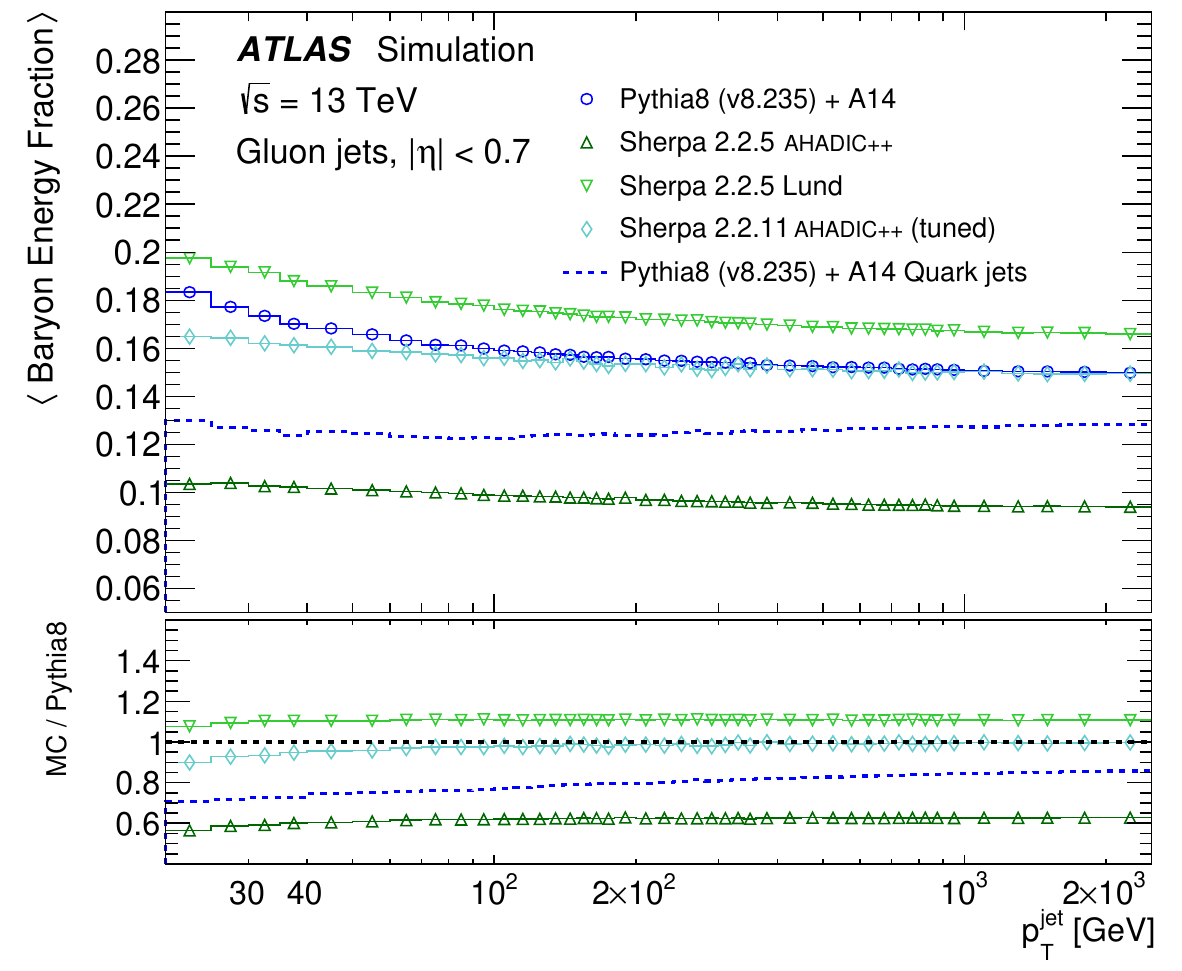}\label{fig:systematics:jes:baryonfrac:b}}
\caption{The average fraction of the gluon-initiated jet's energy carried by baryons as a function of the jet \pt for jets within $|\eta|<0.7$.
The nominal \PYTHIA sample is compared with (a) several \HERWIG samples with different parton shower models, and (b) several \SHERPA samples with different hadronization models and sets of tuned parameters. The dashed line provides a comparison with quark-initiated jets in the nominal \PYTHIA sample.
}
\label{fig:systematics:jes:baryonfrac}
\end{figure}

To further investigate the dependence of the jet response on the particle content of the jet, Figure~\ref{fig:systematics:jes:responseVsFrac} shows the PFlow jet response of gluon- and quark-initiated jets as a function of the fraction of the true jet energy carried by baryons.
It is seen that larger baryon fractions lead to lower jet energy responses for both gluon- and quark-initiated jets.
This is expected, as the majority of these baryons will be protons or neutrons, or their antiparticles.
The average response of these particles is typically lower than that of a mixture of charged and neutral pions, as neutral pions decay mainly into two photons and the calorimeter is calibrated for electromagnetic showers.%
Additionally, while protons, antiprotons and charged pions will have reconstructed tracks during particle flow reconstruction, neutrons and antineutrons will rely solely on calorimeter measurements and thus have a lower response (due to signal leakage, dead material, \emph{etc.}).
Together, these lead to the observed trend that when a larger fraction of the jet's energy is carried by baryons, the jet energy response is lower.

A similar, but smaller, dependence of the jet response on the kaon energy fraction is also observed, and can be explained similarly.

\begin{figure}[tbh]
\centering
\subfigure[]{\includegraphics[width=0.49\textwidth]{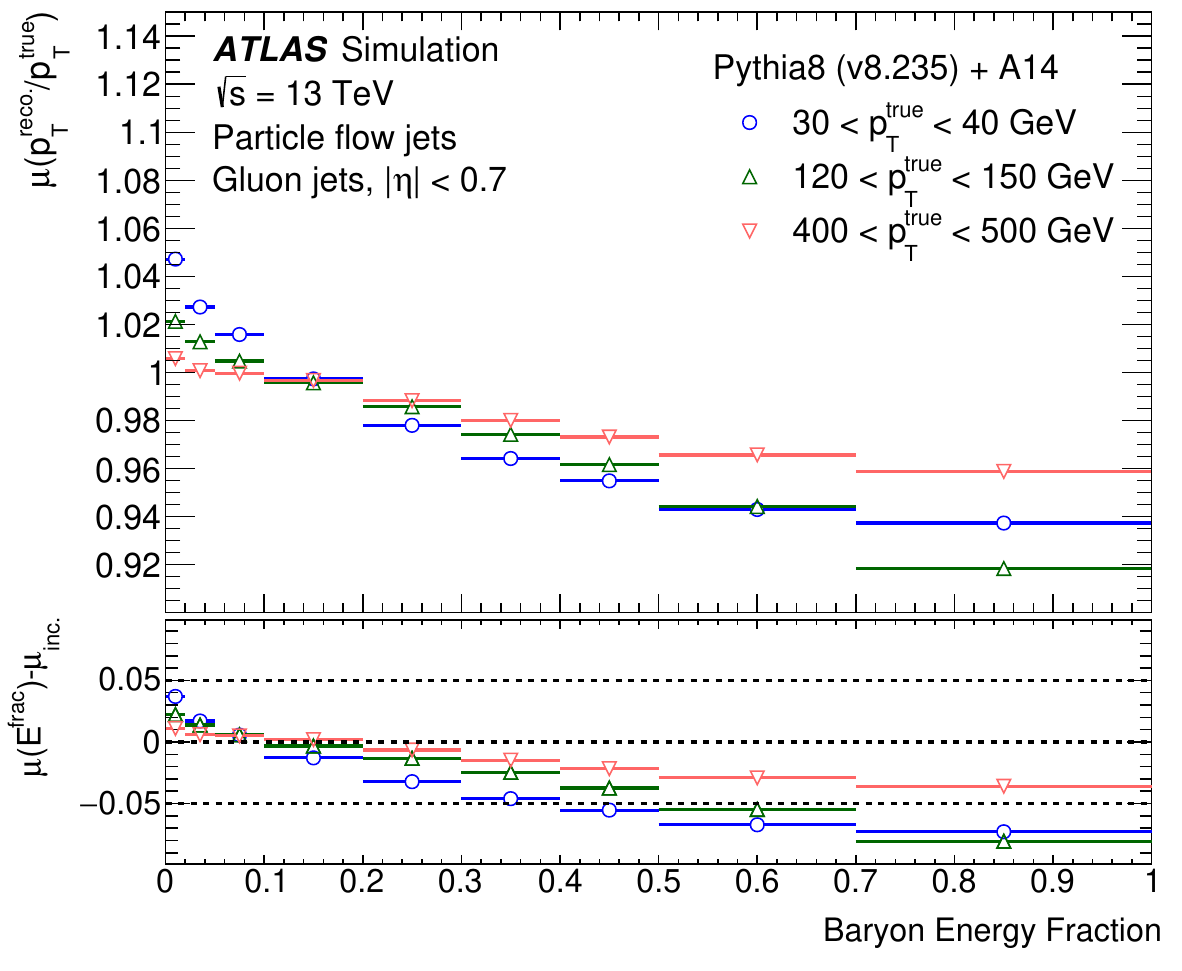}}
\subfigure[]{\includegraphics[width=0.49\textwidth]{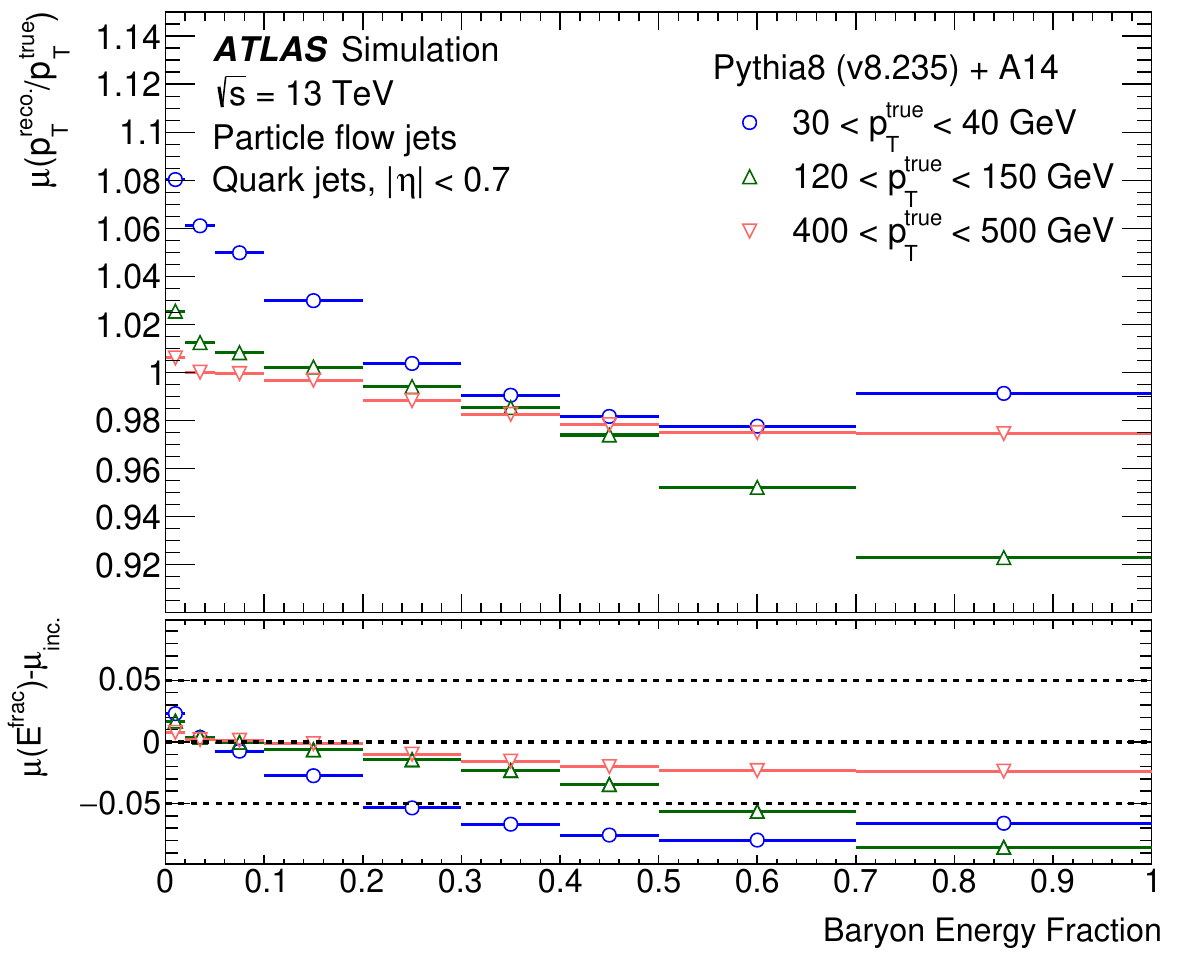}}
\caption{The jet energy response for (a) gluon- and (b) quark-initiated jets as a function of the baryon energy fraction, for various true jet \pt bins in the nominal \PYTHIA sample.}
\label{fig:systematics:jes:responseVsFrac}
\end{figure}

Based on these observations, the treatment of the ATLAS JES flavor response uncertainty was revisited in order to consider differences between the underlying simulated particle spectra more carefully and to use updated MC generator setups.
Instead of the previous two-point comparison of the gluon-initiated jet response between \PYTHIA[8] and \HERWIGpp, three separate uncertainty components are defined by directly comparing MC generator setups that factorize different physical effects:
\begin{enumerate}
\item \textbf{Flavor generator/shower: \PYTHIA[8] vs.\ \SHERPA[2.2.5] w/ Lund string hadronization.} This comparison varies the UE/MPI model and aspects of the PS model (\pt-ordered vs.\ dipole).
\item \textbf{Flavor hadronization: \SHERPA[2.2.11] w/ AHADIC cluster-based hadronization vs.\ \SHERPA[2.2.5] w/ Lund string hadronization.} This comparison varies the nonperturbative hadronization model, between the updated \SHERPA AHADIC cluster-based model and the Lund string model as implemented in \PYTHIA[6].
\item \textbf{Flavor shower: \HERWIG[7.1] w/ angle-ordered PS vs.\ \HERWIG[7.1] w/ dipole PS.} This comparison between the angle-ordered and dipole PS models implemented in \HERWIG ensures that all three major PS schemes used in ATLAS MC generators are considered.
\end{enumerate}
These comparisons ensure that three plausible PS models (\pt-ordered, angle-ordered and dipole) are compared, and that the comparison of nonperturbative models of hadronization physics (AHADIC cluster-based hadronization and Lund string hadronization) does not include model parameterizations that are disfavored by data.
Different PDF sets and color reconnection models are also used for various setups: for complete details of the MC setups used, see Section~\ref{sec:atlas:mc} or Ref.~\cite{ATL-PHYS-PUB-2019-017}.

These uncertainties are derived separately for five different jet flavors ($u$ or $d$, and $s$, $c$, $b$ and $g$) and applied according to a per-jet label in simulated event samples.
This per-jet treatment eliminates the need for the earlier \enquote{jet flavor composition} uncertainty based on the aggregate flavor composition of an analysis' selection relative to the compositions used for \emph{in situ} JES calibrations.

Since the $Z$+jet topology is used directly for the \emph{in situ} JES calibration, the flavor response uncertainty should not vary the response in such events.
In order to maintain a fixed energy scale in the $Z$+jet topology while providing an uncertainty for extrapolation to other flavor compositions, the flavor response uncertainty scales the JES of gluon-initiated jets as $+f_q \Delta(q-g)$ and that of quark-initiated jets as $-(1-f_q)\Delta(q-g)$, where $f_q$ is the $\pT$-dependent fraction of quark-initiated jets in the $Z$+jet sample, and $\Delta(q-g)$ is the difference in the jet response between generators:
\begin{equation*}
\Delta(q-g) = \left( R_q^\text{MC1} - R_g^\text{MC1} \right) - \left(  R_q^\text{MC2} - R_g^\text{MC2}  \right) = \left( R_q^\text{MC1} - R_q^\text{MC2} \right) - \left(  R_g^\text{MC1} - R_g^\text{MC2} \right).
\end{equation*}
This procedure is applied to each of the three MC sample comparisons listed above, resulting in three independent uncertainty components for the JES.
These three components have no effect on samples with the same flavor composition as the $Z+$jet events used in the \emph{in situ} calibration.
The uncertainties in the fractions of quark- and gluon- initiated jets in the simulated samples are evaluated by unfolding data with different MC models as described in Section~\ref{sec:systematics:modeling}.

\subsubsection{Updated single-particle deconvolution uncertainty}\label{sec:systematics:eoverp}

At high jet \pt, the component of the jet energy scale uncertainty determined by the extrapolation of single-particle response measurements to jets via the \enquote{deconvolution} procedure, described in Ref.~\cite{PERF-2011-05,PERF-2015-05}, has been reduced by updating several inputs.
In particular, the response to electromagnetic showers has been updated~\cite{EGAM-2018-01}, and the response to high-\pt pions has been measured \emph{in situ} up to $\pt=250$~\GeV using $W\to\tau(\to\pi\nu)\nu$ events~\cite{JETM-2018-08}, replacing previous test-beam measurements.
The extrapolation of these measurements to other types of hadrons and higher energies is assessed by using updated alternative \GEANT physics lists~\cite{Agostinelli:2002hh} and variations of the detector geometry.
Together, these changes result in a reduction of this uncertainty by roughly a factor of three compared to the uncertainty reported in Ref.~\cite{JETM-2018-05}.

\subsubsection{Summary of \RunTwo jet energy scale uncertainty}

The final \RunTwo JES uncertainty as a function of the jet \pt for central jets is shown separately for gluon- and quark-initiated jets in Figures~\ref{fig:systematics:jes:newflavor:a} and~\ref{fig:systematics:jes:newflavor:b}.
The components due to the flavor generator/shower, shower and hadronization uncertainties are shown, and the overall size of the uncertainty is compared between this updated flavor prescription and the previous one.
Anticorrelations between these uncertainties for gluon- and quark-initiated jets can reduce the aggregate uncertainty for topologies with mixed flavor compositions, such as the dijet flavor composition shown in Figure~\ref{fig:systematics:jes:newflavor:c}.
For both gluon- and quark-initiated jets, the flavor hadronization component is largest for low-\pt jets, up to ${\sim}0.8\%$ below $\pt=100~\GeV$.
The flavor shower component increases for high-\pt jets, up to ${\sim}0.5\%$ for jets above $\pt=400~\GeV$.
The flavor shower/generator component is small (below $0.5\%$) everywhere, but largest for low-\pt jets (below $\pt\sim100~\GeV$).
The updated flavor uncertainty reduces the overall size of the JES uncertainty by up to a factor of two around $\pt\sim100~\GeV$, and renders the component of the JES uncertainty that was previously dominant for $\pt=30{-}400~\GeV$ subdominant everywhere.
The updated single-particle deconvolution uncertainty results in a reduction of this component of the JES at high \pt by roughly a factor of three compared to the uncertainty reported in Ref.~\cite{JETM-2018-05}.

\begin{figure}[t!]
\centering
\subfigure[]{\includegraphics[width=0.49\textwidth]{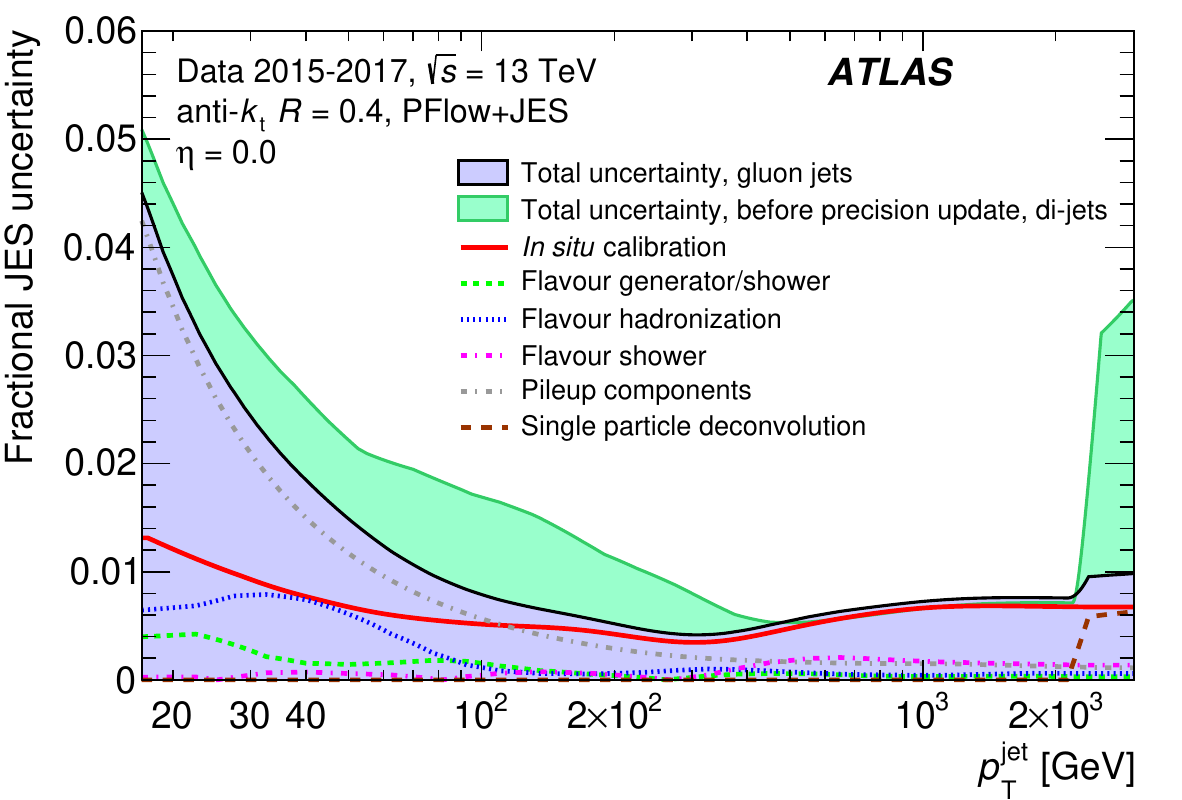}\label{fig:systematics:jes:newflavor:a}}
\subfigure[]{\includegraphics[width=0.49\textwidth]{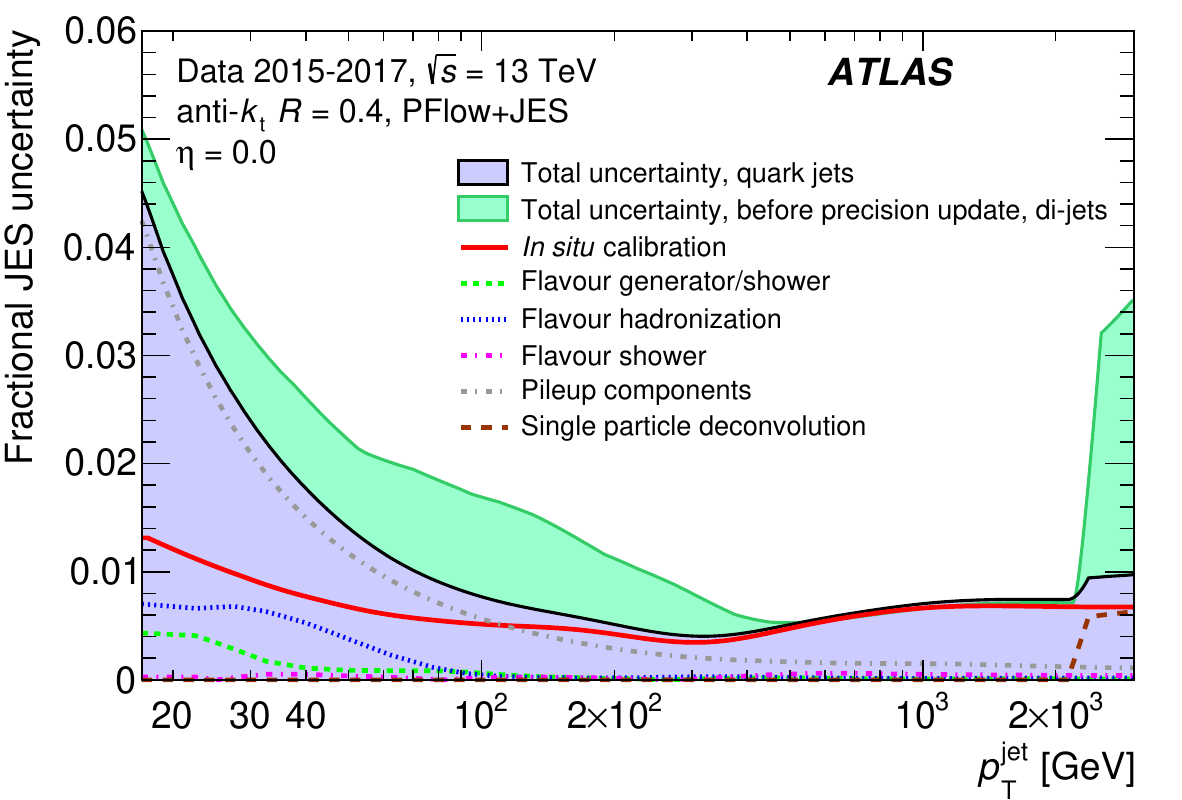}\label{fig:systematics:jes:newflavor:b}}\\
\subfigure[]{\includegraphics[width=0.75\textwidth]{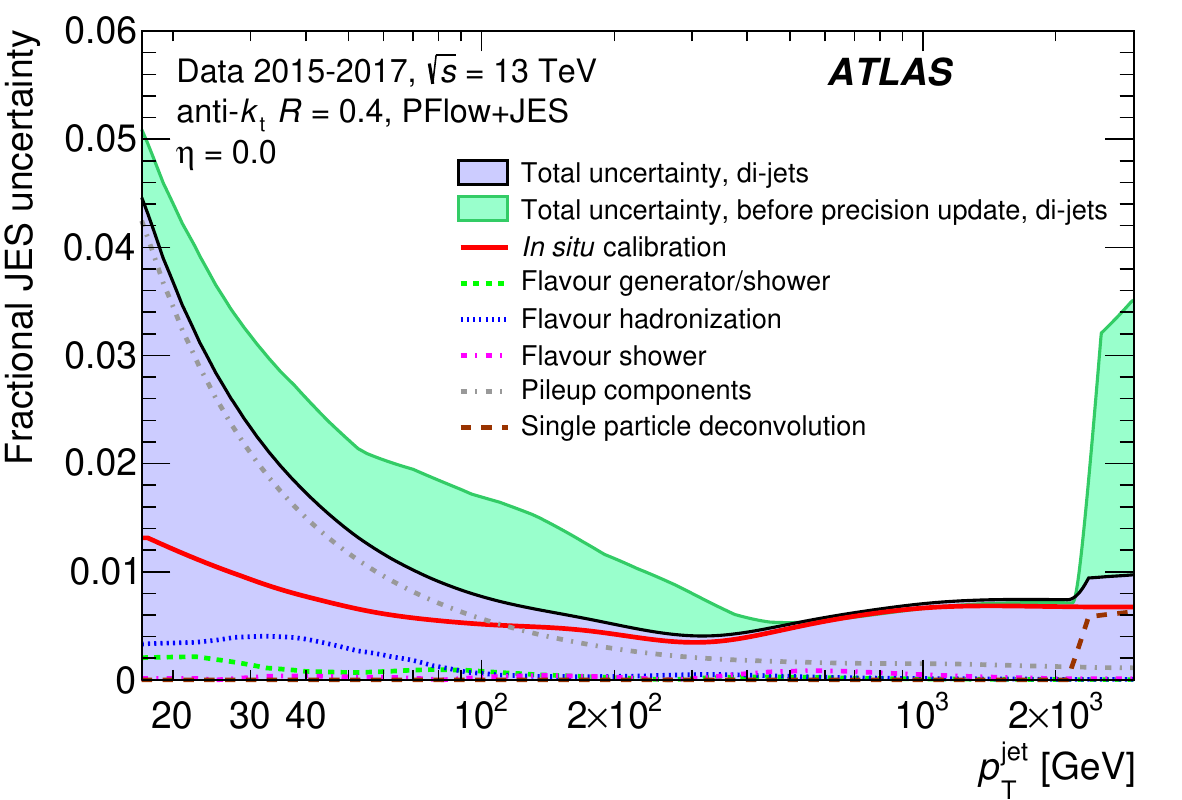}\label{fig:systematics:jes:newflavor:c}}
\caption{The jet energy scale uncertainty for (a) gluon- and (b) quark-initiated jets produced at central pseudorapidities ($|\eta|=0$).
In (c), the overall uncertainty for a dijet flavor composition is shown.
The difference between the total uncertainty obtained from the previous prescriptions documented in Ref.~\cite{JETM-2018-05} and that obtained from the updated uncertainties is indicated by a filled green region, while the new total uncertainty is indicated by the filled blue region.
Subcomponents of the uncertainty originating from the flavor generator/shower, shower and hadronization comparisons are indicated by different lines.
}
\end{figure}

\FloatBarrier
\subsection{Choice of MC models}\label{sec:systematics:modeling}

When unfolding a measurement using iterative Bayesian unfolding, one must select a nominal MC simulation to construct the response matrix that is applied to data.
Different results can be obtained if a different MC model is used to define the unfolding procedure, as the alternative underlying particle spectrum can change the prior, response matrix, fake and efficiency factors.
To account for the uncertainty related to the choice of nominal MC model, the unfolding procedure is repeated with the nominal \PYTHIA prior but a response matrix, fake factors, and efficiency factors constructed using an alternative MC simulation.

The alternative sample used to define this uncertainty is the \SHERPA[2.2.5] sample with cluster-based hadronization, which varies many aspects of the simulation with respect to the nominal \PYTHIA sample (Section~\ref{sec:atlas:mc}), including the PS algorithm and PDF set.
Despite the differences between these setups, both provide good descriptions of the measured data.
In order to factorize modeling effects due to the JES and reduce double-counting, the average JES of the alternative \SHERPA samples was recalibrated to match that of the \PYTHIA sample when evaluating this uncertainty.
This MC-to-MC correction procedure significantly reduces the size of the MC modeling systematic uncertainty, from a few percent to less than 1\% for most of the measurement.

When changing the MC model, the effects on the analysis efficiencies, acceptance and unfolding response matrix are considered individually.
The three components of this uncertainty are summed in quadrature to obtain the total modeling uncertainty.

\FloatBarrier
\subsection{Unfolding methodology: statistical uncertainties and nonclosure}%

Statistical uncertainties arise from the finite MC and data sample sizes in the measurement, and are estimated during the unfolding procedure with Poissonian pseudo-experiments, as described in Ref.~\cite{ATL-PHYS-PUB-2021-011}.
For the MC simulation, pseudo-experiments are used to vary the response matrix used for the unfolding procedure.
The input MC prior is unfolded with each varied response matrix.
The efficiencies and acceptances are allowed to vary during this process.
For the data statistical uncertainty, pseudo-experiments are generated to vary the input data spectrum (\enquote{prior}) for the unfolding procedure, and are then unfolded using the nominal \PYTHIA response matrix.
One hundred pseudo-experiments are generated in both cases; using larger numbers of pseudo-experiments did not alter the results significantly.
The 68\% interquantile range of the output distributions generated as a result of these variations is taken as the corresponding statistical uncertainty.

The nonclosure uncertainty in the unfolding procedure is evaluated using a data-driven reweighting procedure~\cite{Malaescu:2009dm}.
The detector-level \PYTHIA spectrum is reweighted to match the observed data spectrum, and then unfolded with the nominal \PYTHIA response matrix.
The difference between this unfolded result and the nominal \PYTHIA particle-level spectrum is taken as a systematic uncertainty.

\FloatBarrier
\subsection{Other experimental uncertainties}%

Other uncertainties related to experimental effects are accounted for in this analysis.
They are typically small, but can occasionally be significant in certain measurement bins.

The uncertainty in the absolute luminosity measurement is applied as a $0.83\%$ variation of the normalization of the data~\cite{DAPR-2021-01}.
This uncertainty is negligible for the measurement of cross-section ratios.

Uncertainties due to the mismodeling of \pileup events are included by reweighting the distribution of the average number of \pileup interactions and are found to be negligible throughout the measurement.

During certain \RunTwo data-taking periods, specific modules of the tile calorimeter were disabled due to technical problems.
Some of these modules are also disabled in the simulated events corresponding to a given data-taking period, while other modules that were temporarily disabled during data-taking were not disabled in the simulation.
No additional correction is applied to the \pt of jets which may have deposited energy in disabled tile modules.
The impact of the disabled tile modules on the unfolded distributions is evaluated by repeating the measurement and vetoing events with jets directed at disabled modules in either data or the nominal \PYTHIA sample.
Differences between these results with vetoed events and the nominal set are taken as a source of systematic uncertainty.

\subsection{Summary of experimental uncertainties}

A breakdown of the experimental uncertainties for two representative distributions, \HTTwo with \ptthree$>60$~\GeV~and \dymax, for the cross-sections and their ratios is shown in Figure~\ref{fig:systematics:example}.
For \HTTwo, the JES uncertainties dominate everywhere, highlighting the importance of the JES uncertainty reductions.
The MC-to-MC correction improves the consistency of the jet \pt modeling between MC generators, rendering this a subleading source of uncertainty for \HTTwo.
For \dymax, the JES uncertainties dominate everywhere except the smallest rapidity differences, where the modeling uncertainty dominates.

\begin{figure}[t!]
\centering
\subfigure[]{\includegraphics[width=0.49\textwidth]{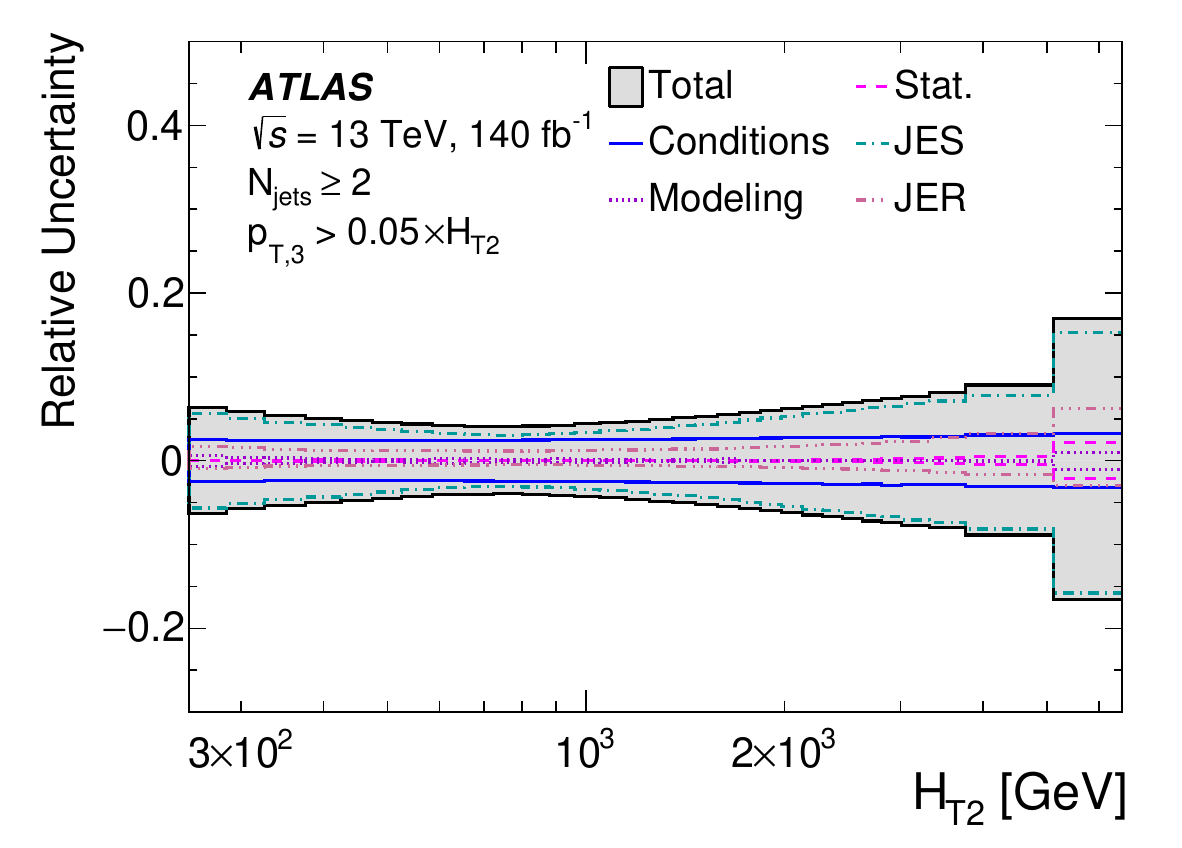}}
\subfigure[]{\includegraphics[width=0.49\textwidth]{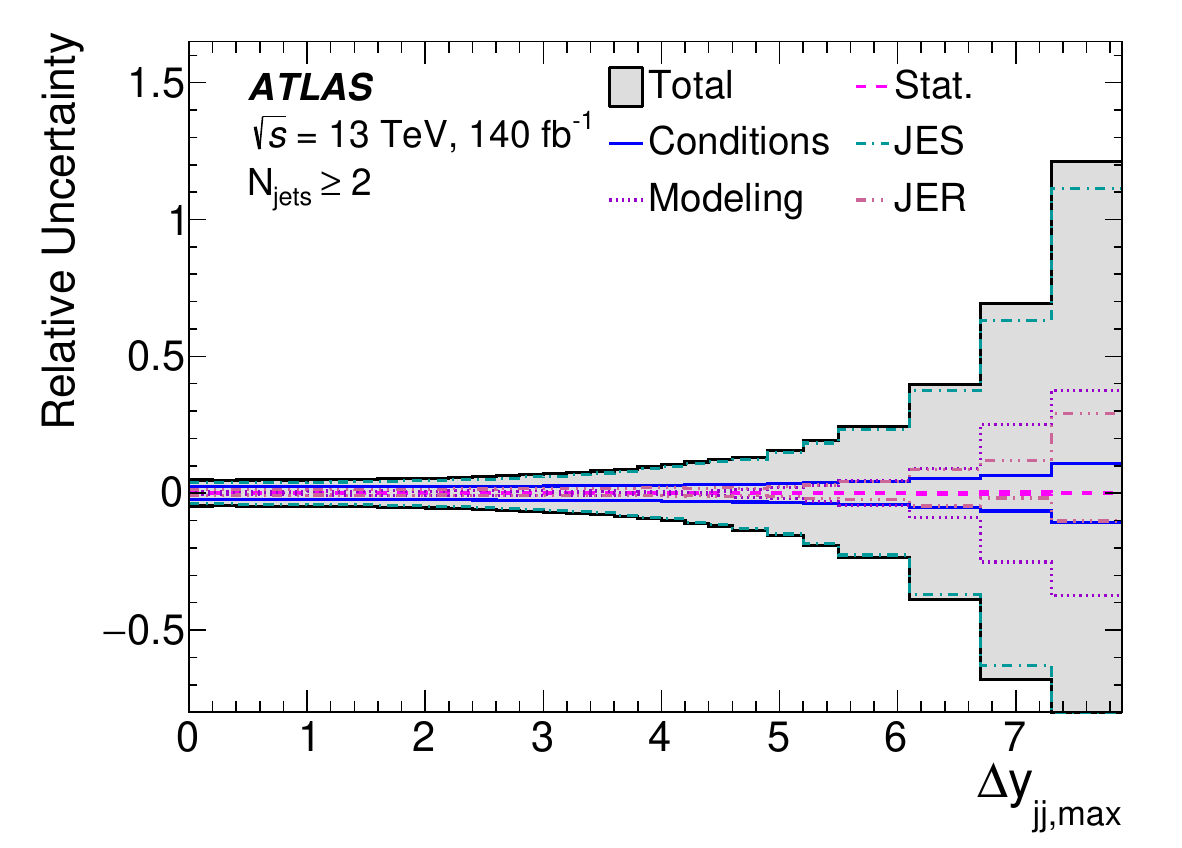}}\\
\subfigure[]{\includegraphics[width=0.49\textwidth]{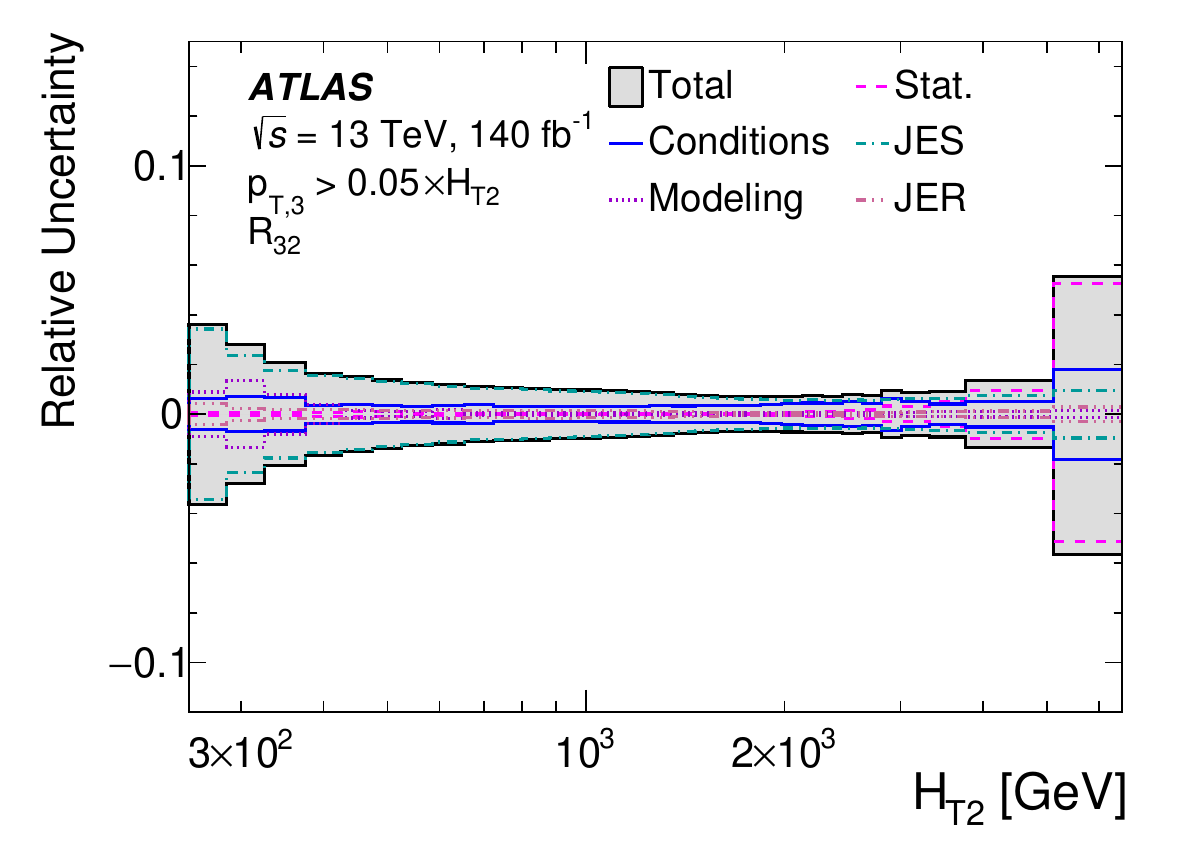}}
\subfigure[]{\includegraphics[width=0.49\textwidth]{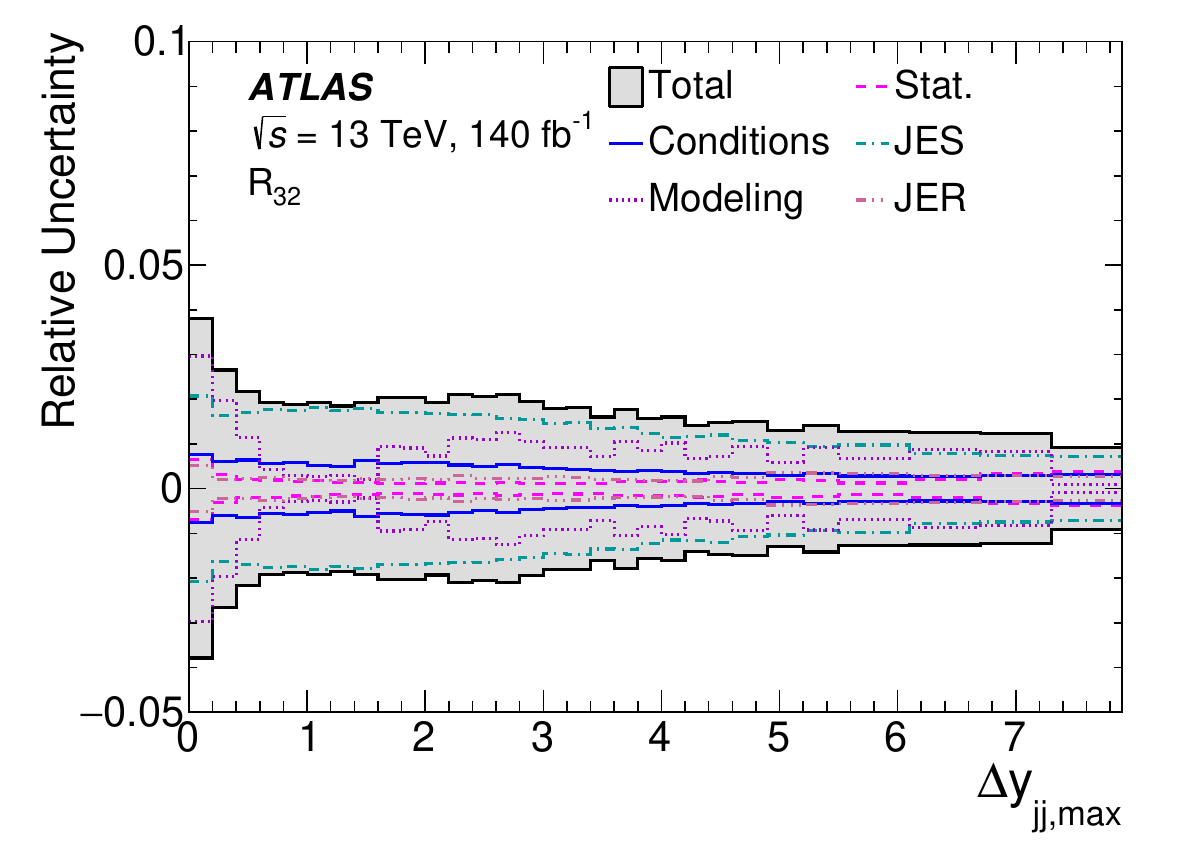}}\\
\caption{The breakdown of the experimental uncertainties for the (a, b) inclusive two-jet cross-section measurement and (c, d) \RThreeTwo measurement, differential in (a, c) \HTTwo with \ptthree$>60$~\GeV, and (b, d) \dymax.
}
\label{fig:systematics:example}
\end{figure}


%

\FloatBarrier
\section{Fixed-order QCD predictions}
\label{sec:theory}

\subsection{NLO prediction}

The theoretical predictions for three-jet and two-jet cross-sections are calculated at NLO in perturbative QCD using the \NLOjet program~\cite{nlojet1, nlojet2}.
The partonic cross-section is convolved with NNLO PDFs obtained from the LHAPDF interfaces~\cite{lhapdf} to CT18~\cite{Hou:2019efy}, \NNPDF[4.0]~\cite{nnpdf40}, MSHT20~\cite{msht20} and ATLASpdf21~\cite{STDM-2020-32}.
The PDFs are based on the $\NF = 5$ scheme, where \NF is the number of parton flavors.
The value of $\alphas(m_{Z})$ is set consistently between the partonic matrix-element calculation and the PDF; the central value is taken to be $\alphas(m_{Z})=0.118$.
The partonic events are clustered with the \akt algorithm ($R=0.4$) before the phase-space requirements of this measurement are applied.
The renormalization and factorization scales (\muR and \muF, respectively) are set to the scalar sum of the \pt of all partons in the final state, as recommended in Ref.~\cite{Currie:2015}:
\begin{equation}\label{eqn:scale}
\muR = \muF = \hat{H}_\mathrm{T} = \sum_i p_{\mathrm{T},i},
\end{equation}
where $i$ is the parton index.

The uncertainty of the CT18 PDF set is smaller than 2\% throughout the fiducial volume of the measurement.
It covers the differences between the studied PDF sets, with the exception of \NNPDF[4.0], which differs by around 2$\sigma$ (1$\sigma$) at low (high) \HTTwo.

Renormalisation and factorisation scale uncertainties are estimated by varying \muR and \muF up and down by a factor of two, avoiding configurations in which the scales are varied in different directions.
The envelope of results from this seven-point scale variation is taken as the uncertainty, which tends to be 5\% or smaller throughout the measurement.
For comparisons with ratios of jet cross-section measurements, the NLO prediction is defined as the ratio of the NLO prediction for three-jet production to the NLO prediction for two-jet production.

\FloatBarrier
\subsection{NNLO prediction}

Fixed-order predictions for the (differential) \RThreeTwo ratios are obtained at NNLO in perturbative QCD using the computational framework in Refs.~\cite{Czakon:2014oma, Czakon:2019tmo, Czakon:2021mjy, Alvarez:2023fhi}.
The evaluation of scattering amplitudes makes use of \textsc{Avhlib} \cite{Bury:2015dla}, \textsc{OpenLoops2}~\cite{Buccioni:2019sur}, \textsc{FivePointAmplitudes}~\cite{Abreu:2021oya} and \textsc{PentagonFunctions++}~\cite{Chicherin:2020oor}.
The partonic cross-sections are convolved with PDFs provided by the LHAPDF package, using the NNLO MSHT20 PDF set as the nominal one.
The perturbative QCD calculations are performed with $\NF = 5$ massless quark flavors, \emph{i.e.}\ without top-quark contributions to scattering amplitudes.
The contribution from top-quark pair production is estimated to be below 0.3\% in the relevant phase space.
The value of $\alphas(m_{Z})$ used in the partonic matrix-element calculation and PDFs is chosen to be $0.118$.
The partonic events are clustered with the \akt algorithm ($R=0.4$) before the phase-space requirements of this measurement are applied.
The renormalization and the factorization scale for each event are chosen to be the scalar sum of the \pt of all partons in the final state (Eq.~(\ref{eqn:scale})), and varied with the same seven-point scheme used for the NLO prediction.
For comparisons with ratios of jet cross-section measurements, the NNLO prediction is defined as the ratio of the NNLO prediction for three-jet production to the NNLO prediction for two-jet production.

\subsection{Nonperturbative corrections for fixed-order predictions}

In order to compare the theoretical predictions with the measured data (Section~\ref{sec:results}), nonperturbative QCD effects from hadronization and the underlying event (UE) must be included.
To determine the size of these corrections, MC predictions are obtained at hadron-level including the UE, and compared with parton-level distributions where the UE contribution is disabled in the MC generator.
The ratio of these two predictions is applied as a bin-by-bin correction to the theoretical prediction. The nonperturbative corrections are typically found to deviate from unity by about 2\% for the two- and three-jet selections separately, and generally by about $0.5\%$ for the \RThreeTwo ratio itself.

The uncertainty in this correction is estimated by changing the set of tuned parameters used for the \PYTHIA MC generator.
While the nominal correction makes use of the A14-\NNPDF[3.1nlo] ATLAS tune~\cite{ATL-PHYS-PUB-2014-021}, an alternative correction factor is calculated using the \textsc{Monash} tune~\cite{Skands:2014pea}, which uses a larger value of \alphas for final-state radiation ($0.1365$ vs.\ $0.1270$).
The difference between the two correction factors defines the systematic uncertainty of the nonperturbative corrections, and is generally smaller than 0.5\%. %

\subsection{HEJ Prediction}

The High Energy Jets (HEJ) framework~\cite{Andersen:2009nu,Andersen:2011hs,Andersen:2020yax} calculates the tower of leading logarithmic QCD corrections in $\hat{s}/\pt^2$ (where $\hat{s}$ is the parton center-of-mass energy) to all orders in the strong coupling \alphas for all relevant Standard Model processes.
These corrections are relevant in regions of phase space where jets span a large range of rapidity or where pairs of jets have a large invariant mass.
The predictions from HEJ contain both the resummation of logarithmic corrections and the matching of these to fixed-order accuracy.
This includes matching of all processes of $pp\rightarrow2j$, $3j$, $4j$, $5j$, and $6j$ to tree-level accuracy point-by-point in phase space.
All the predictions are obtained using a renormalization and factorization scale of $\hat{H}_\mathrm{T}/2$, with an independent seven-point variation of the scales by factors of two.
The PDF set is \NNPDF[3.1nlo] ($\alphas=0.118$), and $R=0.4$ and $p_{\mathrm{T, min}}=60$~\GeV are used throughout the \antikt algorithm.


%

%
\section{Results}
\label{sec:results}

\FloatBarrier
\subsection{Measured cross-sections}\label{sec:results:xs}

The unfolded cross-section measurements for the different observables studied in this analysis are shown in Figures~\ref{fig:xs:HT2}--\ref{fig:xs:mjj}.

The differential cross-section as a function of \HTTwo is compared to several MC generator predictions in Figure~\ref{fig:xs:HT2}, for various requirements on the inclusive jet multiplicity and the transverse momentum of the third jet (\ptthree).
No single MC prediction is able to describe the data across all \HTTwo and multiplicity bins.
The \PYTHIA prediction has an approximately constant offset relative to the data, with the offset decreasing at larger values of \ptthree.
The two \SHERPA models have nearly identical behavior, since the \HTTwo cross-section is not significantly impacted by the hadronization model.
The \HERWIG model underestimates the two-jet cross-section, but provides a good description of higher multiplicities, except for the highest \ptthree bin, where no models provide a good description.

\begin{figure}[tbh]
\centering
\subfigure[]{\includegraphics[width=0.41\textwidth]{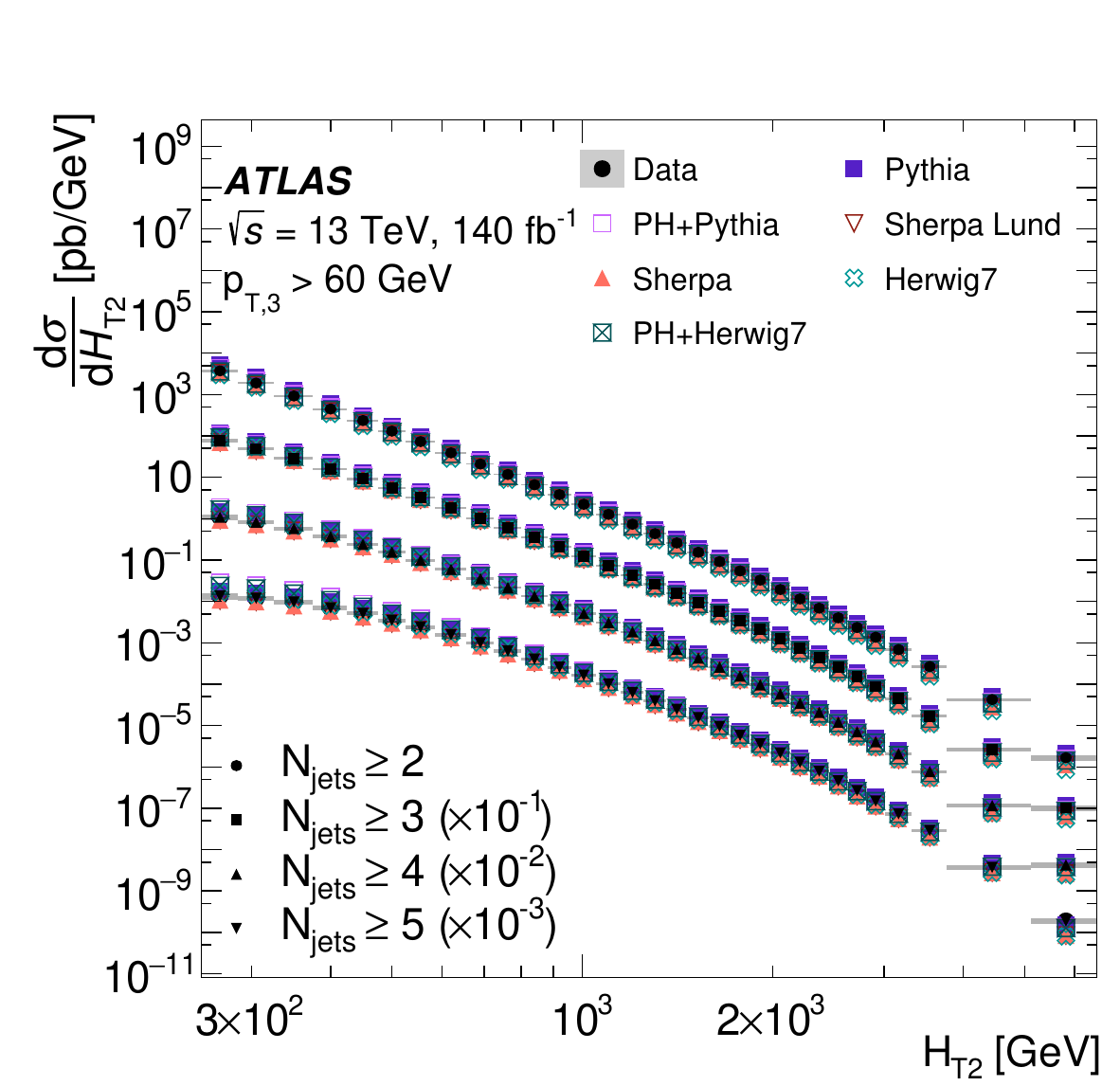}}
\subfigure[]{\includegraphics[width=0.41\textwidth]{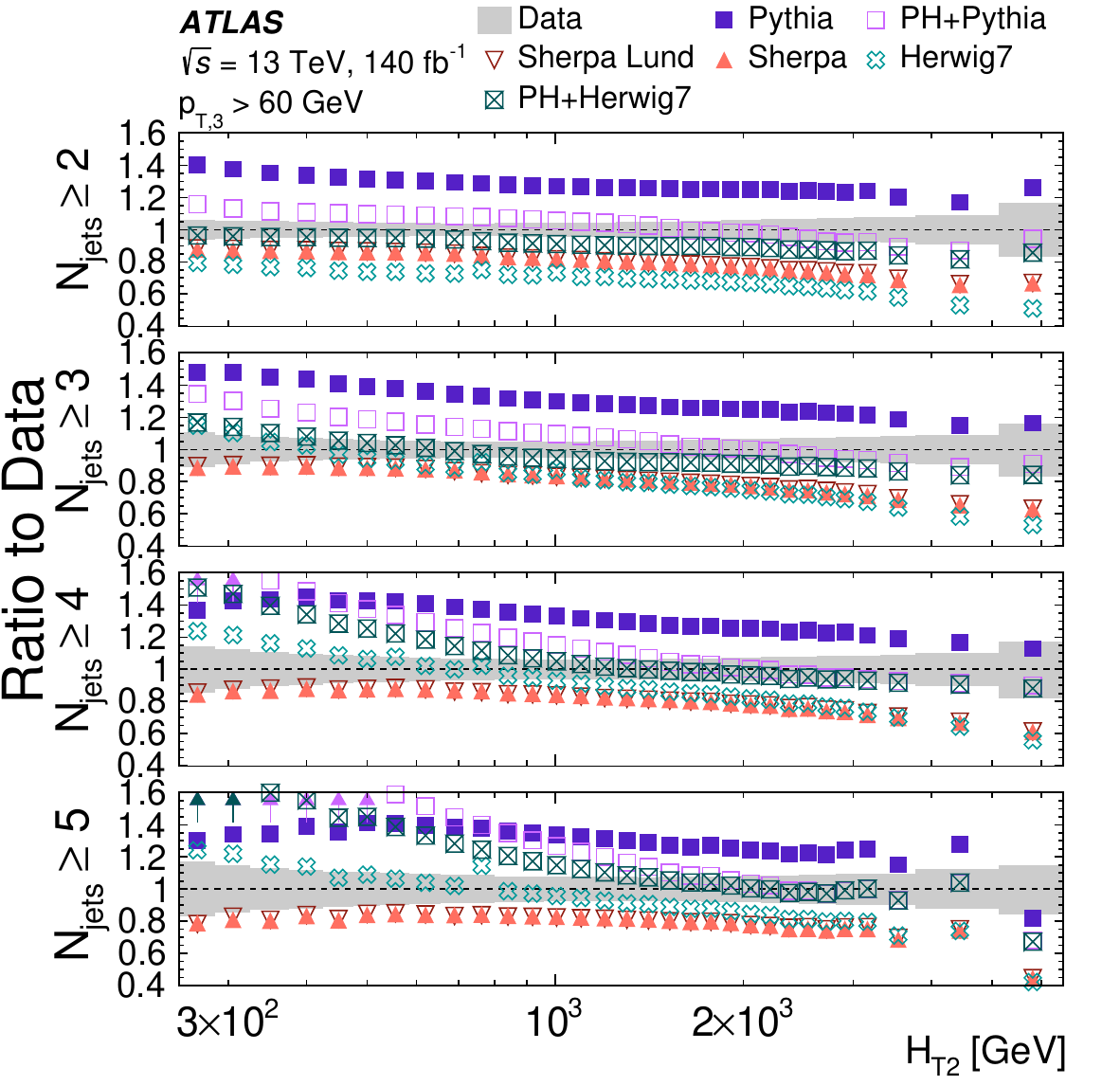}}
\subfigure[]{\includegraphics[width=0.41\textwidth]{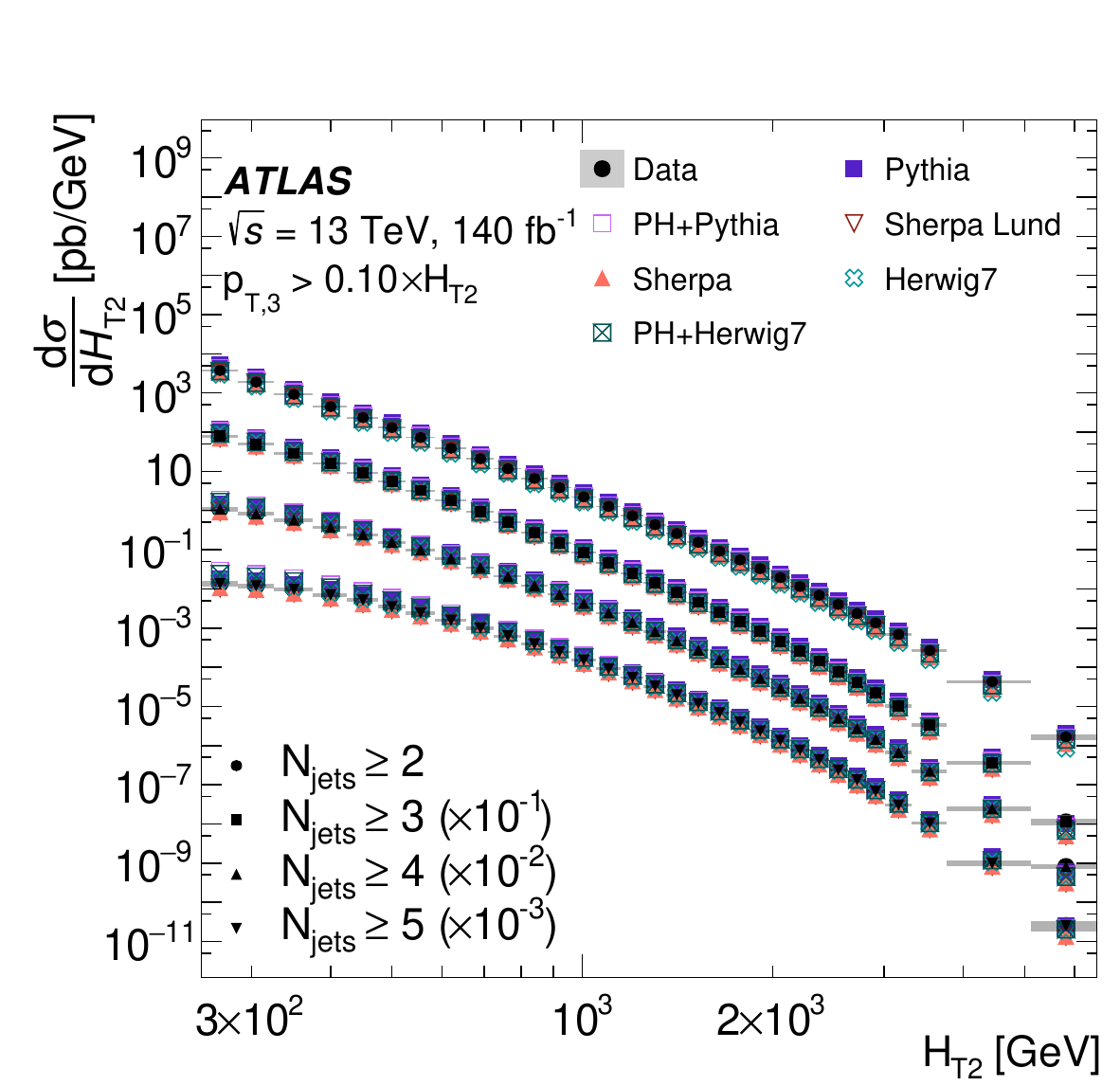}}
\subfigure[]{\includegraphics[width=0.41\textwidth]{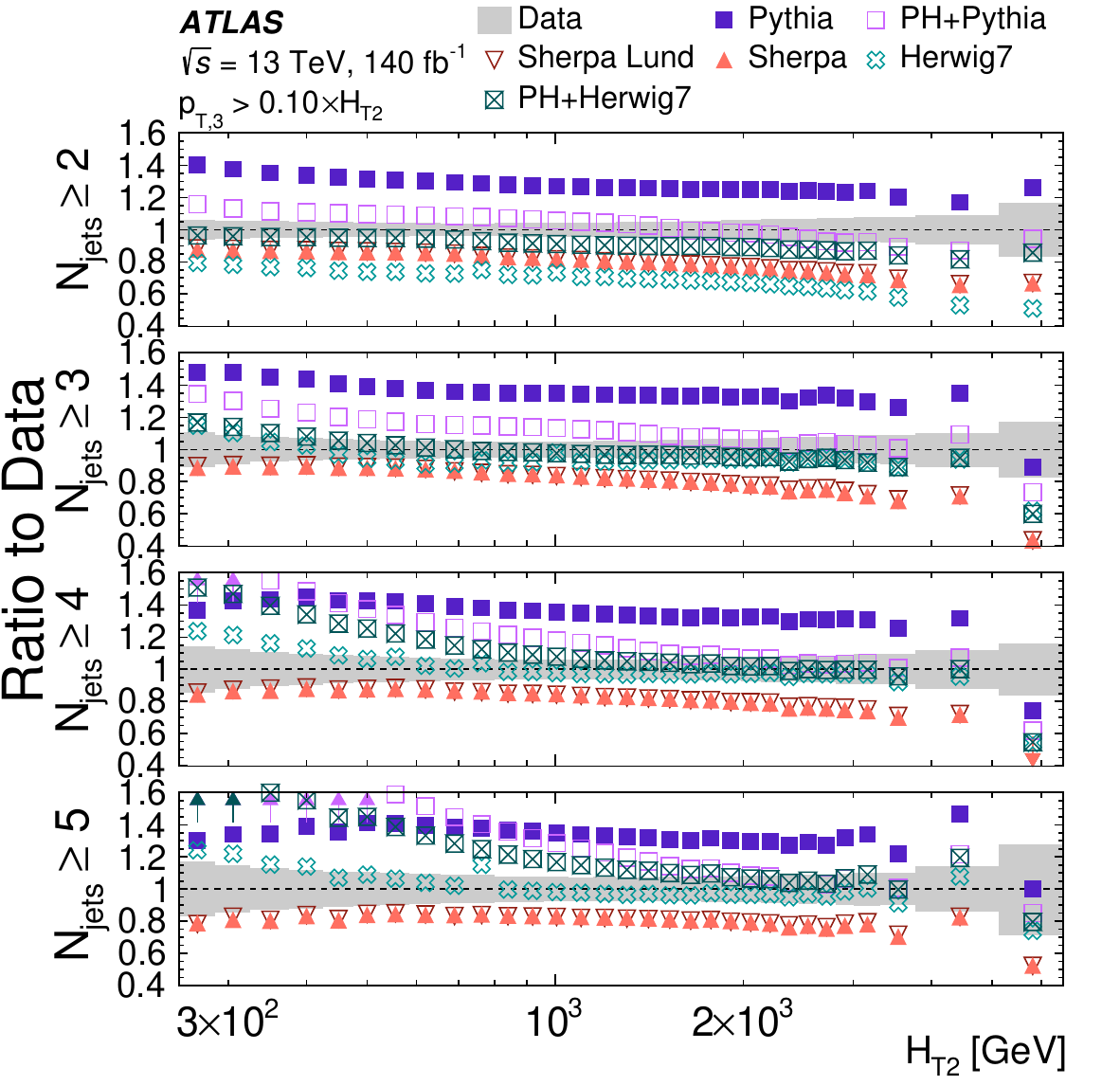}}
\subfigure[]{\includegraphics[width=0.41\textwidth]{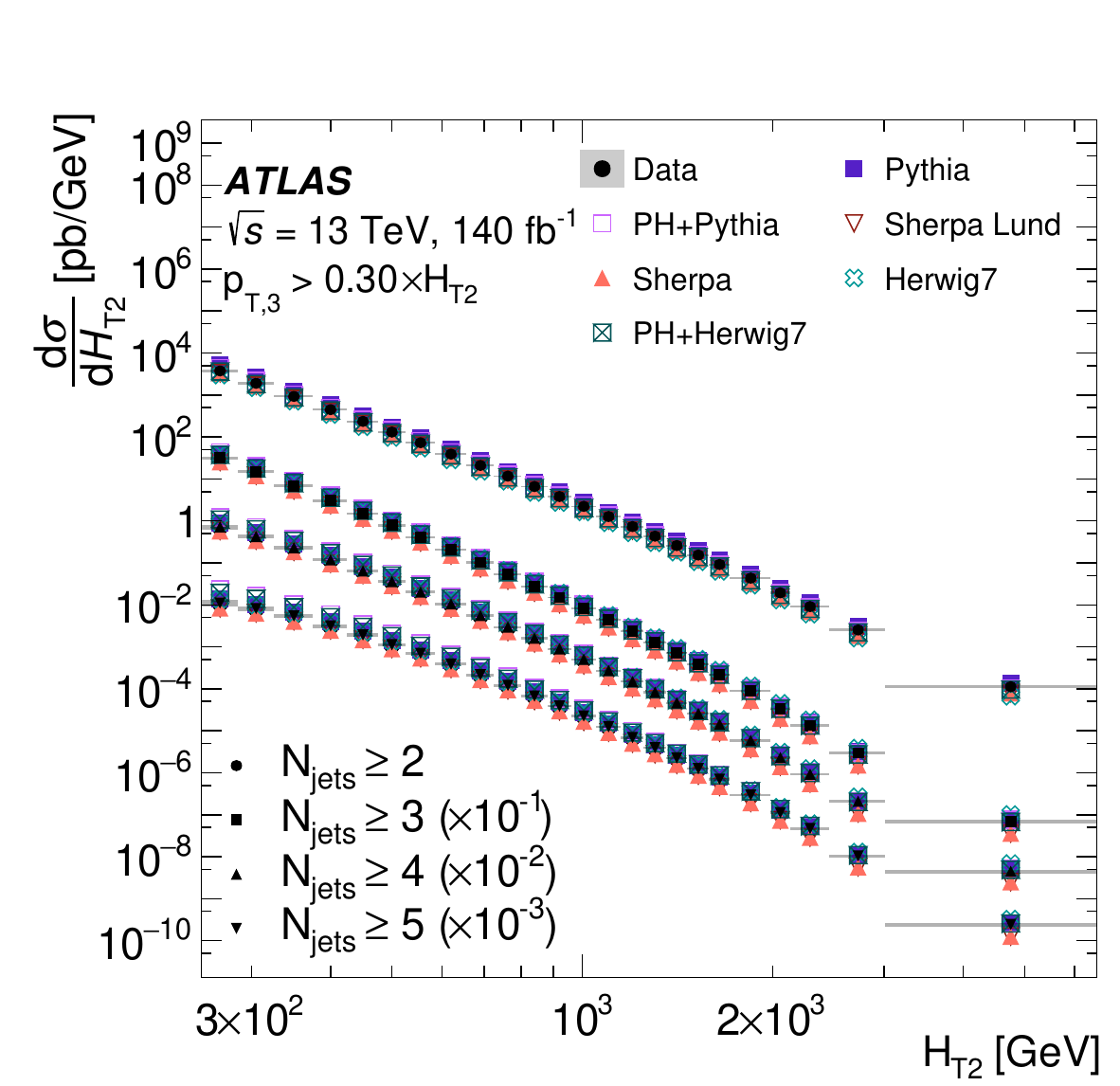}}
\subfigure[]{\includegraphics[width=0.41\textwidth]{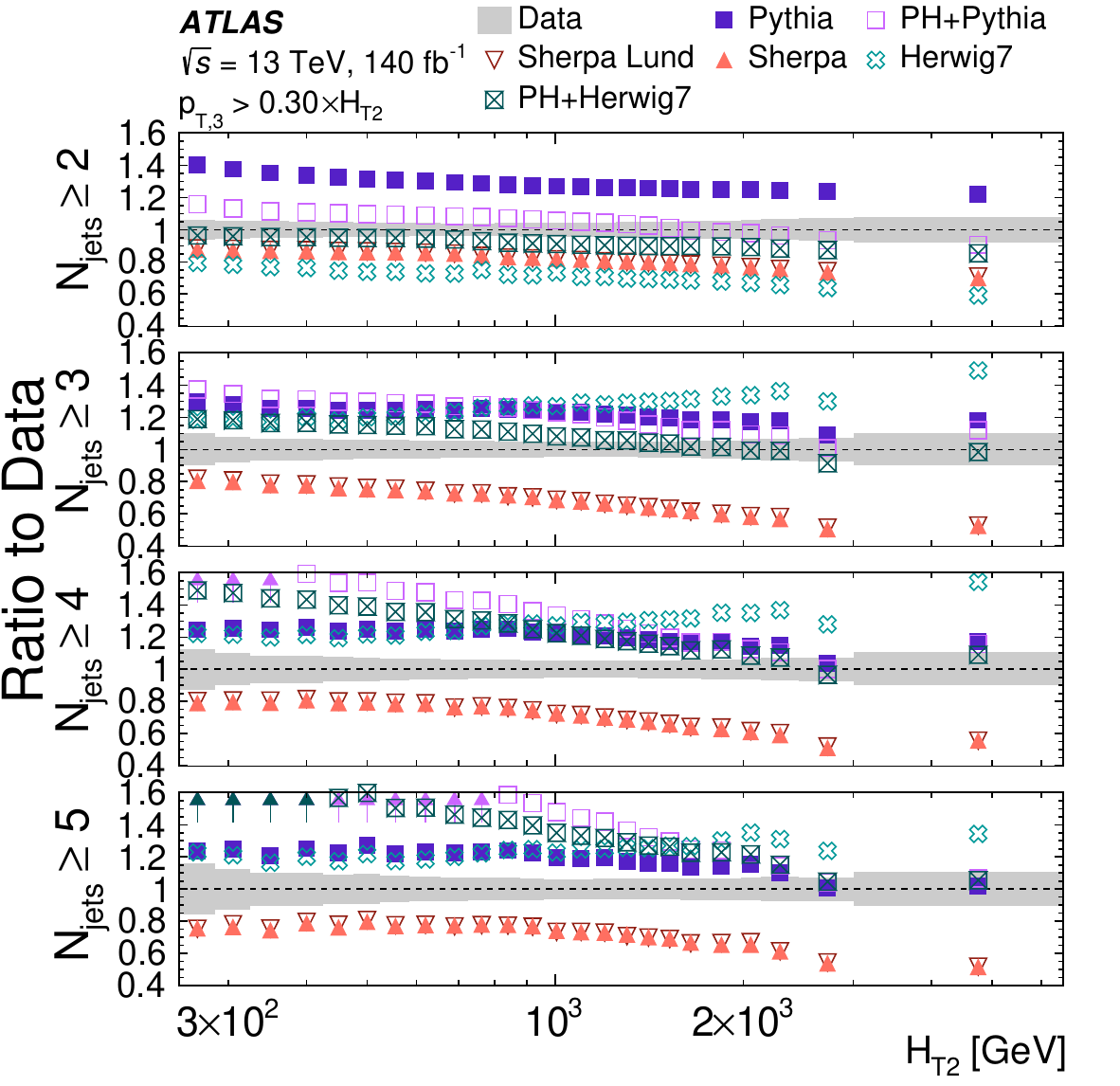}}
\caption{ (a, c, e) The differential cross-section as a function of  \HTTwo, in inclusive bins of \njets, and (b, d, f) the ratios of MC predictions to the measured data distribution vs. (a, b) $\ptthree > 60$~\GeV, (c, d) $\ptthree > 0.10\times\HTTwo$, and (e, f)  $\ptthree > 0.30\times\HTTwo$.
The data error bands show the statistical and systematic components summed in quadrature. Arrows are used to indicated cases where the ratio falls outside of the plotted ratio range.}
\label{fig:xs:HT2}
\end{figure}

The differential cross-section as a function of \ptnincl is shown in Figure~\ref{fig:xs:nincl}, differentially in bins of the inclusive number of jets.
Because of the event selection requirement that $\HTTwo > 250$~\GeV combined with the minimum \pt cut of $60$~\GeV, each event has at least two jets with \pt around 125~\GeV, resulting in a sharp downturn in \ptnincl around half of the value of the \HTTwo cut.
The MC predictions have an offset in the cross-section, which is generally constant as a function of \pt, except for $\pt < 100$~\GeV where there is also a shape difference in the predictions.
The exception to this is the \SHERPA predictions, which do not show this shape difference at low \pt.

\begin{figure}[tbh]
\centering
\subfigure[]{\includegraphics[width=0.49\textwidth]{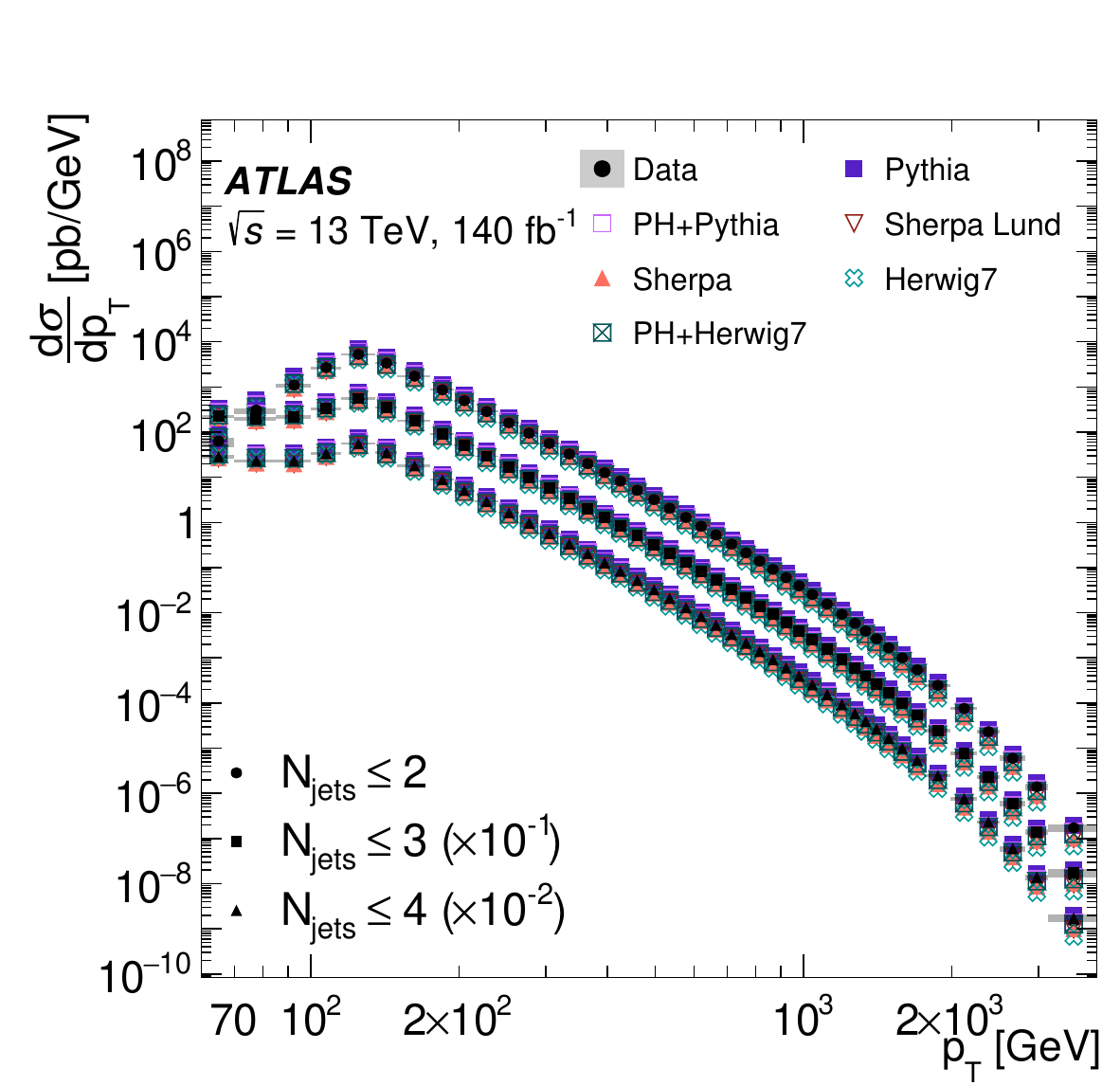}}
\subfigure[]{\includegraphics[width=0.49\textwidth]{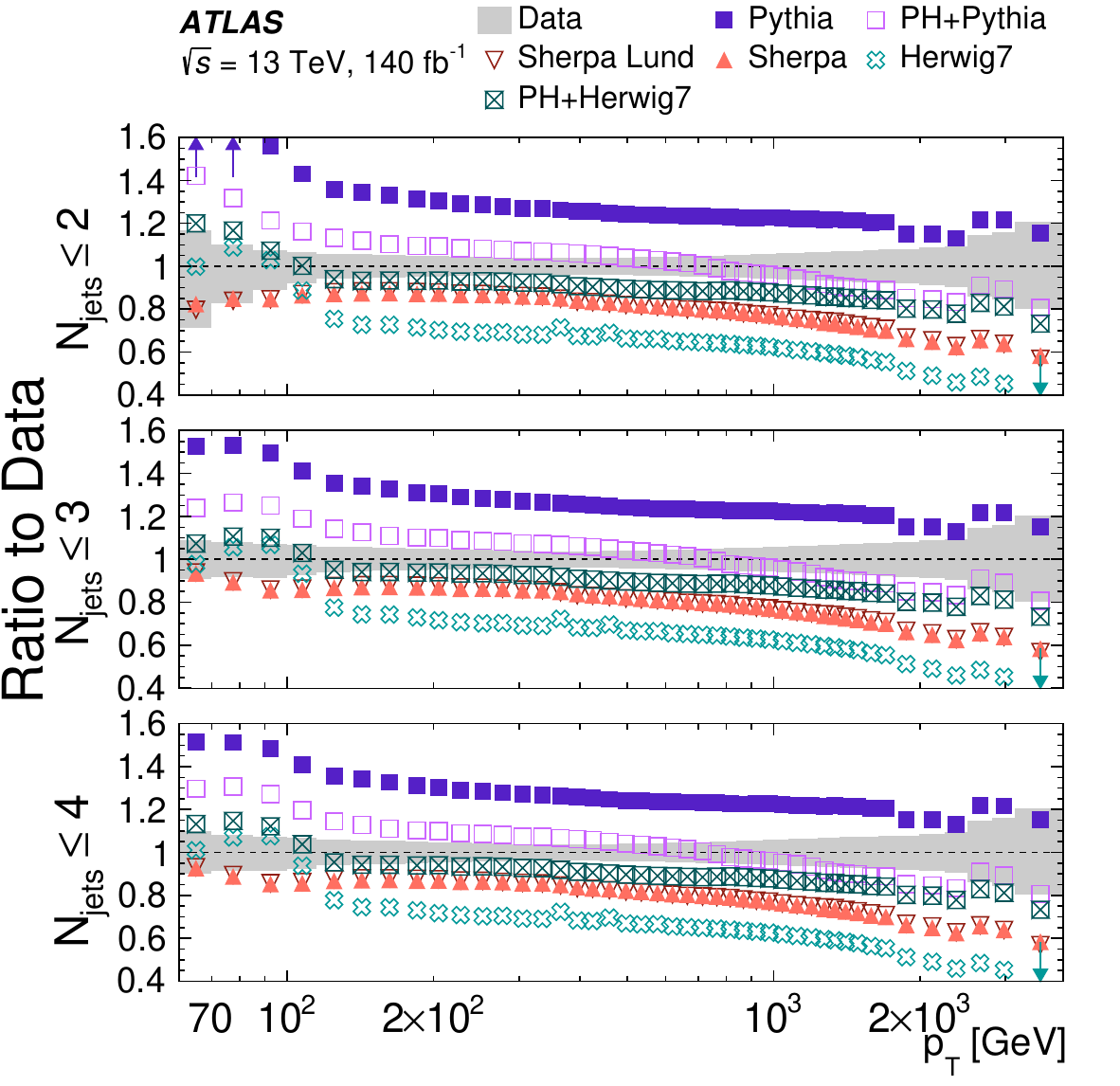}}
\caption{(a) The differential cross-section is shown as a function of \ptnincl, in inclusive bins of \njets, and (b) the ratios of MC predictions to the measured data distribution. The data error bands show the statistical and systematic components summed in quadrature. Arrows are used to indicated cases where the ratio falls outside of the plotted ratio range.}
\label{fig:xs:nincl}
\end{figure}

The differential cross-section as a function of \dyjj and \mjj is presented in Figure~\ref{fig:xs:dy}, in bins of inclusive jet multiplicity.
For both observables, \SHERPA provides the best description of the data.
\HERWIG models the data well for small rapidity differences and small dijet masses, but its performance quickly deteriorates at larger rapidity differences and masses.
\PYTHIA and \POWPY overestimate the data everwhere, while \POWHER provides a reasonable description of the data for low jet multiplicities and small values of \dyjj and \mjj, with a poor description of the data elsewhere.

\begin{figure}[tbh]
\centering
\subfigure[]{\includegraphics[width=0.49\textwidth]{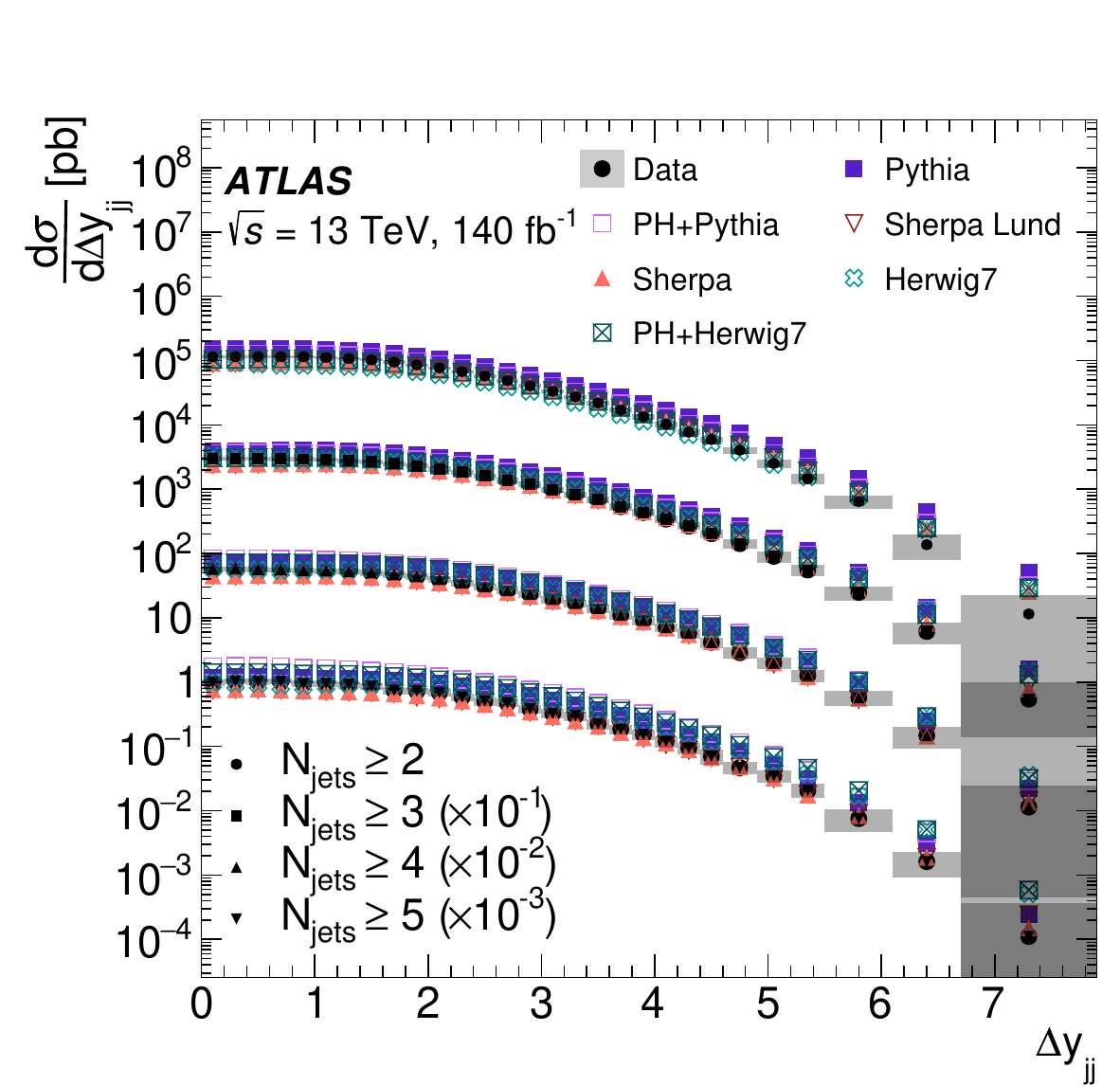}}
\subfigure[]{\includegraphics[width=0.49\textwidth]{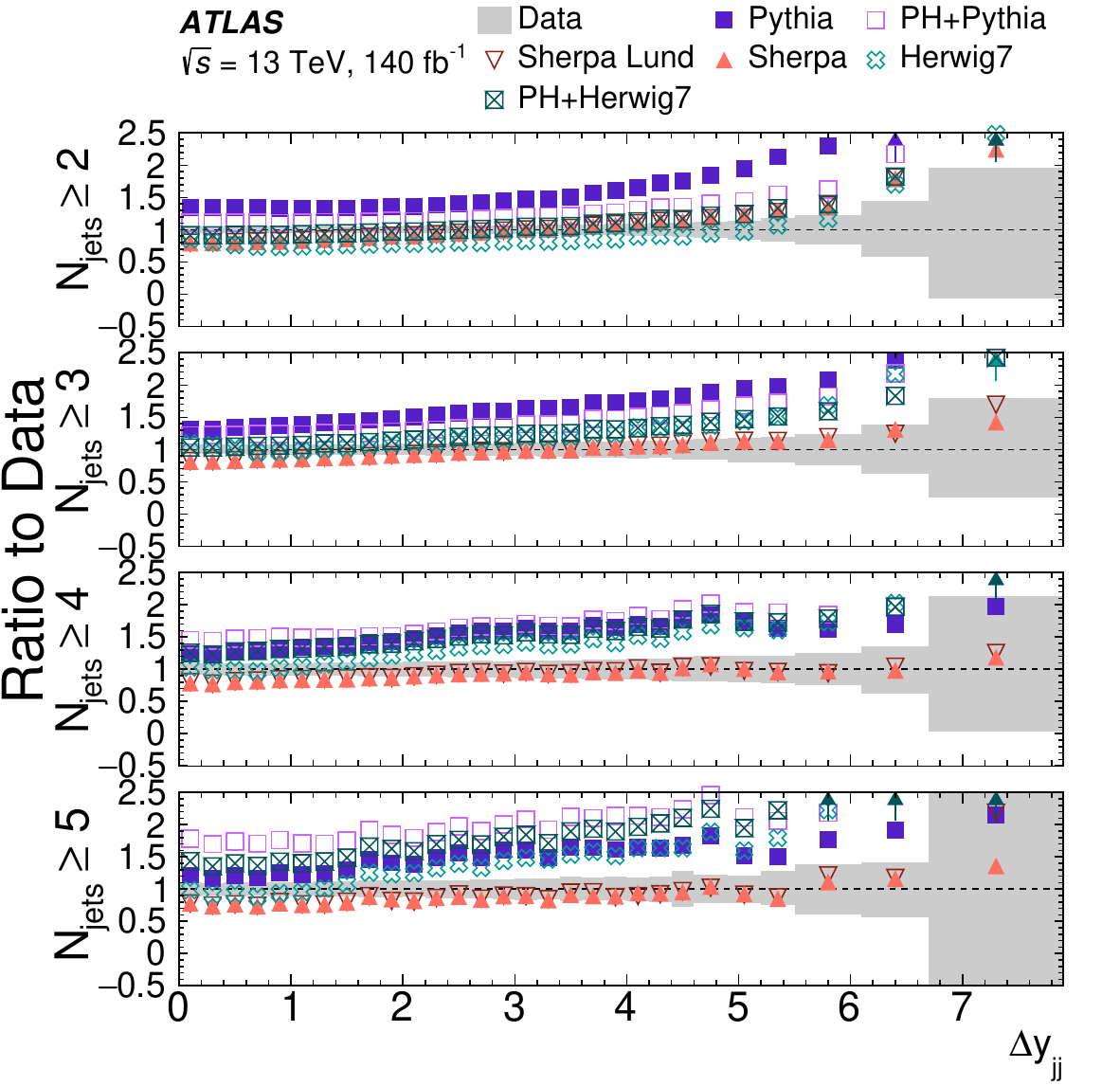}}
\subfigure[]{\includegraphics[width=0.49\textwidth]{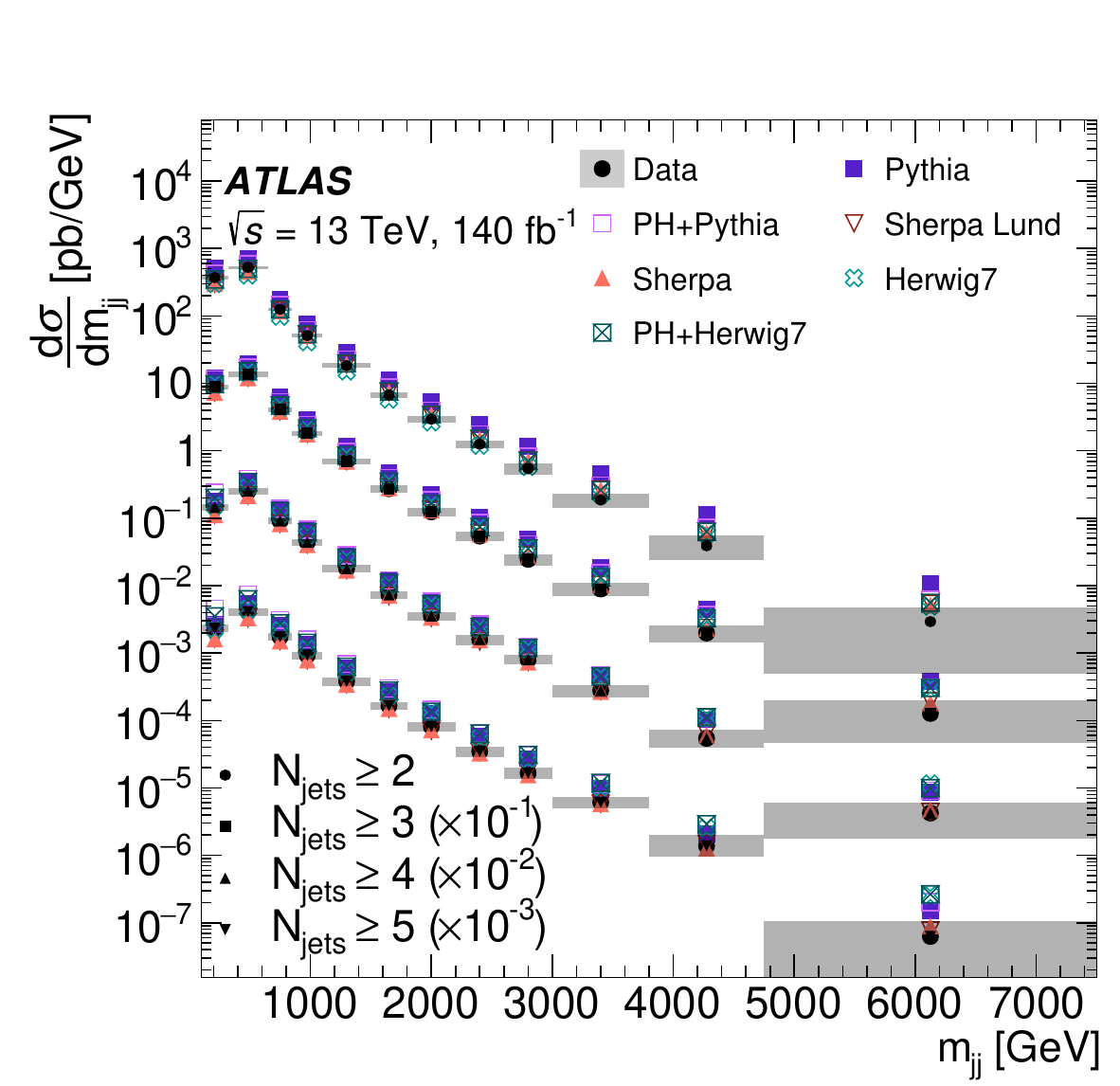}}
\subfigure[]{\includegraphics[width=0.49\textwidth]{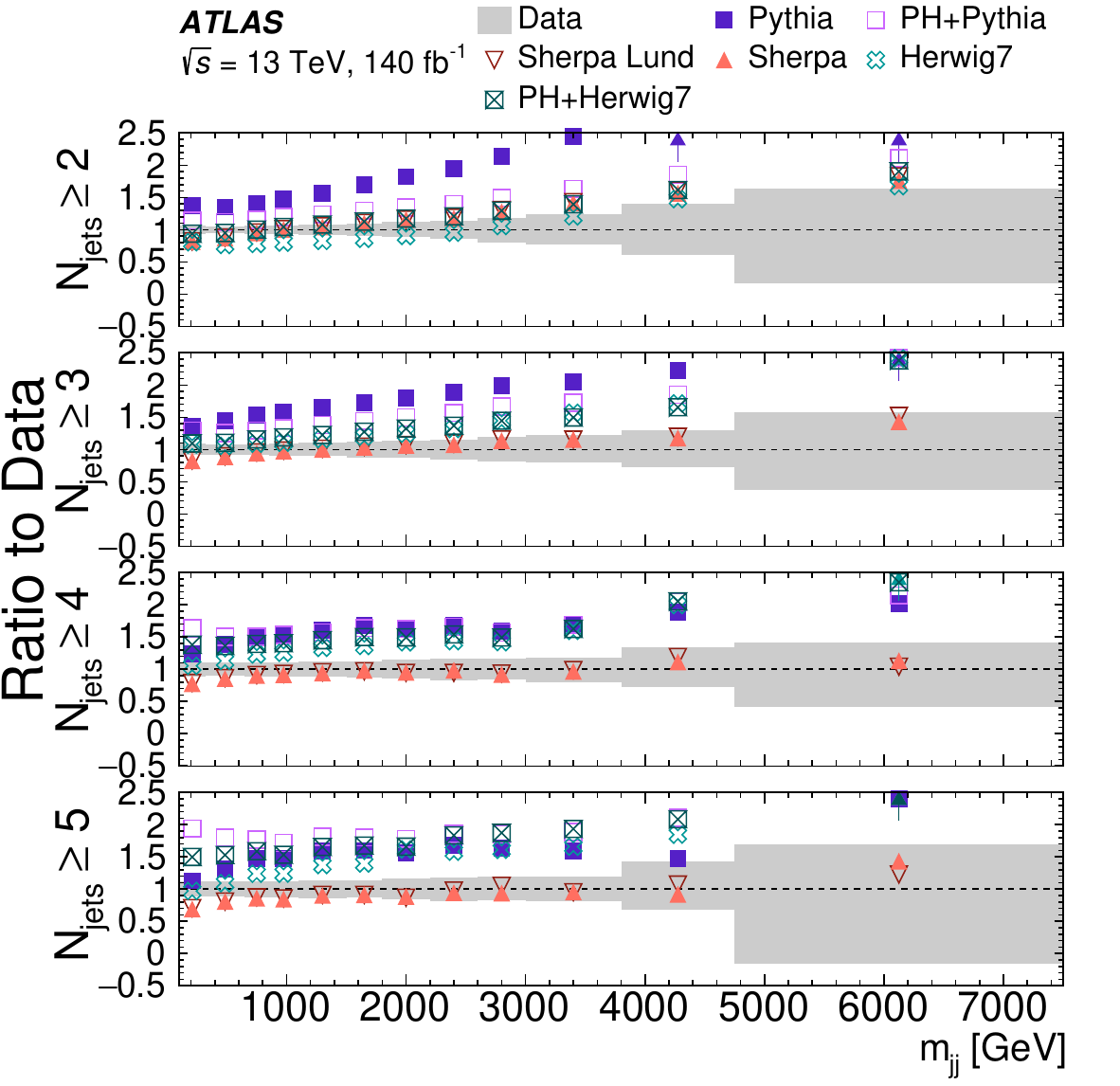}}
\caption{ The differential cross-section as a function of (a) \dyjj and (c) \mjj, in inclusive bins of \njets, and the ratios of MC predictions to the measured data distribution in bins of \njets vs. (b) \dyjj and (d) \mjj. The data error bands show the statistical and systematic components summed in quadrature. Arrows are used to indicated cases where the ratio falls outside of the plotted ratio range.
}
\label{fig:xs:dy}
\end{figure}

Finally, the differential cross-section as a function of \dymax and \mjjmax is presented in Figure~\ref{fig:xs:mjj}, in bins of inclusive jet multiplicity.
Overall, \SHERPA provides the best description of the data, but it underestimates the cross-section at low \dymax and \mjjmax.
\POWHER describes the data well at low multiplicities, but significantly overestimates the cross-section at high multiplicities, particularly for large \dymax and \mjjmax.
For \mjjmax, \PYTHIA and \POWPY overestimate the cross-section everywhere, with agreement worsening at high \mjjmax.
Similar behavior is observed for \dymax, except in the low \dymax region, where there is fair agreement.
The \HERWIG prediction differs from the data in overall shape for both observables and does not provide a good description of the data for any multiplicity.

\begin{figure}[tbh]
\centering
\subfigure[]{\includegraphics[width=0.49\textwidth]{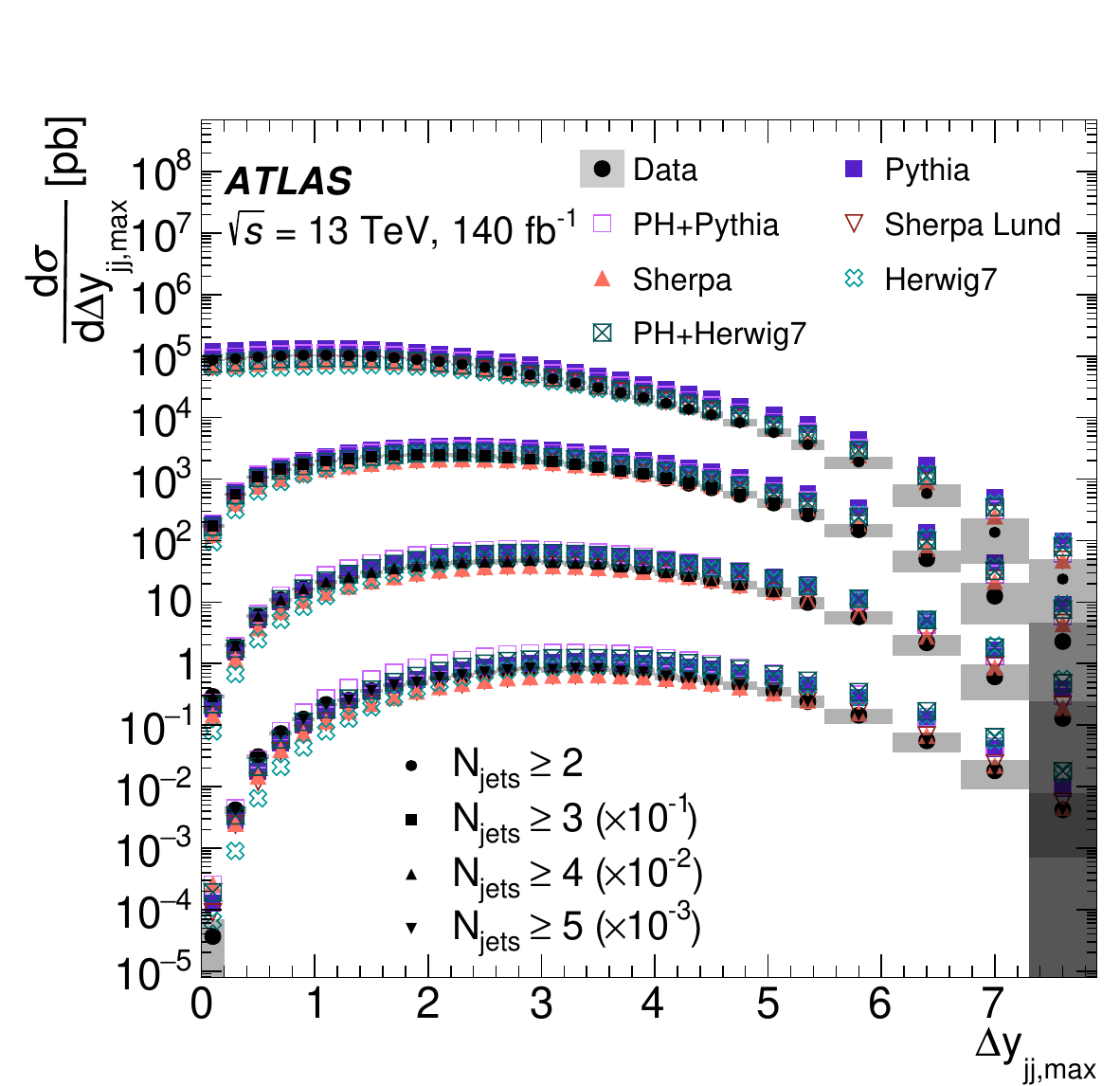}}
\subfigure[]{\includegraphics[width=0.49\textwidth]{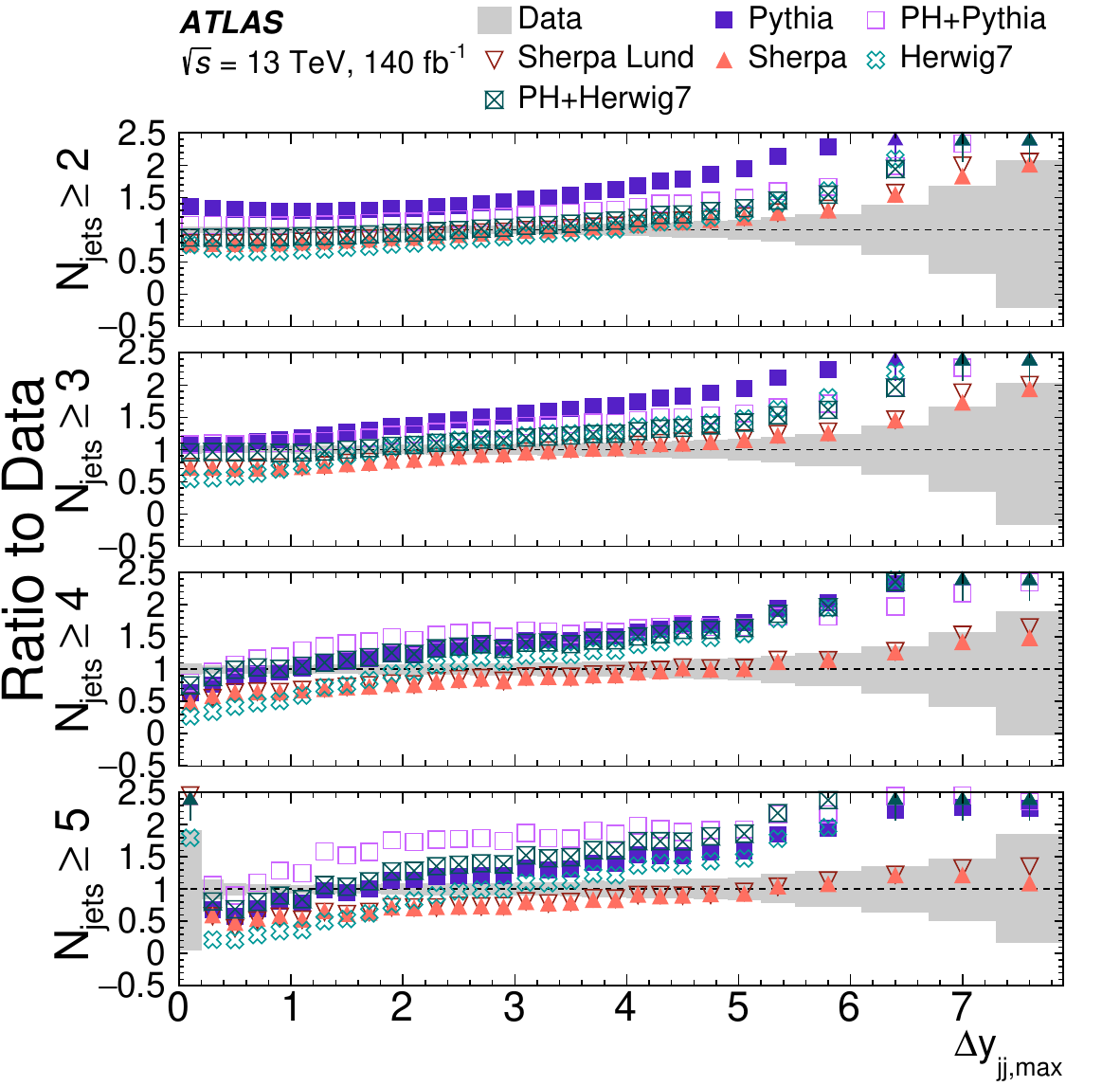}}
\subfigure[]{\includegraphics[width=0.49\textwidth]{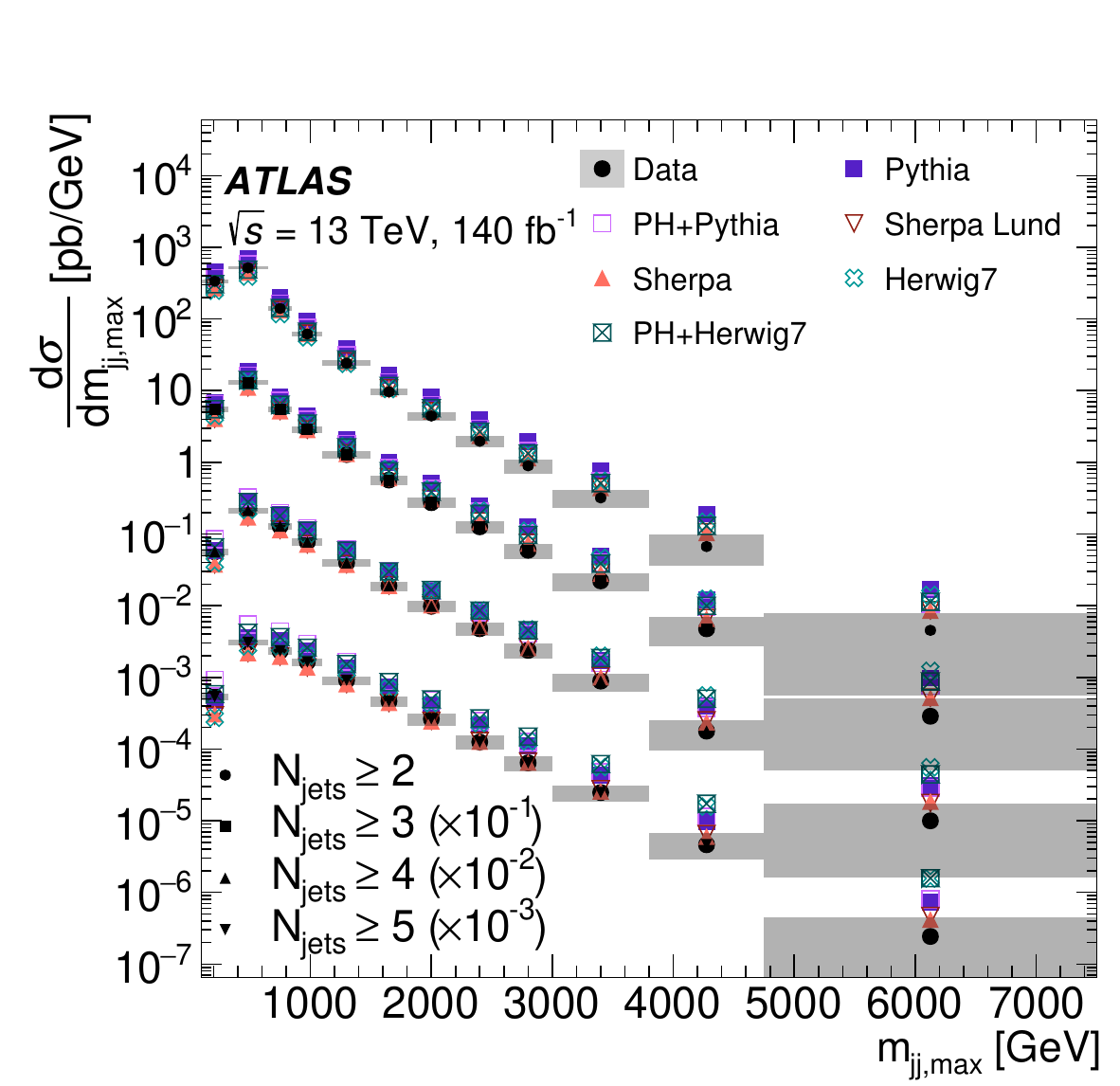}}
\subfigure[]{\includegraphics[width=0.49\textwidth]{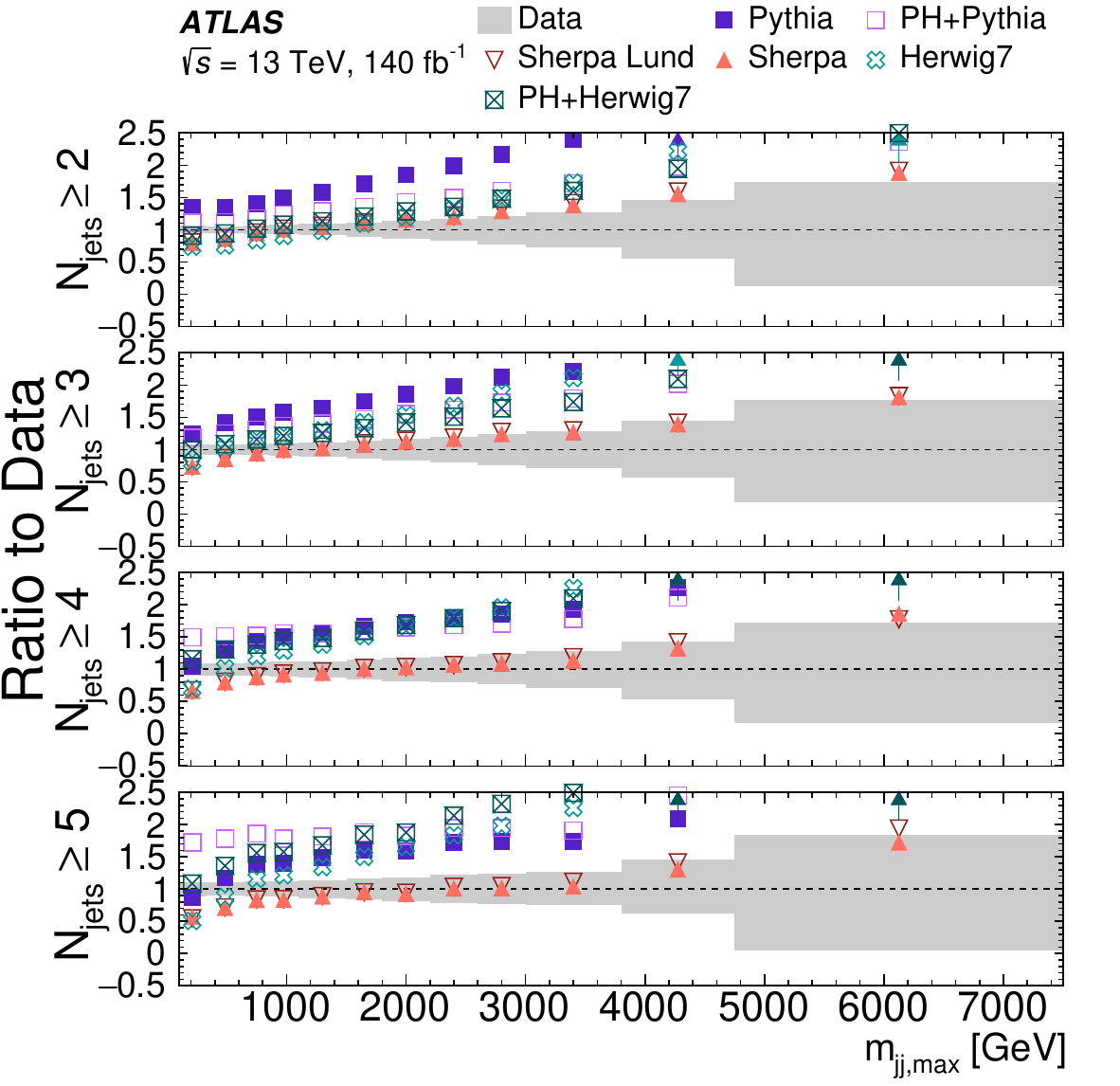}}
\caption{ The differential cross-section as a function of (a) \dymax and (c) \mjjmax, in inclusive bins of \njets, and the ratios of MC predictions to the measured data distribution in bins of \njets vs. (b) \dymax and (d) \mjjmax.
The data error bands show the statistical and systematic components summed in quadrature. Arrows are used to indicated cases where the ratio falls outside of the plotted ratio range.}
\label{fig:xs:mjj}
\end{figure}

\subsection{Cross-section ratios}\label{sec:results:ratios}

Ratios of the measured observables between different bins of inclusive jet multiplicity are presented in this section.
When compared with the cross-section measurements, the uncertainties are generally reduced because correlated systematic variations in the numerator and denominator partially cancel out.

The ratios of the measured \HTTwo distributions are shown in Figure~\ref{fig:ratios:HT2}.
The shape of the \RThreeTwo distribution changes as the \ptthree threshold is varied, because the \ptthree requirement depends on the event's \HTTwo value.
When the \ptthree cut does not depend on \HTTwo (Figure~\ref{fig:ratios:HT2:a}), the \RThreeTwo ratio increases smoothly until dropping slightly at the highest \HTTwo values, as the probability to emit a third hard parton increases with the energy scale of the event .
At the highest values of \HTTwo, events with a soft third jet are often in a back-to-back configuration: the third jet can be merged with one of the leading two, causing the ratio to decrease slightly~\cite{STDM-2019-02, STDM-2020-20}.
When the \ptthree cut is made to depend on \HTTwo, a feature related to this dependence appears at a value of \HTTwo corresponding to the ratio of the third jet's \pt threshold (60~\GeV in this case) to the fraction of \HTTwo that \ptthree must satisfy in that bin: for example, in Figure~\ref{fig:ratios:HT2:b} where $\ptthree / \HTTwo > 0.10$, the distribution turns over at a value of $60~\GeV / 0.10 = 600~\GeV$.
This turn-over point shifts to lower values as the fractional \HTTwo requirement is increased in higher \ptthree bins.
The \RThreeTwo value decreases after this point because of the steeply falling \pt spectrum of the third jet.

In general, agreement between the data and predictions worsens as the third jet's \pt cut is increased, and the \RFourThree and \RFiveFour ratios tend to be better modeled than the \RThreeTwo and \RFourTwo ratios.
\PYTHIA tends to predict slightly higher values of \RThreeTwo and \RFourTwo than seen in data, with better agreement at larger \ptthree, and agrees fairly well with the data for \RFourThree and \RFiveFour.
\POWPY overestimates the value of all four ratios, particularly at low \HTTwo.
Both \SHERPA predictions describe the data well for small values of \ptthree, but tend to underestimate the values of \RThreeTwo and \RFourTwo for large values of \ptthree, particularly at high \HTTwo.
For \RThreeTwo and \RFourTwo, \HERWIG predicts significantly fewer two-jet events than are seen in data, particularly for large values of \ptthree,
while \POWHER gives a better description of the data, particularly for large values of \HTTwo.
Both \HERWIG and \POWHER provide poorer descriptions of the data than \PYTHIA and \SHERPA for \RThreeTwo and \RFourTwo.
For \RFourThree and \RFiveFour, \HERWIG provides a relatively good description of the data, while \POWHER tends to overestimate both of these ratios.

\begin{figure}[tbh]
\centering

\subfigure[]{\includegraphics[width=0.32\textwidth]{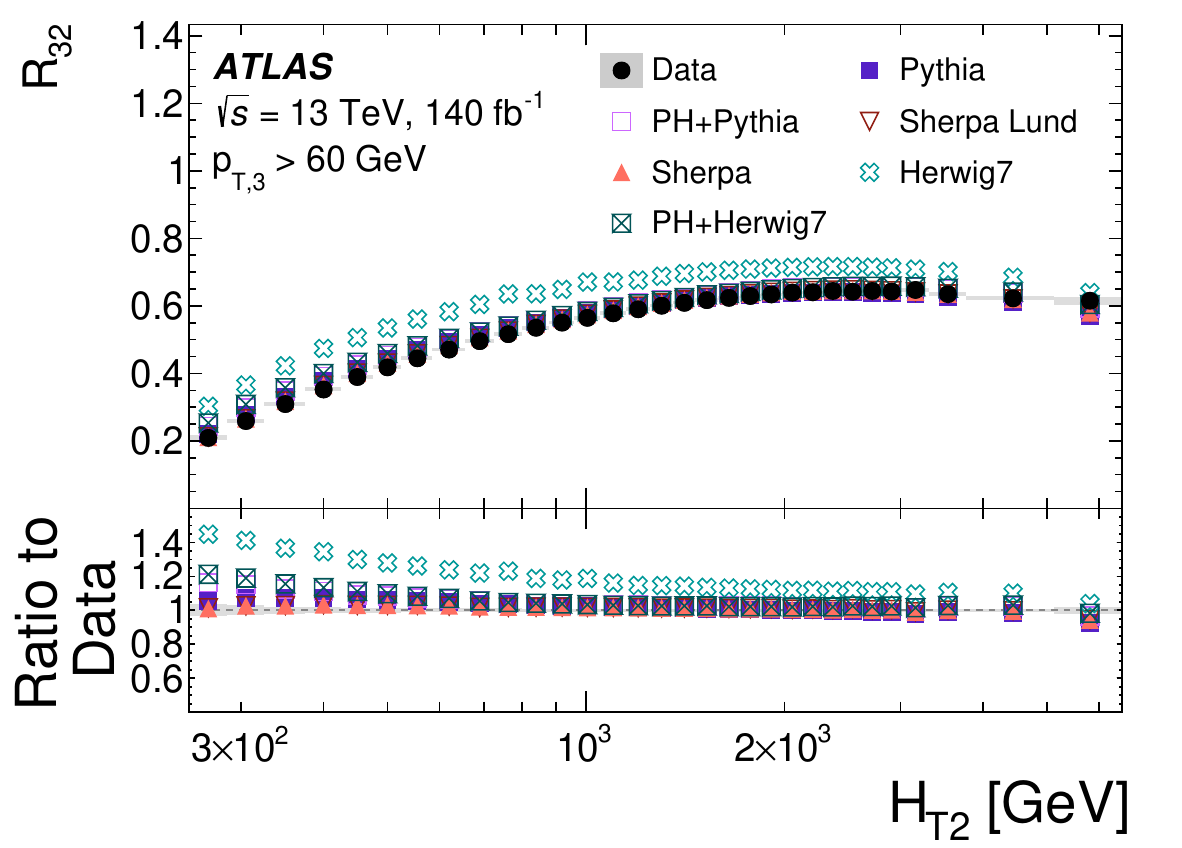}\label{fig:ratios:HT2:a}}
\subfigure[]{\includegraphics[width=0.32\textwidth]{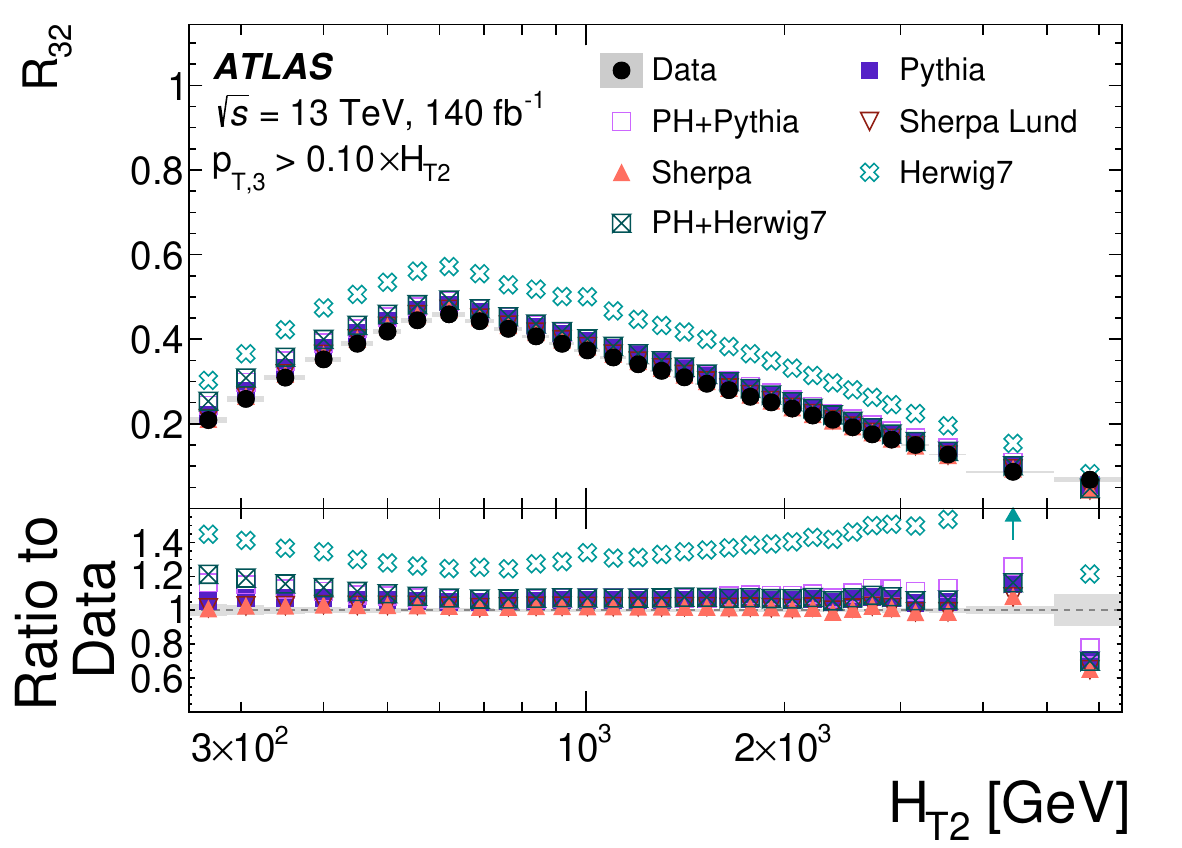}\label{fig:ratios:HT2:b}}
\subfigure[]{\includegraphics[width=0.32\textwidth]{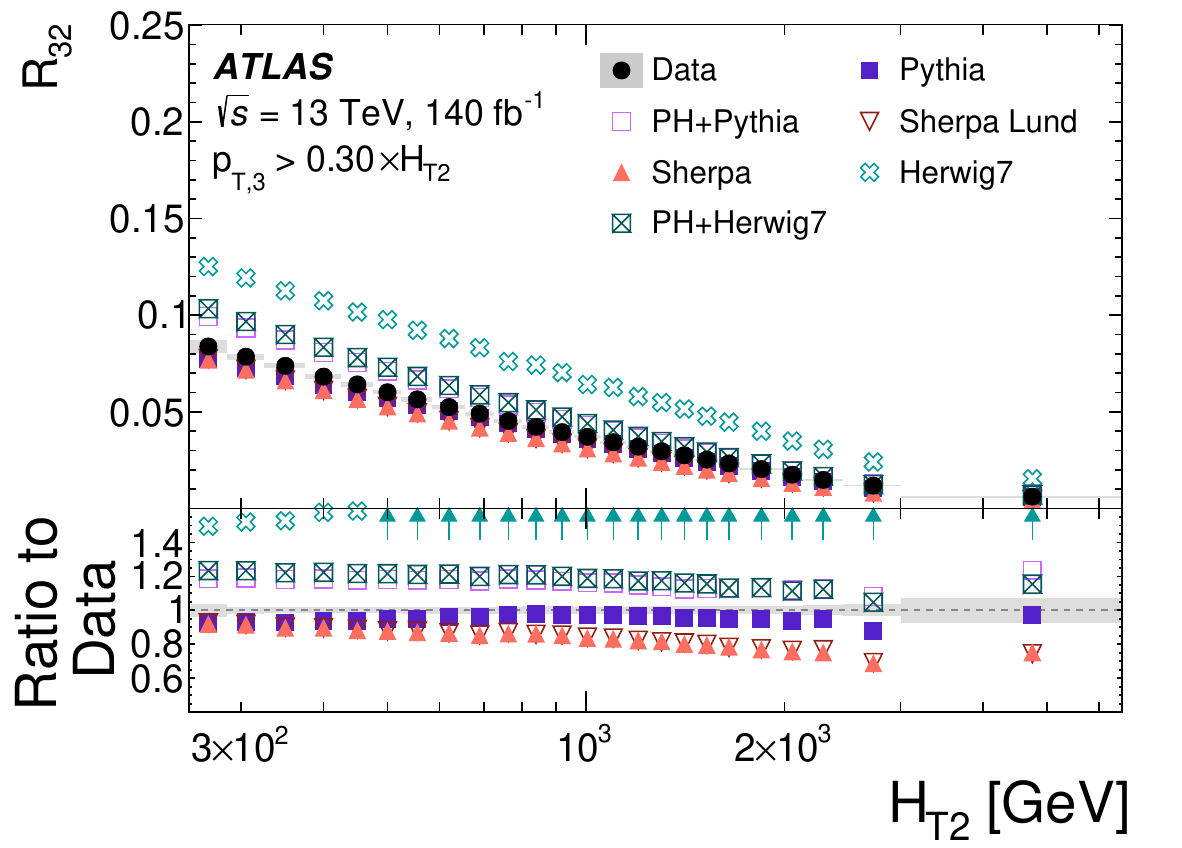}}
\subfigure[]{\includegraphics[width=0.32\textwidth]{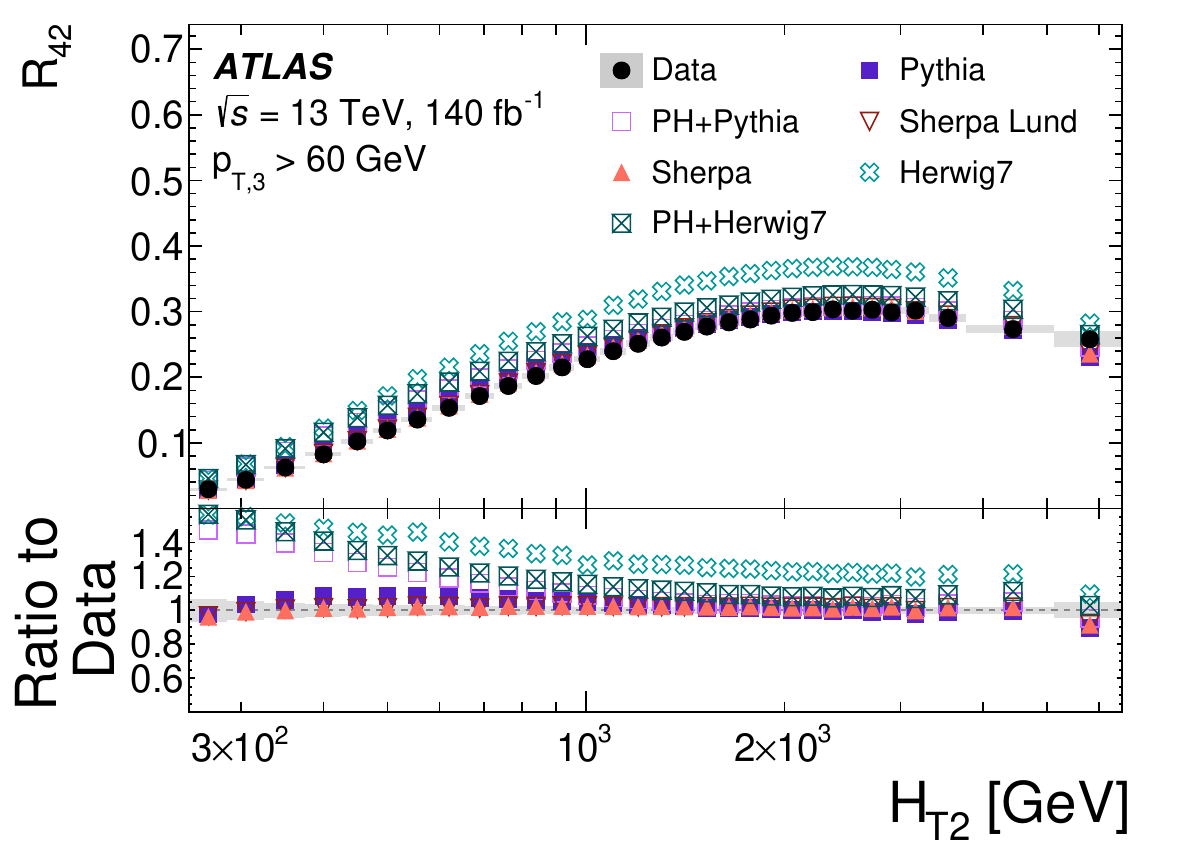}}
\subfigure[]{\includegraphics[width=0.32\textwidth]{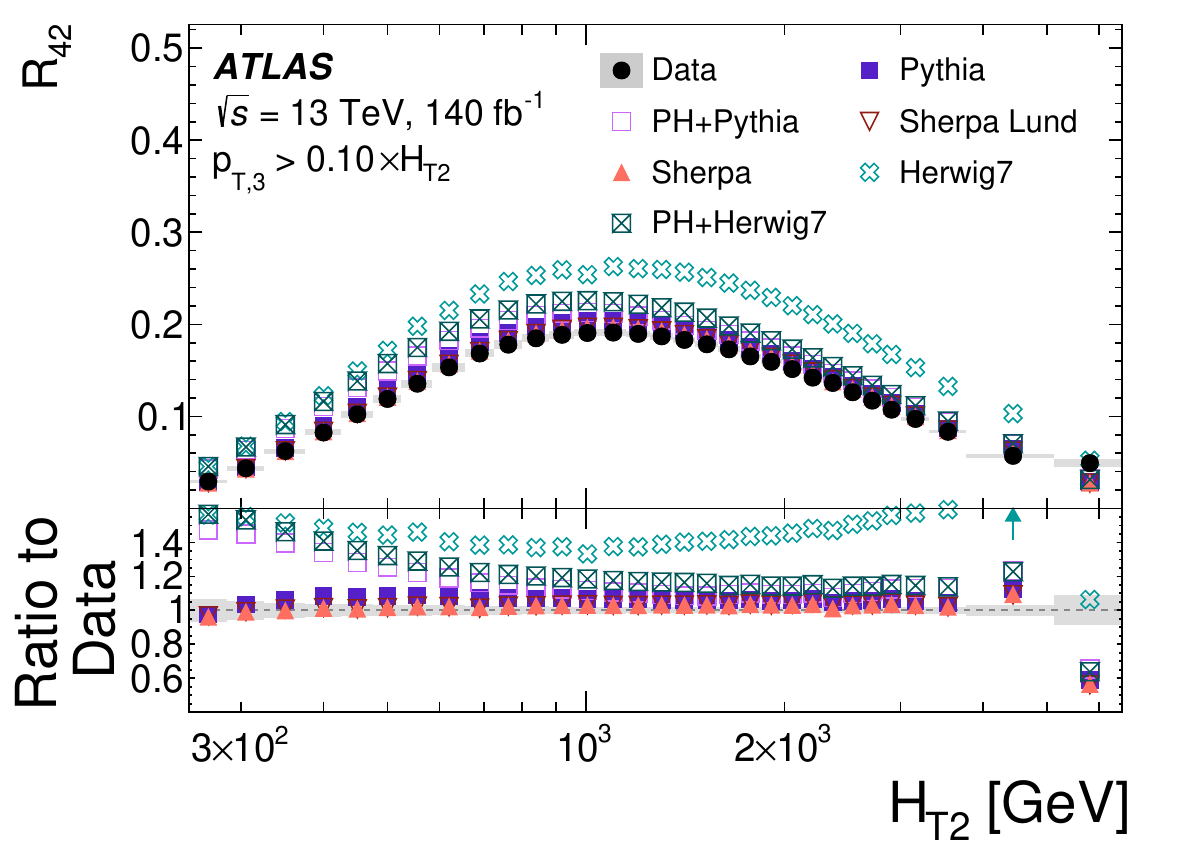}}
\subfigure[]{\includegraphics[width=0.32\textwidth]{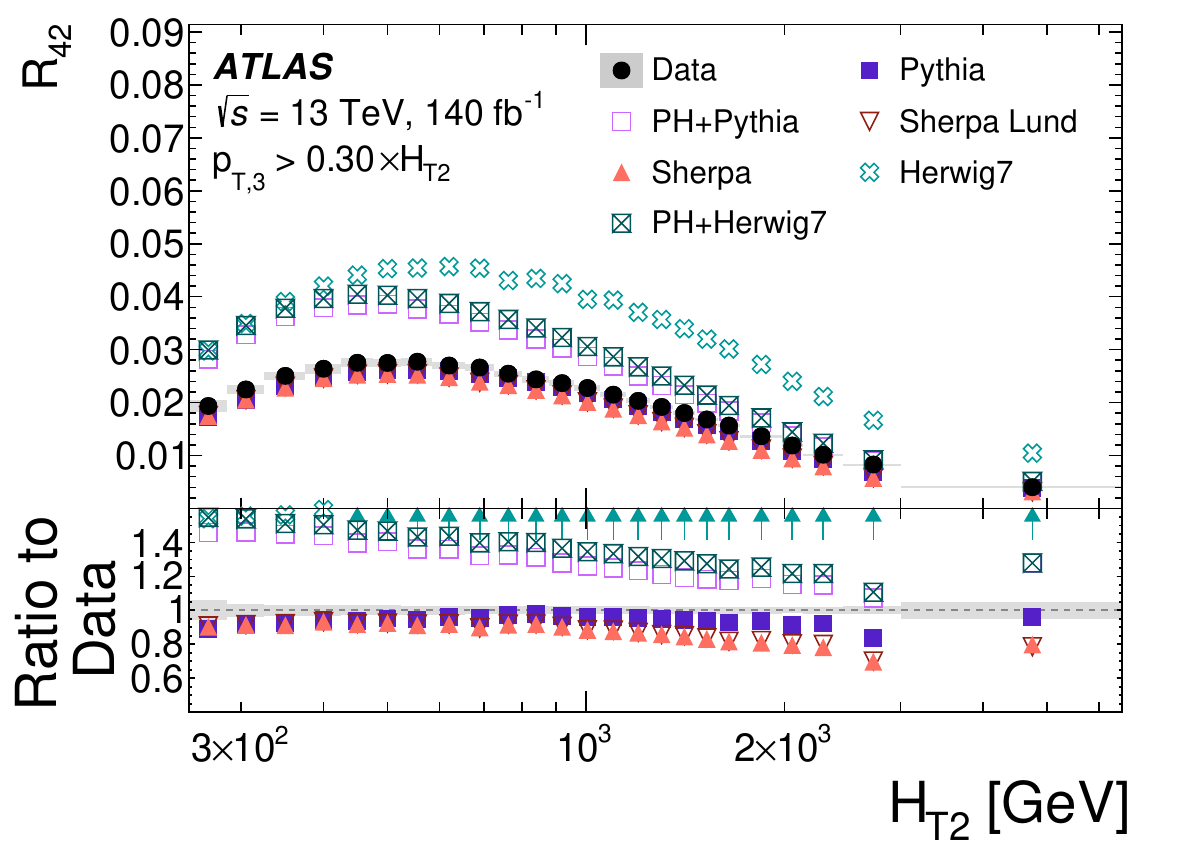}}
\subfigure[]{\includegraphics[width=0.32\textwidth]{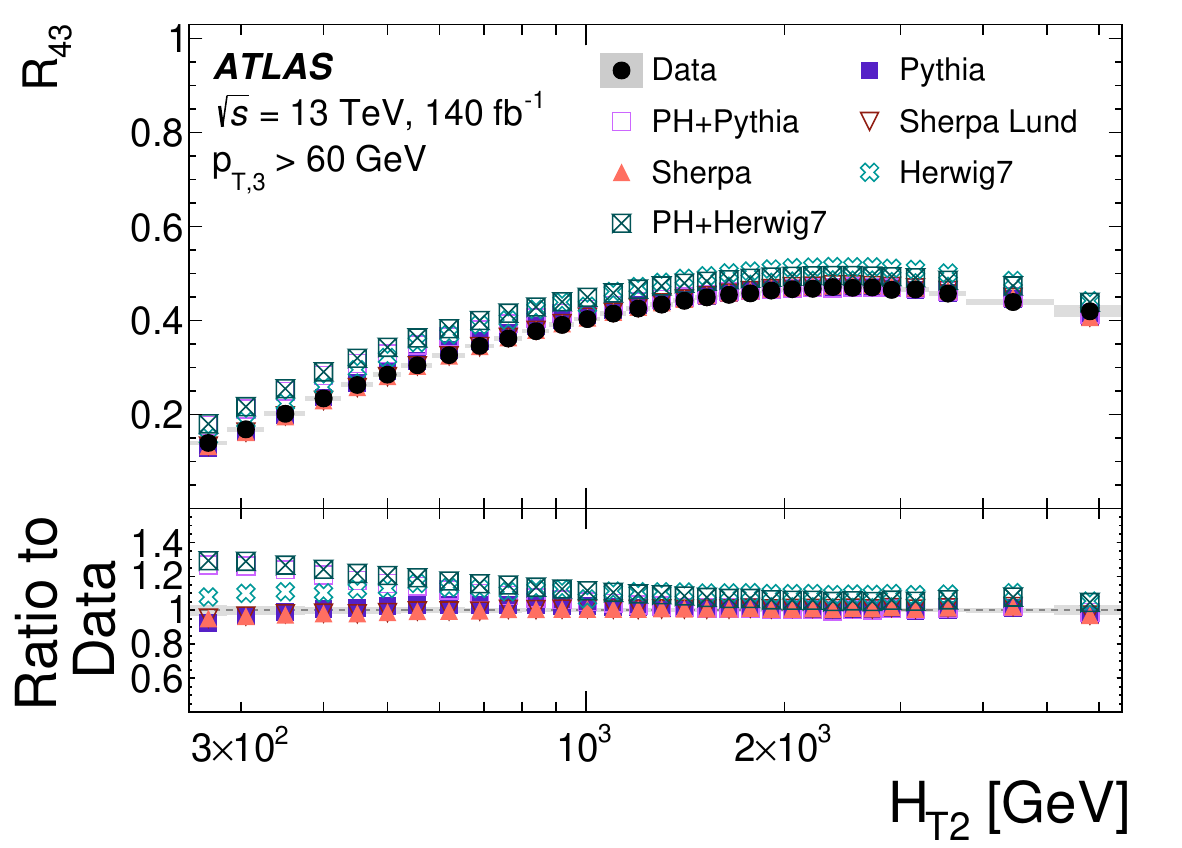}}
\subfigure[]{\includegraphics[width=0.32\textwidth]{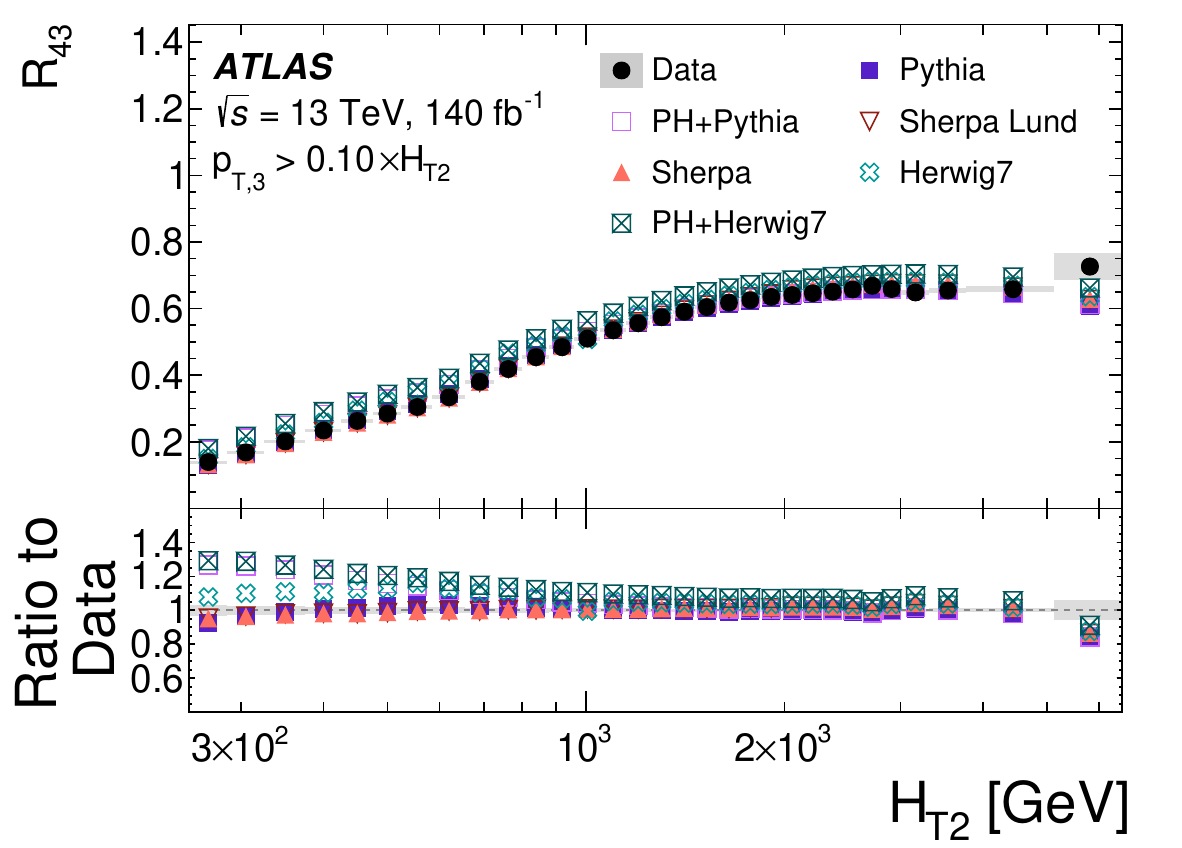}}
\subfigure[]{\includegraphics[width=0.32\textwidth]{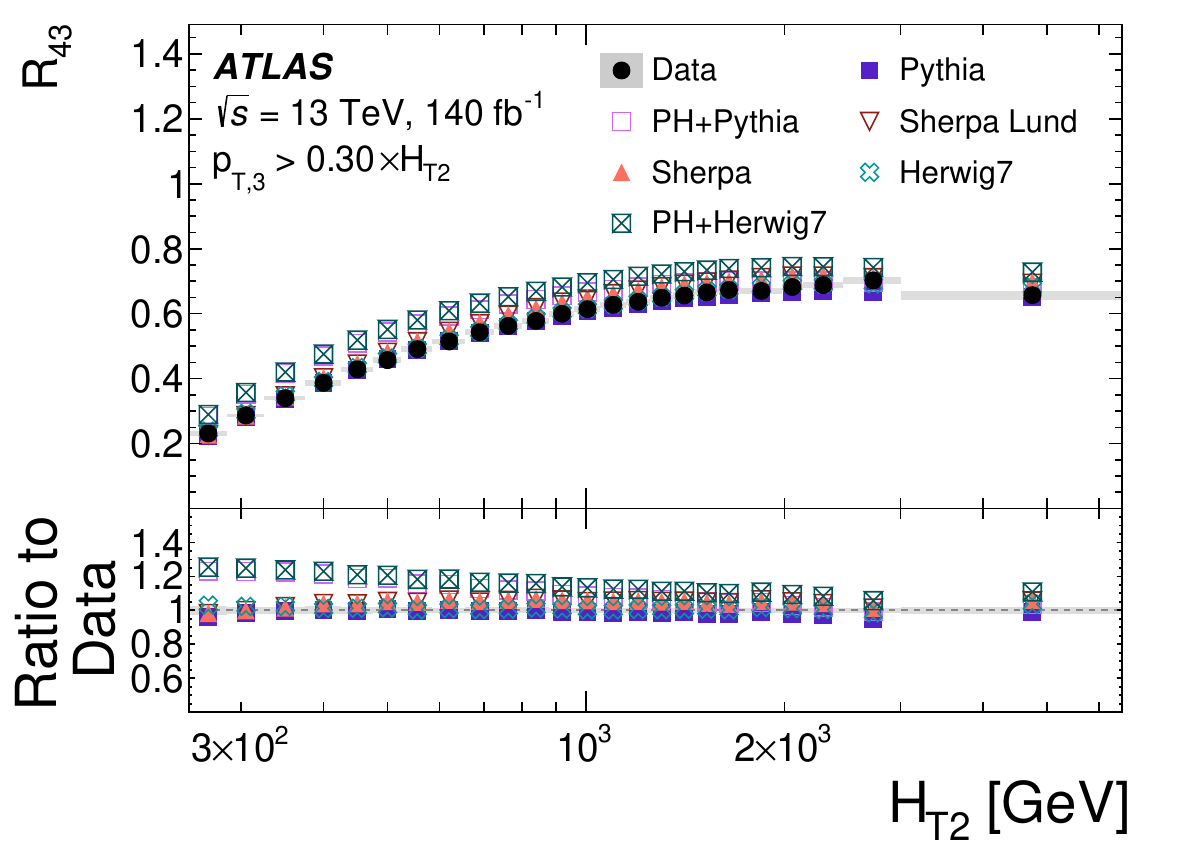}}
\subfigure[]{\includegraphics[width=0.32\textwidth]{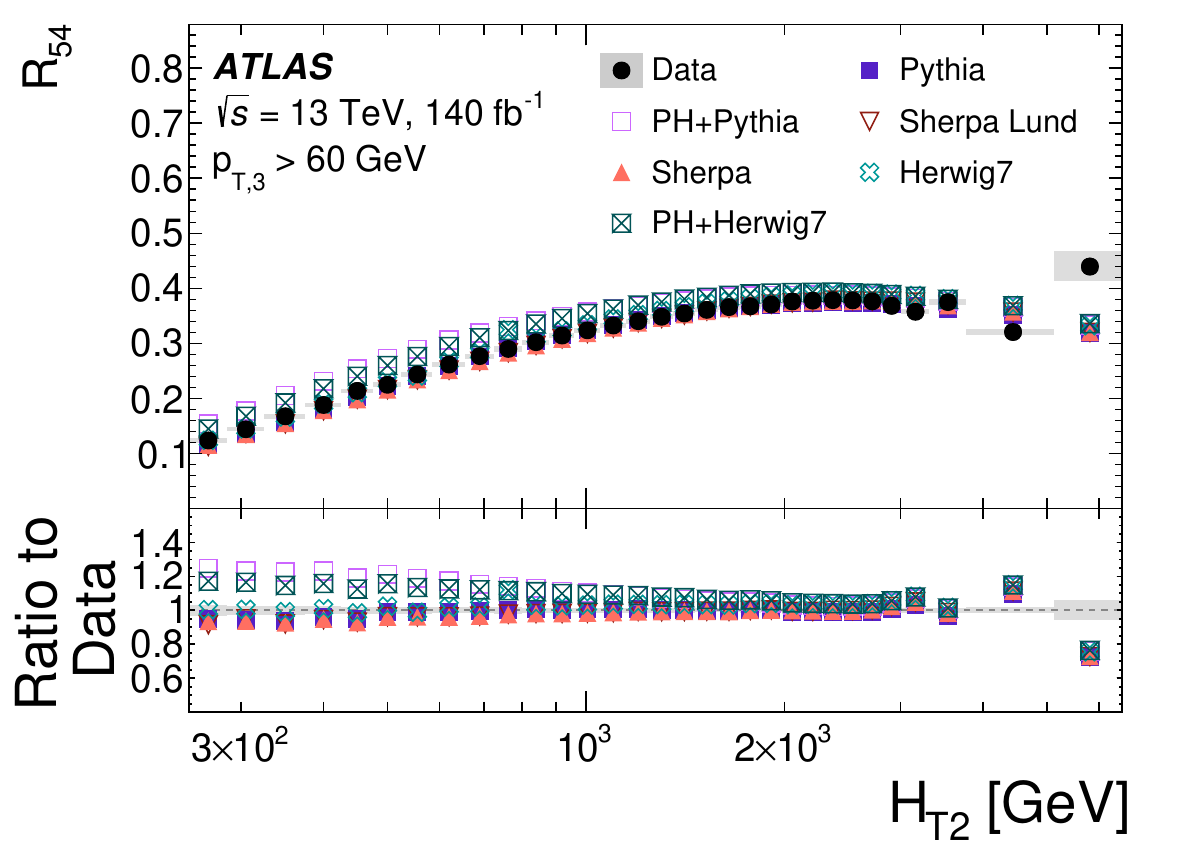}}
\subfigure[]{\includegraphics[width=0.32\textwidth]{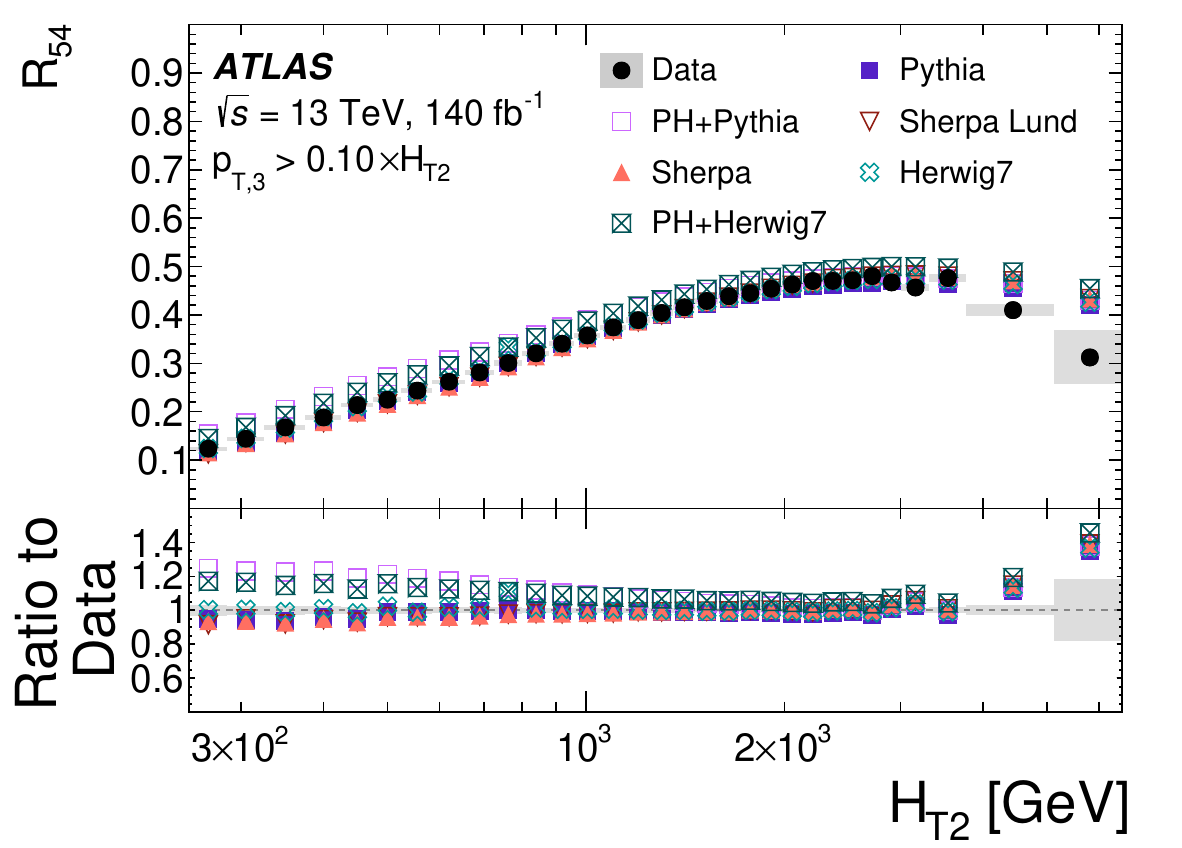}}
\subfigure[]{\includegraphics[width=0.32\textwidth]{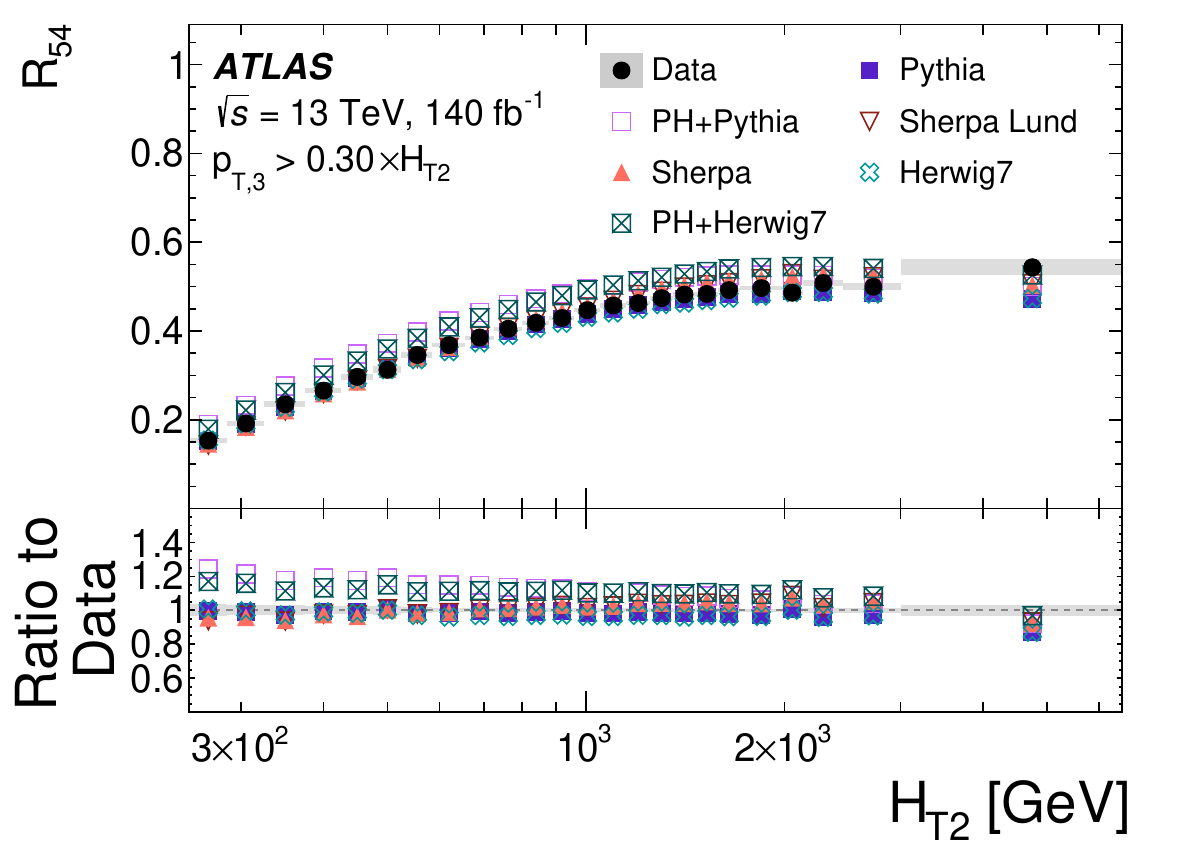}}
\caption{(a, b, c) \RThreeTwo vs.\ \HTTwo, (d, e, f) \RFourTwo vs.\ \HTTwo, (g, h, i) \RFourThree vs.\ \HTTwo, and (j, k, l) \RFiveFour vs.\ \HTTwo, with (a, d, g, j) \ptthree > 60~\GeV, (b, e, h, k) $\ptthree > 0.10\times\HTTwo$ and (c, f, i, l) $\ptthree > 0.3\times\HTTwo$.
The data error bands show the statistical and systematic components summed in quadrature.
The lower figure panels provide ratios of the MC predictions to the unfolded data. Arrows are used to indicated cases where the ratio falls outside of the plotted ratio range.}
\label{fig:ratios:HT2}
\end{figure}

The ratios of the measured \ptnincl distributions are shown in Figure~\ref{fig:ratios:nincl}.
The ratios tend towards one at high \pt, since very few events have more than two jets with \pt above a few hundred \GeV.
While the uncertainties cancel out significantly for the entire \pt distribution, the differences between data and the MC predictions are generally covered by the uncertainties.

\begin{figure}[tbh]
\centering
\subfigure[]{\includegraphics[width=0.45\textwidth]{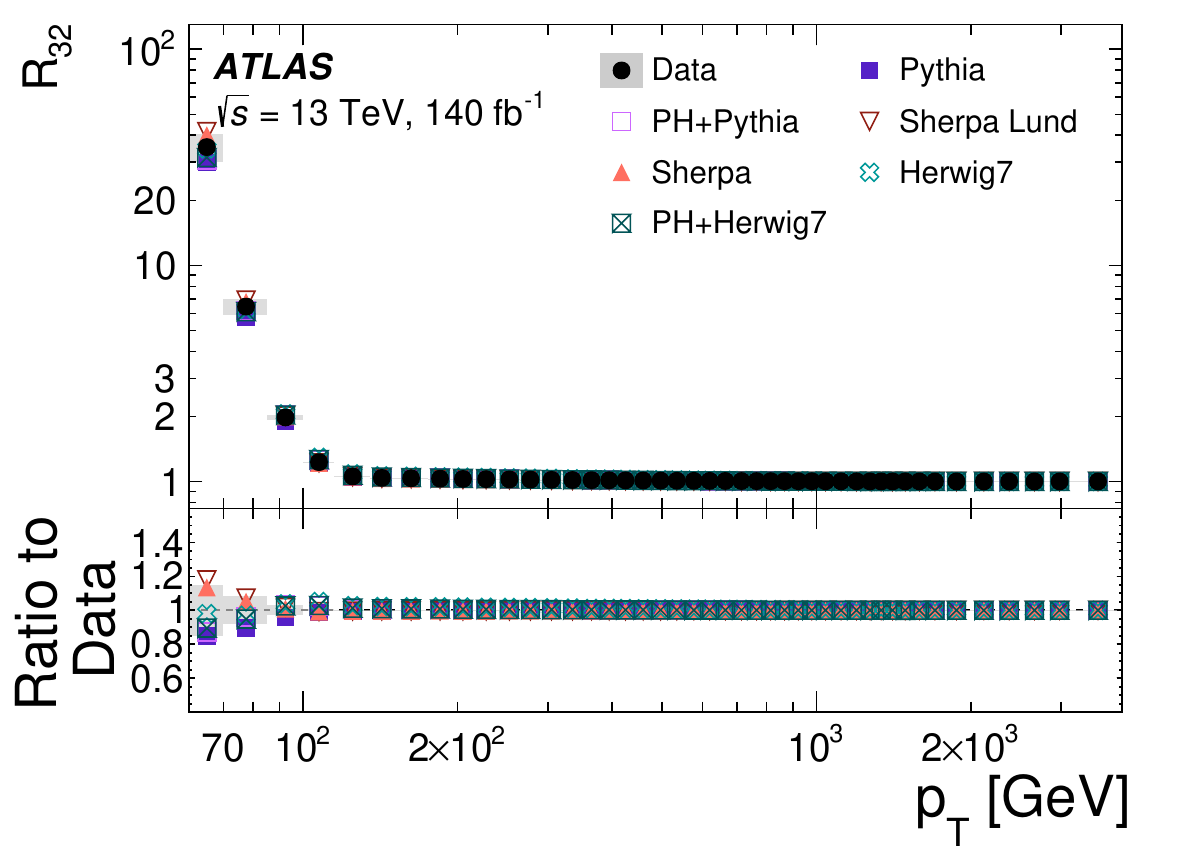}}
\subfigure[]{\includegraphics[width=0.45\textwidth]{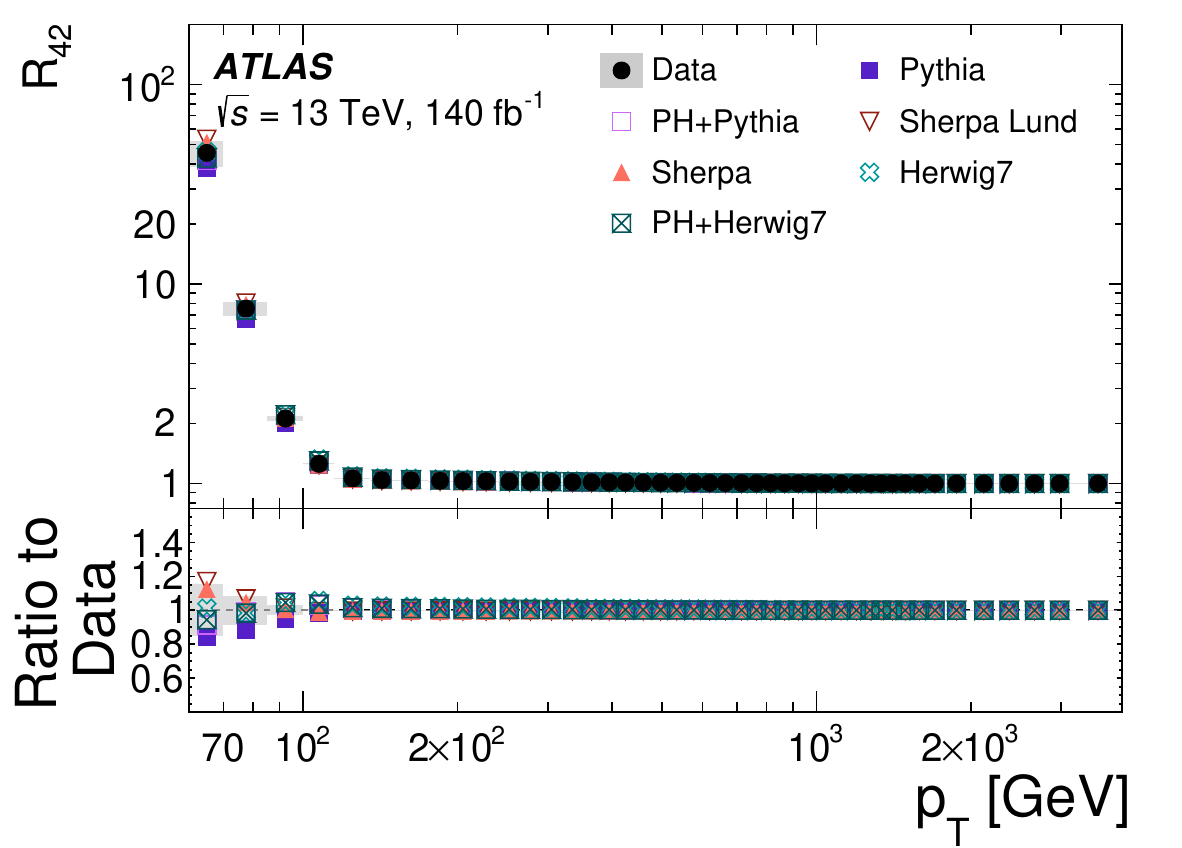}}\\
\subfigure[]{\includegraphics[width=0.45\textwidth]{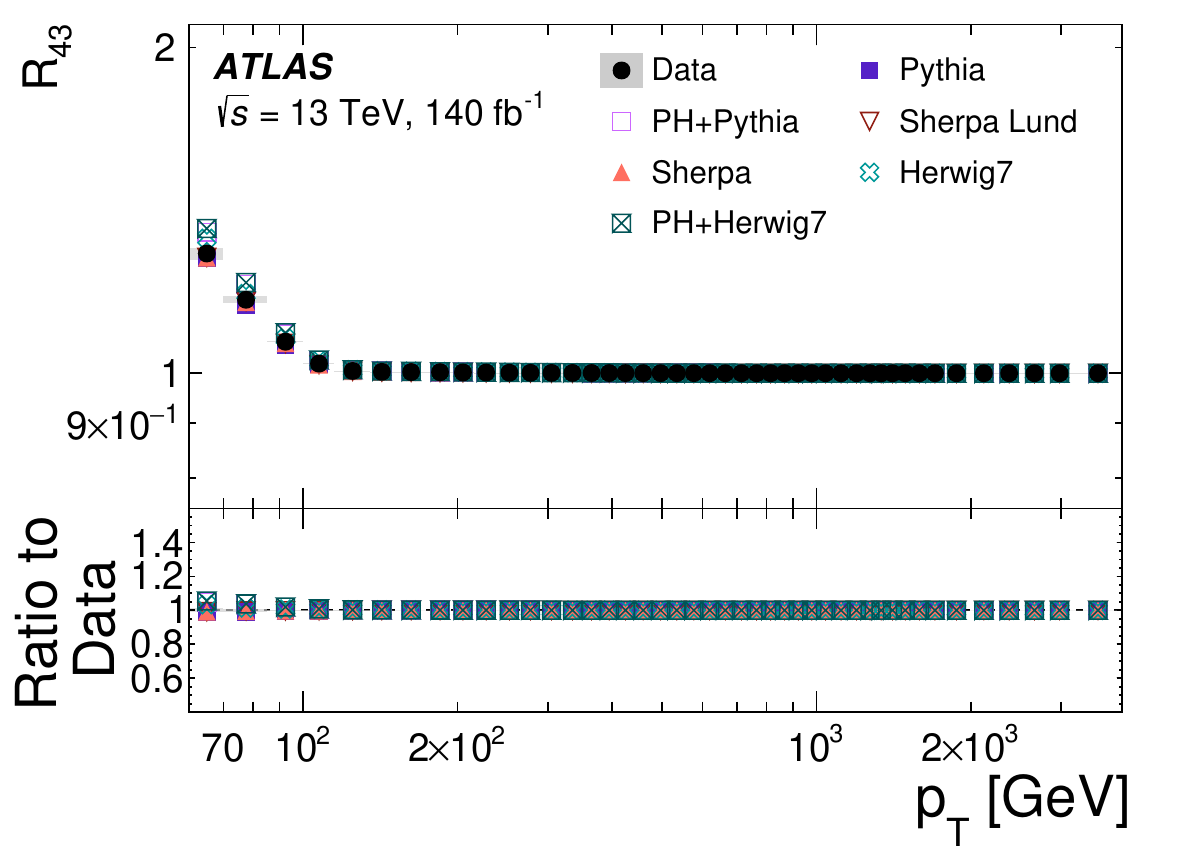}}
\caption{(a) \RThreeTwo vs.\ \ptnincl, (b) \RFourTwo vs.\ \ptnincl, and (c) \RFourThree vs.\ \ptnincl. The data error bands show the statistical and systematic components summed in quadrature.
The lower figure panels provide ratios of the MC predictions to the unfolded data. Arrows are used to indicated cases where the ratio falls outside of the plotted ratio range.}
\label{fig:ratios:nincl}
\end{figure}

The ratios of the measured \dyjj and \mjj distributions are shown in Figures~\ref{fig:ratios:dy} and~\ref{fig:ratios:mjj}, respectively.
For both observables, the \HERWIG, \POWHER, and \POWPY predictions significantly overestimate all four ratios for all rapidity differences.
The \PYTHIA and \SHERPA predictions provide a good description of the data for all four ratios for rapidity differences $\dyjj < 6$ and dijet masses above 2~\TeV, while at larger rapidity differences and larger dijet masses, they underestimate the ratios, with the exception of \RFiveFour, which is modeled well across all bins.

\begin{figure}[tbh]
\centering
\subfigure[]{\includegraphics[width=0.49\textwidth]{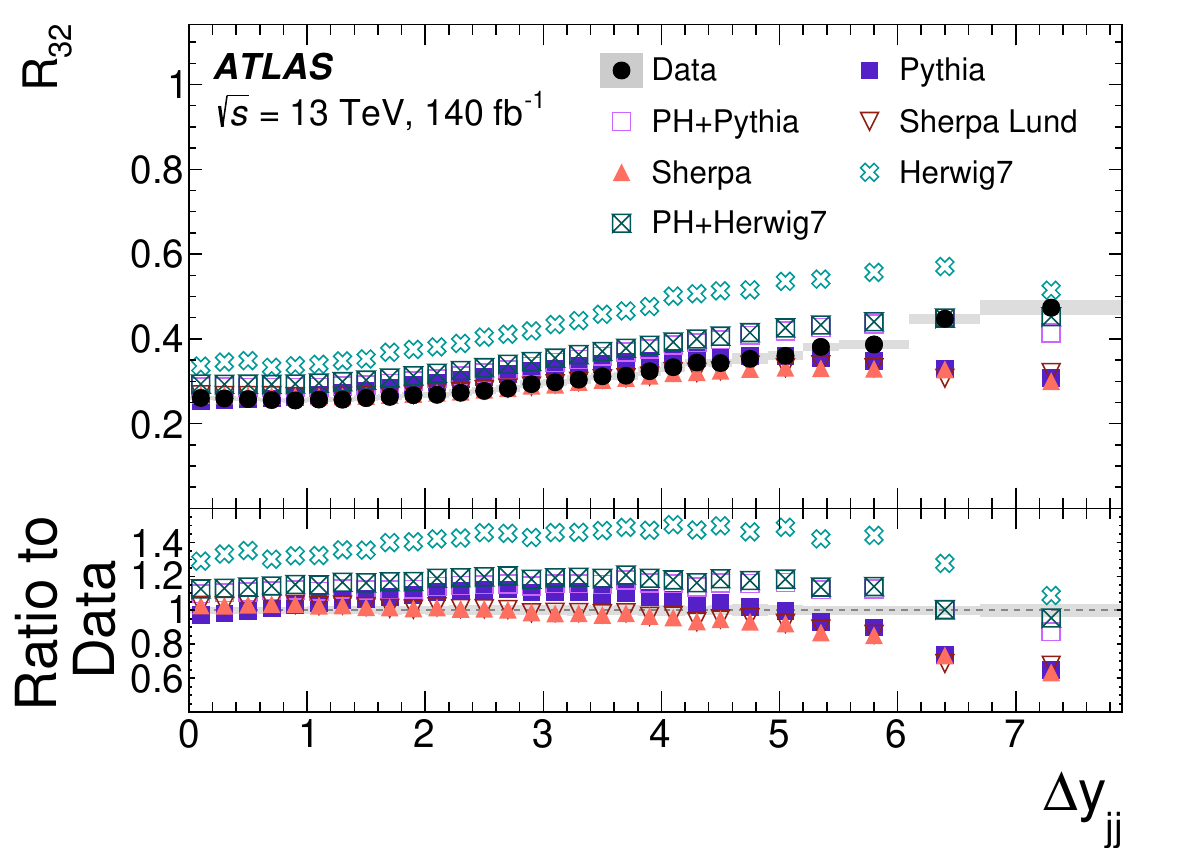}}
\subfigure[]{\includegraphics[width=0.49\textwidth]{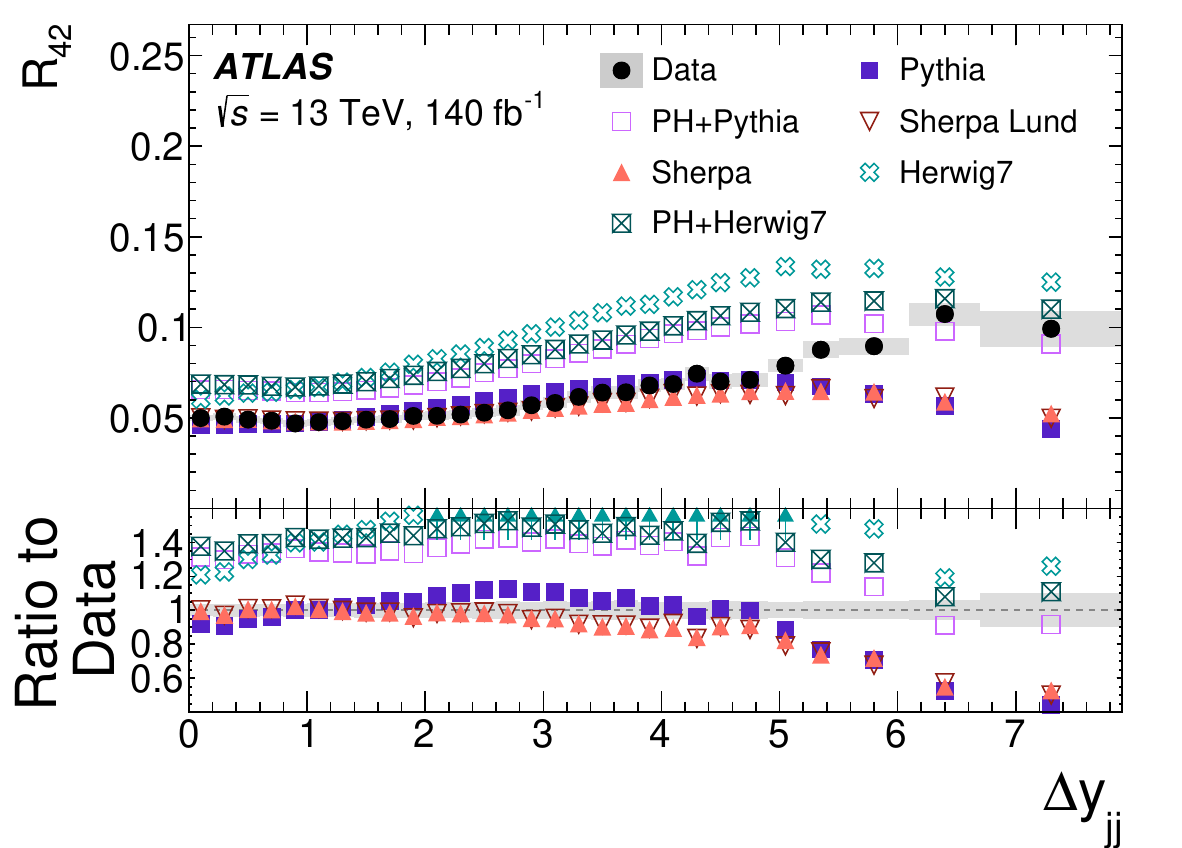}}
\subfigure[]{\includegraphics[width=0.49\textwidth]{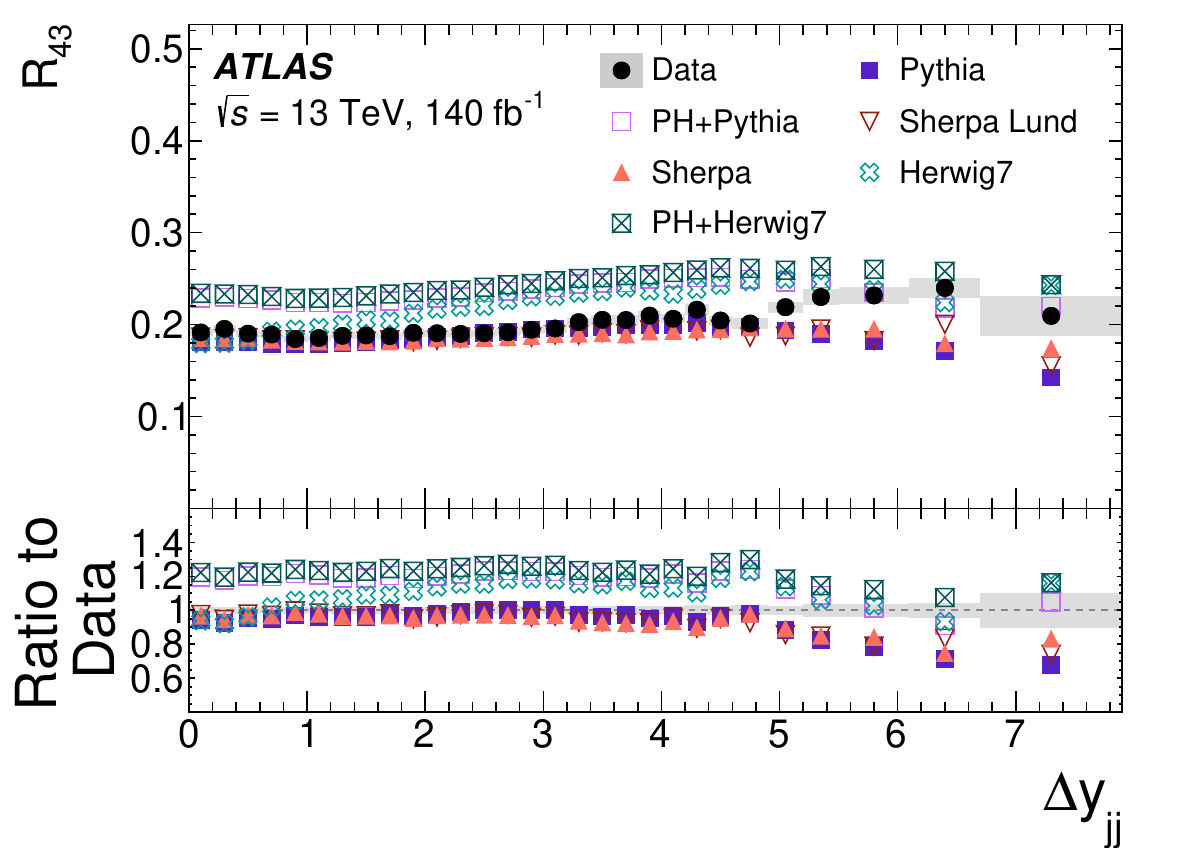}}
\subfigure[]{\includegraphics[width=0.49\textwidth]{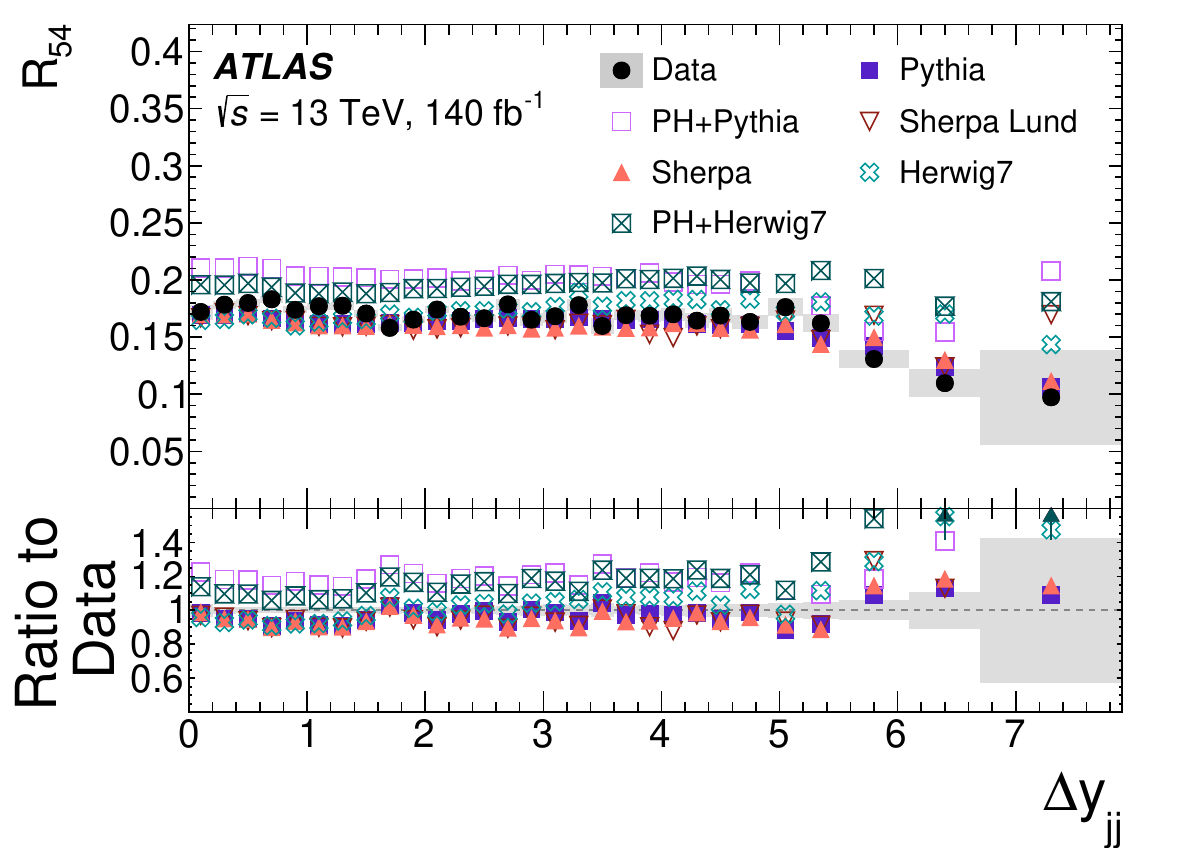}}
\caption{(a) \RThreeTwo, (b) \RFourTwo, (c) \RFourThree, and (d) \RFiveFour vs. \dyjj with $\ptthree > 60~\GeV$. The data error bands show the statistical and systematic components summed in quadrature.
The lower figure panels provide ratios of the MC predictions to the unfolded data. Arrows are used to indicated cases where the ratio falls outside of the plotted ratio range.}
\label{fig:ratios:dy}
\end{figure}

\begin{figure}[tbh]
\centering
\subfigure[]{\includegraphics[width=0.49\textwidth]{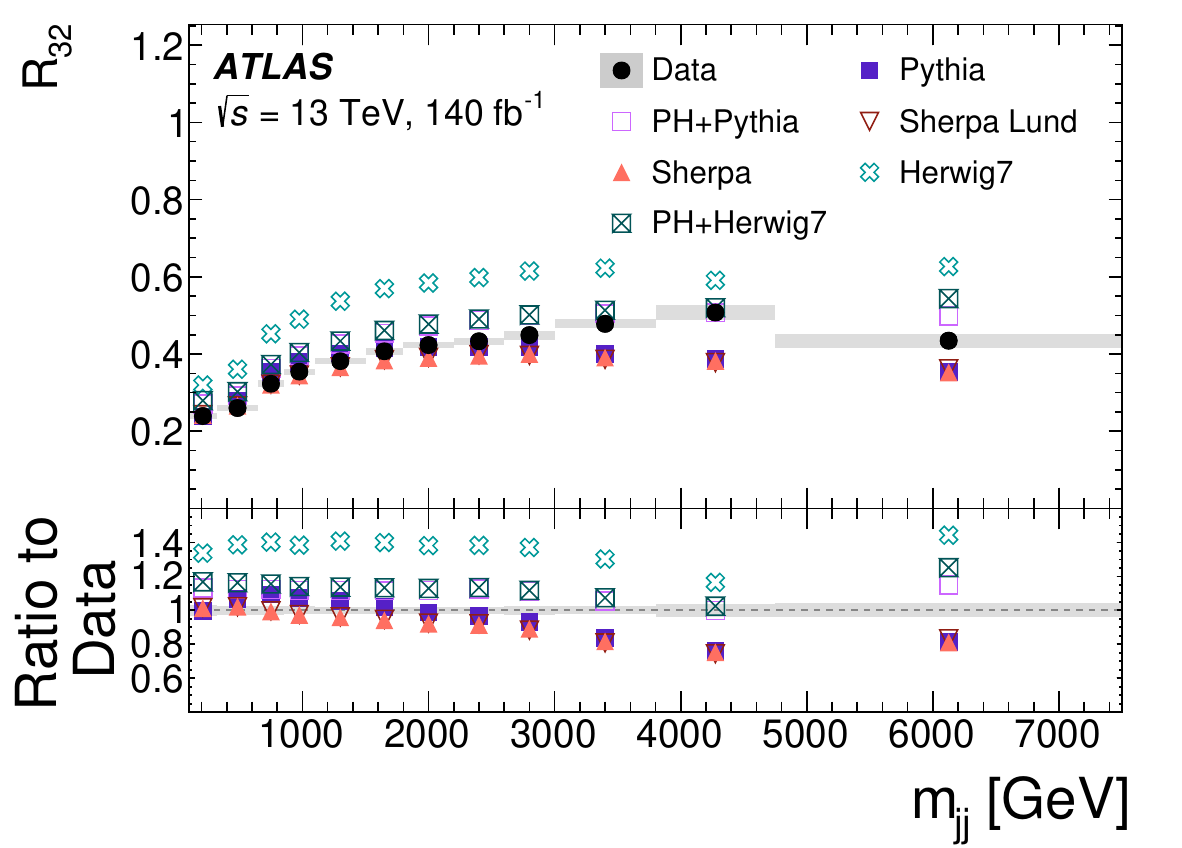}}
\subfigure[]{\includegraphics[width=0.49\textwidth]{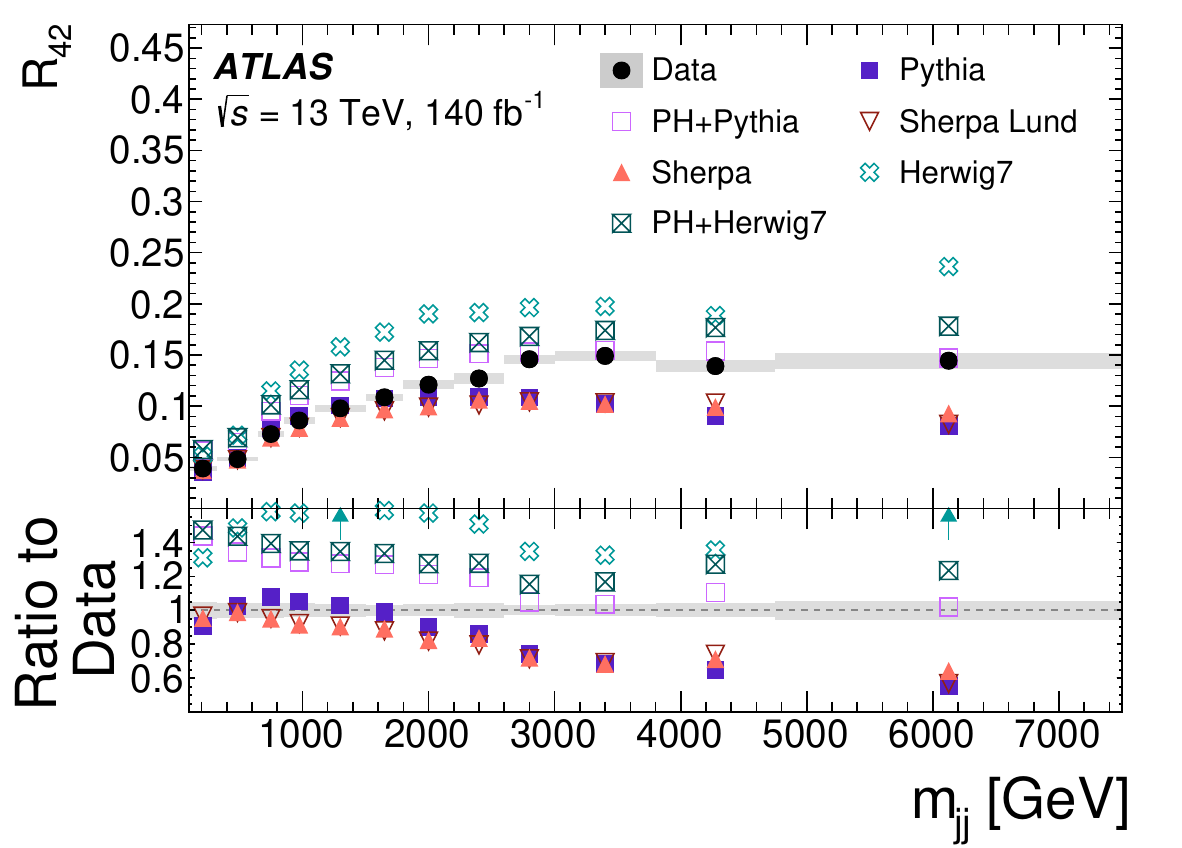}}
\subfigure[]{\includegraphics[width=0.49\textwidth]{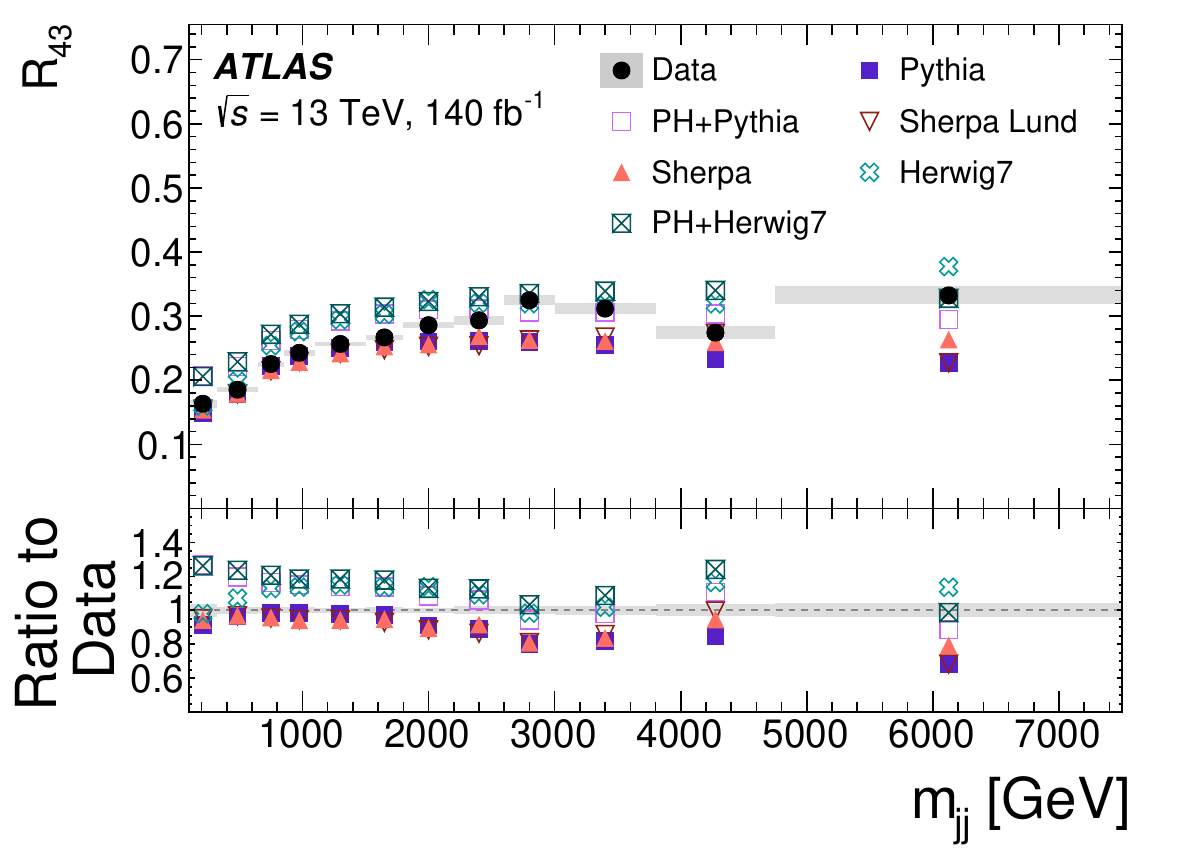}}
\subfigure[]{\includegraphics[width=0.49\textwidth]{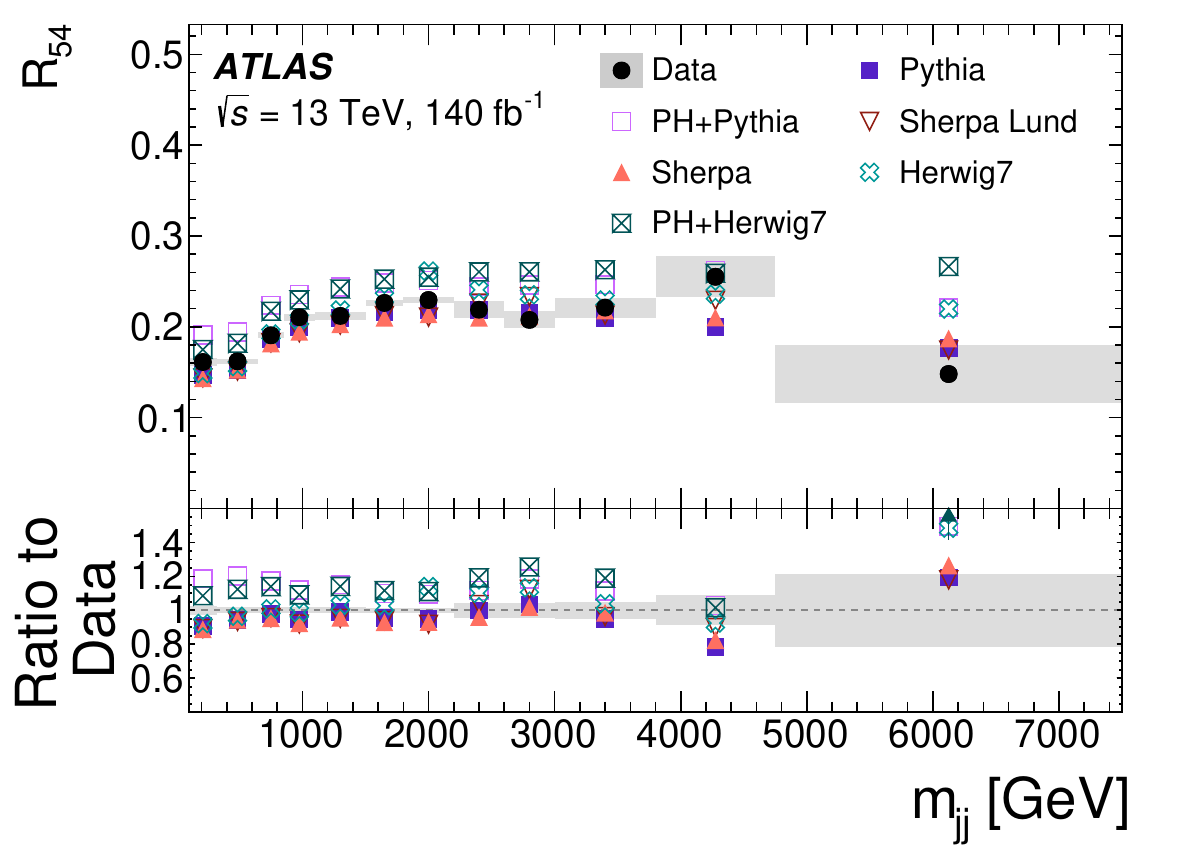}}
\caption{(a) \RThreeTwo, (b) \RFourTwo, (c) \RFourThree, and (d) \RFiveFour vs. \mjj with $\ptthree > 60~\GeV$. The data error bands show the statistical and systematic components summed in quadrature.
The lower figure panels provide ratios of the MC predictions to the unfolded data. Arrows are used to indicated cases where the ratio falls outside of the plotted ratio range.}
\label{fig:ratios:mjj}
\end{figure}

The ratios of the measured \dymax and \mjjmax distributions are shown in Figures~\ref{fig:ratios:dyjj} and~\ref{fig:ratios:mjjmax},  respectively.
For the \dymax ratios, the \PYTHIA and \SHERPA predictions model the data well, except in the low and high rapidity-difference regions.
Similar features are seen for \mjjmax, although the mismodeling at low dijet masses is only seen in the lowest \mjjmax bin.
\POWHER and \POWPY do not model any of the ratios well, with the smallest disagreement seen for \RThreeTwo.
\HERWIG does not describe \RThreeTwo or \RFourTwo well for either observable, but provides a reasonable description for \RFourThree and \RFiveFour, except for the low rapidity-difference region, where it underestimates the ratio.

\begin{figure}[tbh]
\centering
\subfigure[]{\includegraphics[width=0.49\textwidth]{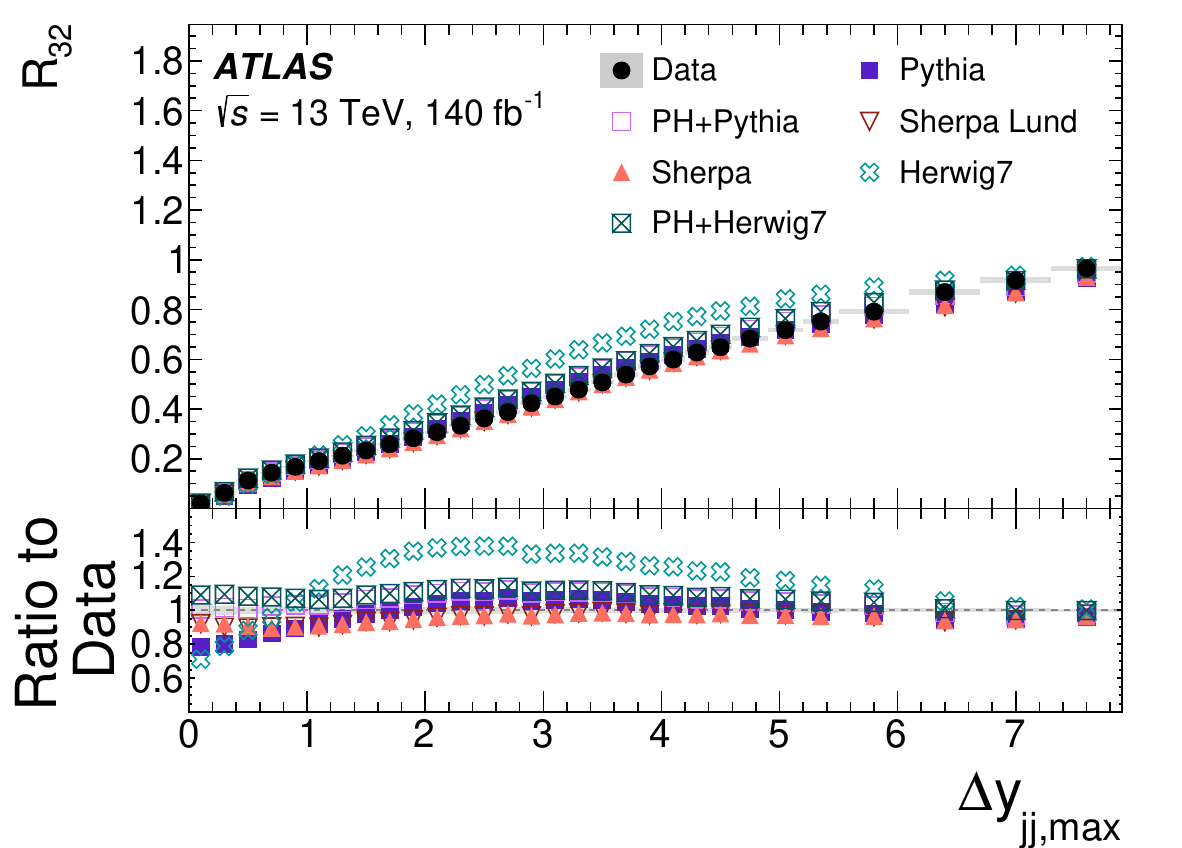}}
\subfigure[]{\includegraphics[width=0.49\textwidth]{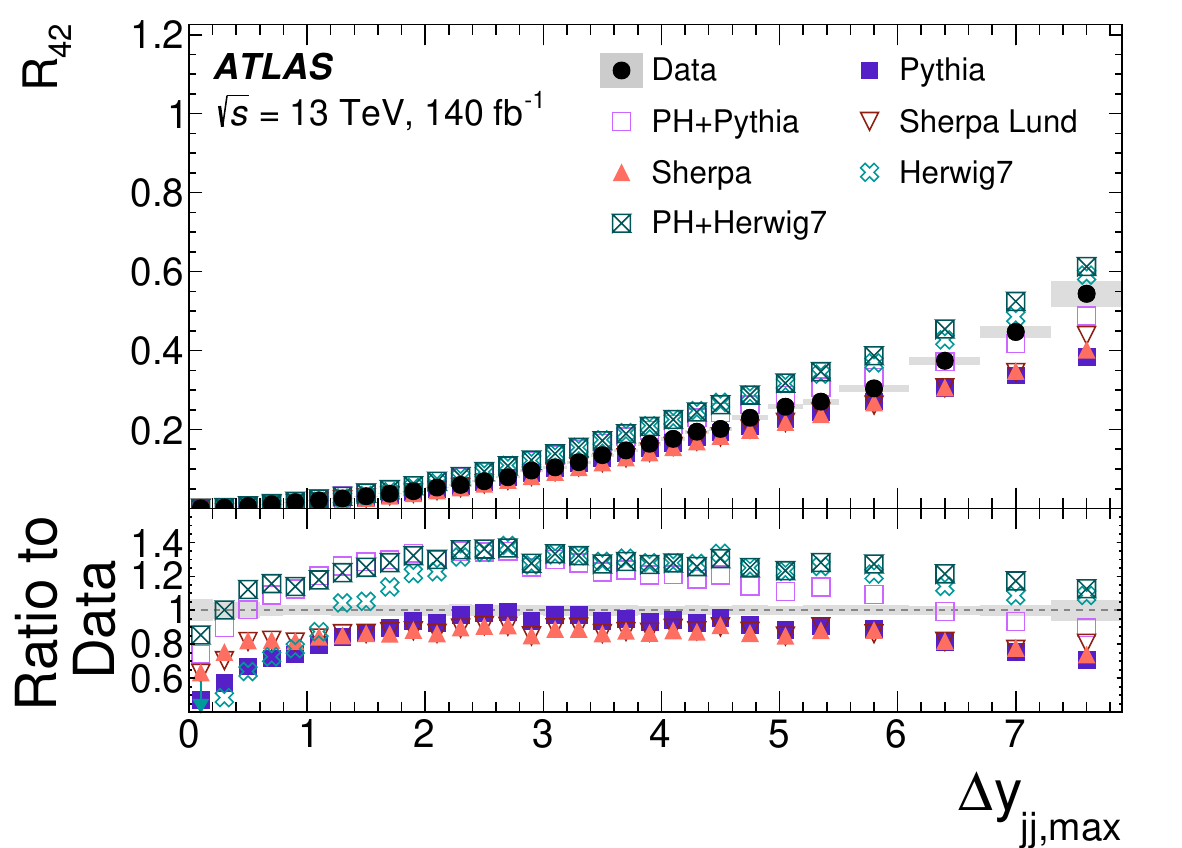}}
\subfigure[]{\includegraphics[width=0.49\textwidth]{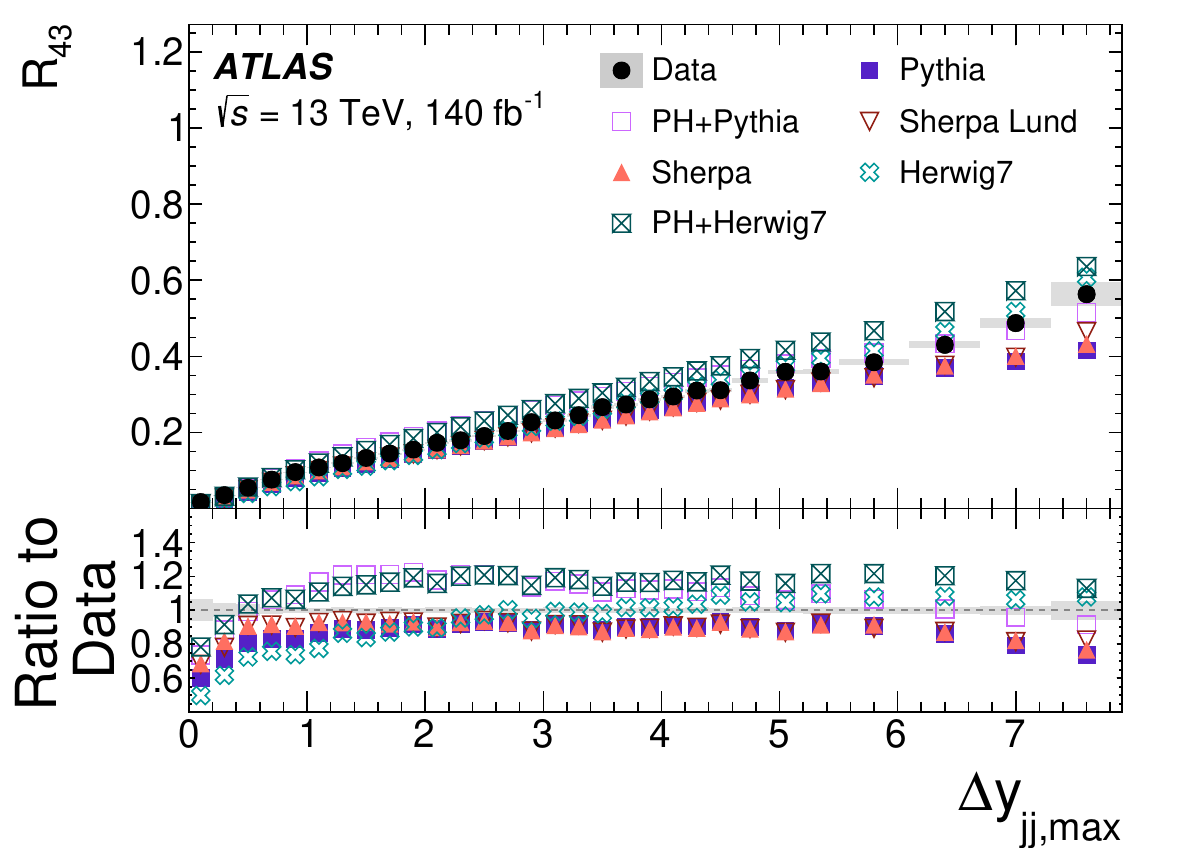}}
\subfigure[]{\includegraphics[width=0.49\textwidth]{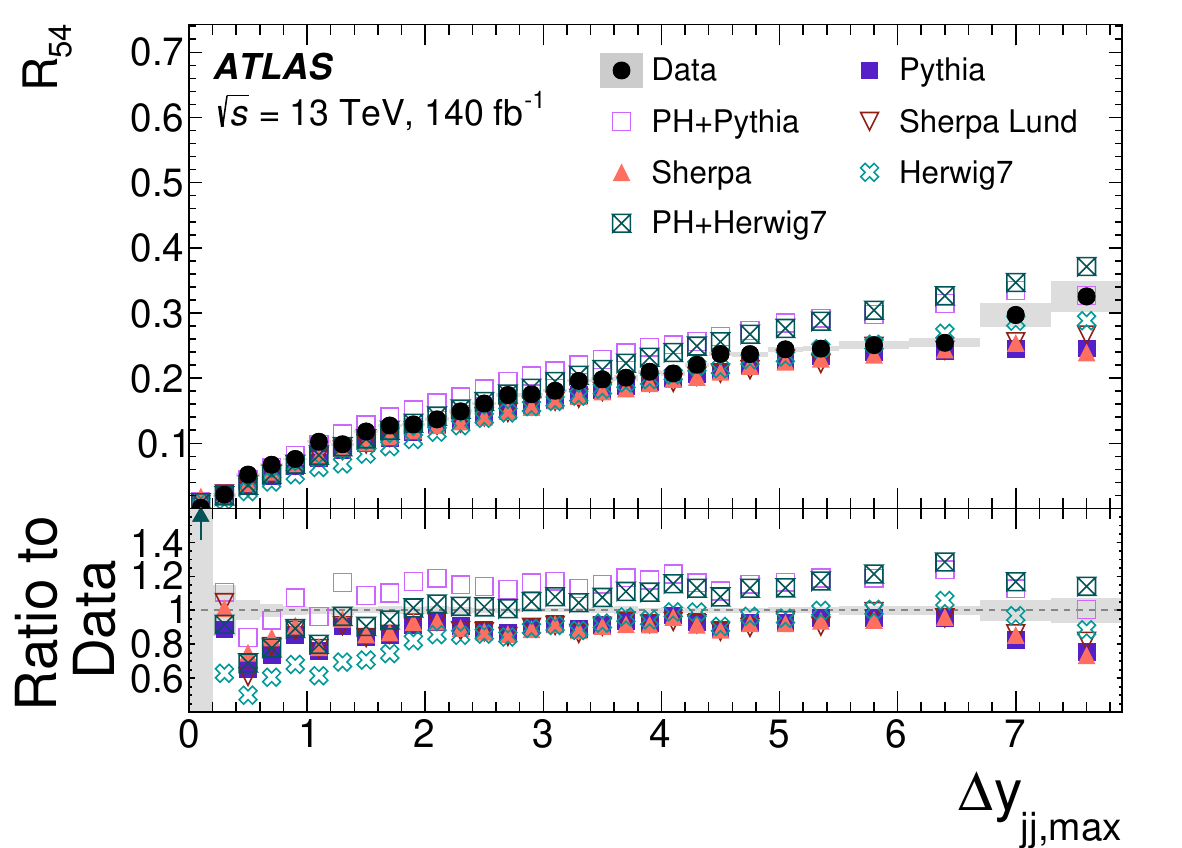}}
\caption{(a) \RThreeTwo, (b) \RFourTwo, (c) \RFourThree, and (d) \RFiveFour vs. \dymax with $\ptthree > 60~\GeV$. The data error bands show the statistical and systematic components summed in quadrature.
The lower figure panels provide ratios of the MC predictions to the unfolded data. Arrows are used to indicated cases where the ratio falls outside of the plotted ratio range.}
\label{fig:ratios:dyjj}
\end{figure}

\begin{figure}[tbh]
\centering
\subfigure[]{\includegraphics[width=0.49\textwidth]{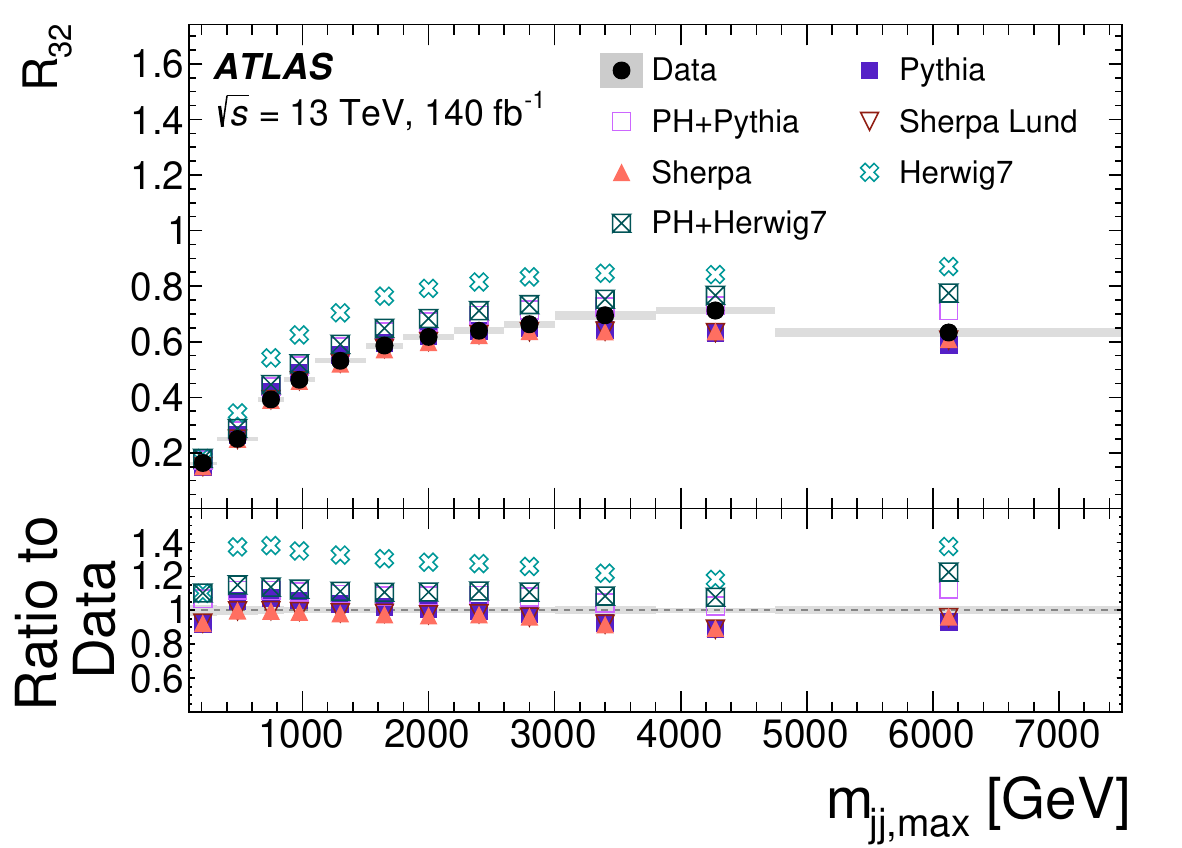}}
\subfigure[]{\includegraphics[width=0.49\textwidth]{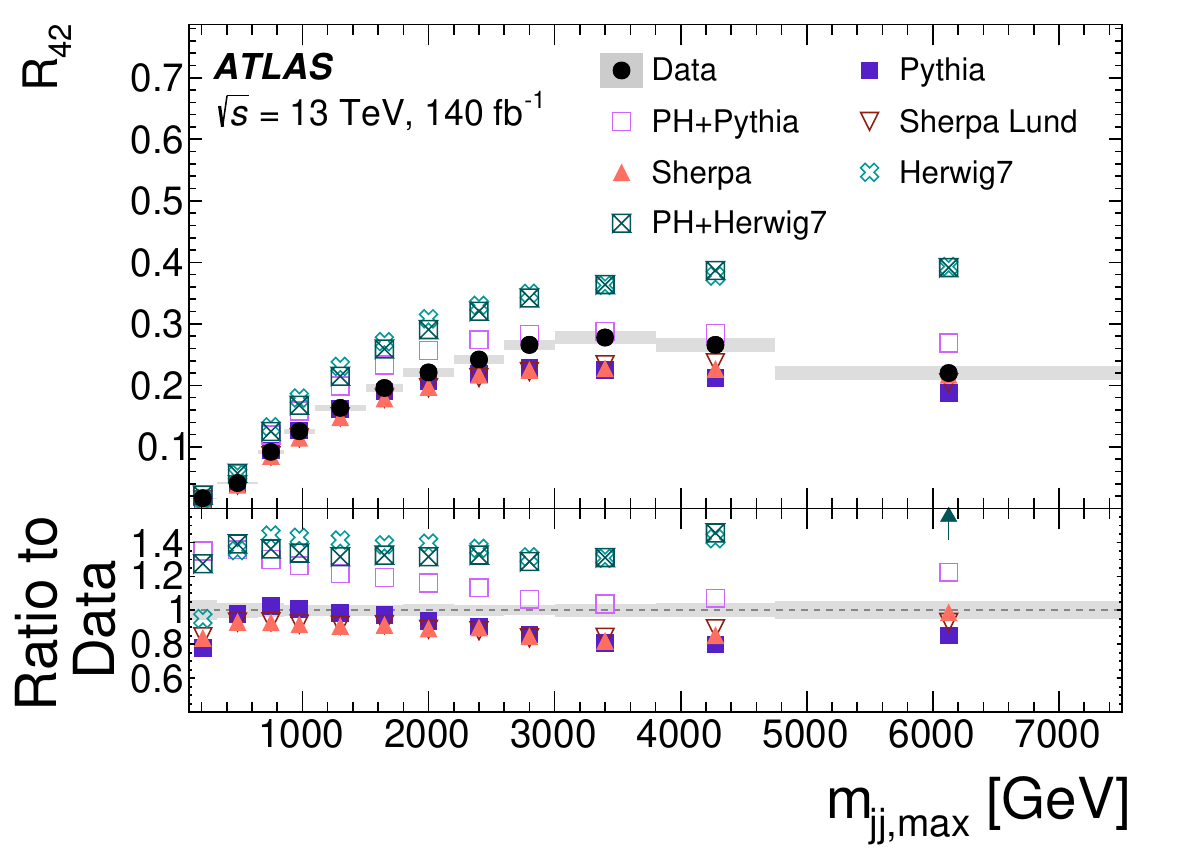}}
\subfigure[]{\includegraphics[width=0.49\textwidth]{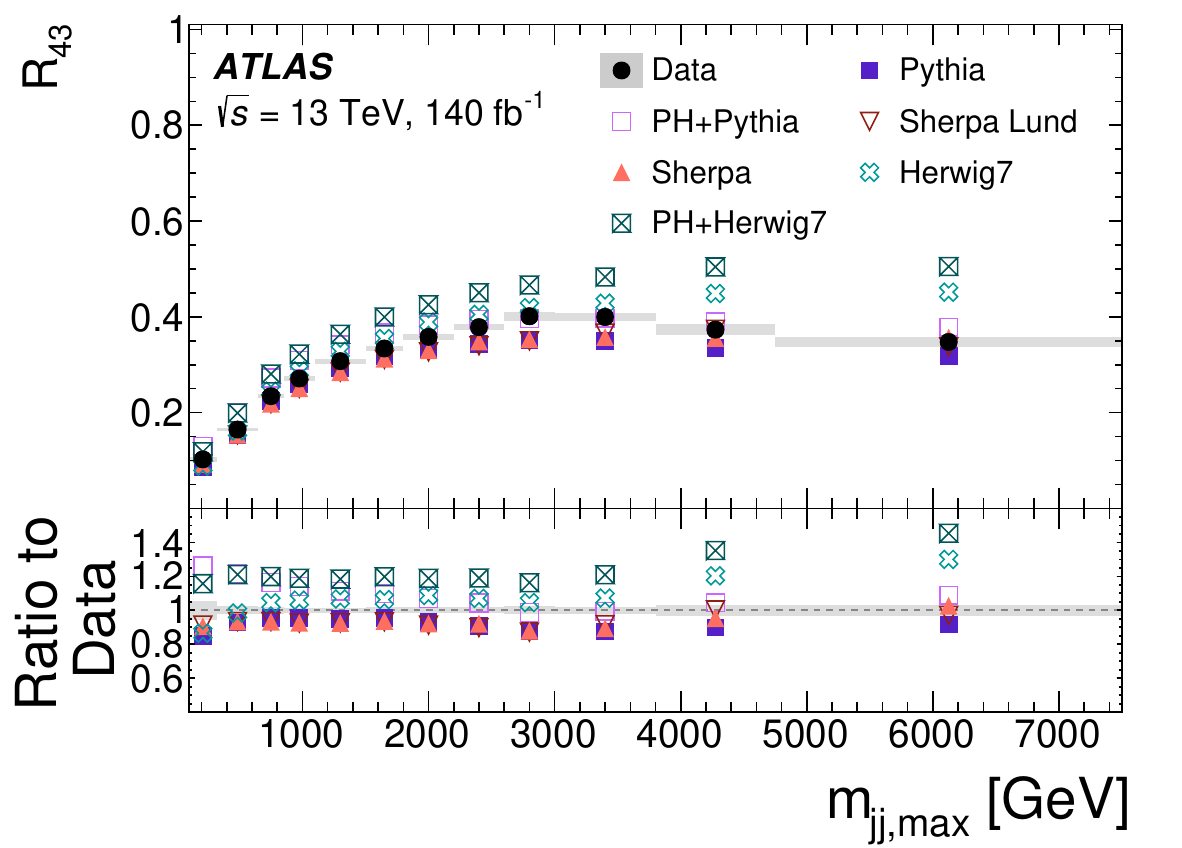}}
\subfigure[]{\includegraphics[width=0.49\textwidth]{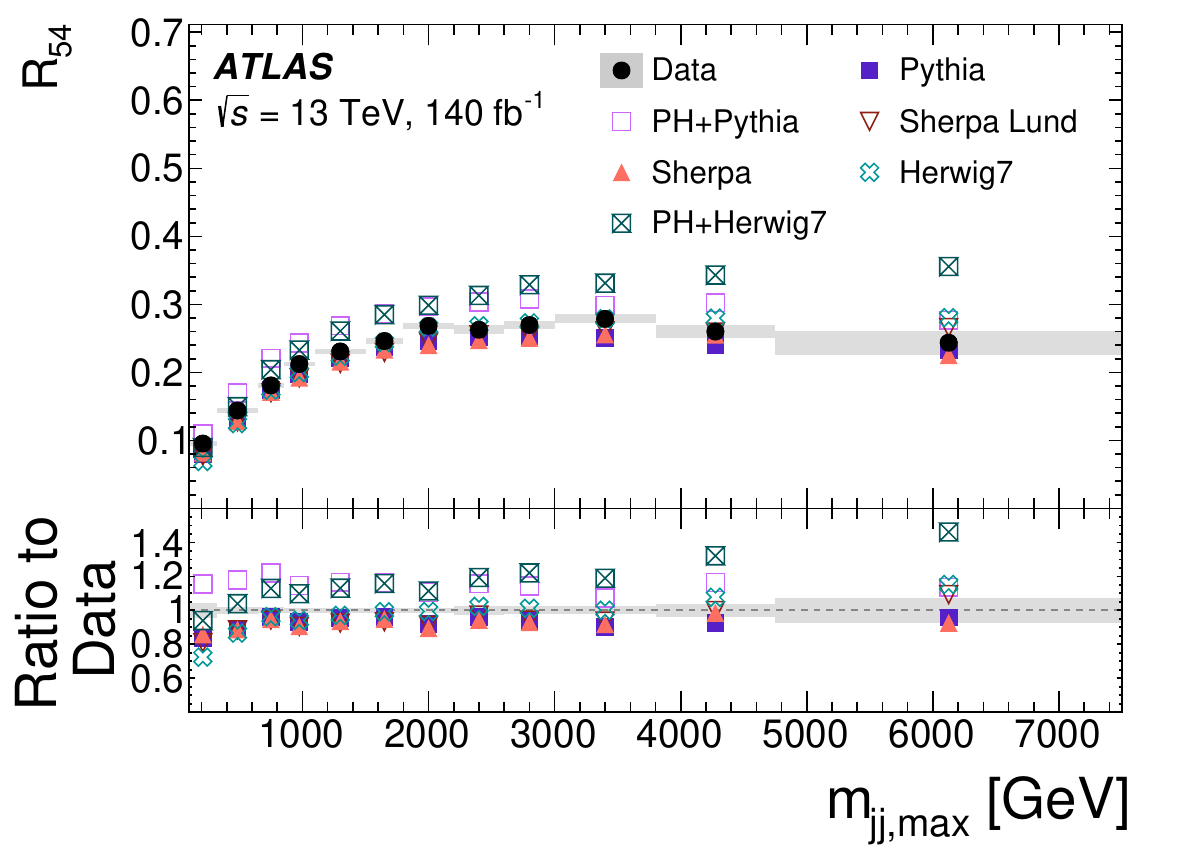}}
\caption{(a) \RThreeTwo, (b) \RFourTwo, (c) \RFourThree, and (d) \RFiveFour vs. \mjjmax with $\ptthree > 60~\GeV$. The data error bands show the statistical and systematic components summed in quadrature.
The lower figure panels provide ratios of the MC predictions to the unfolded data. Arrows are used to indicated cases where the ratio falls outside of the plotted ratio range.}
\label{fig:ratios:mjjmax}
\end{figure}

\subsection{Comparisons with fixed-order and resummed calculations}

The ratios \RThreeTwo of the measured \HTTwo distributions are shown in Figure~\ref{fig:ratios:httwo:pt3:theory}, compared with the NLO and NNLO predictions.
The NNLO prediction provides an accurate description of the value and shape of \RThreeTwo for all the different \ptthree bins of the measurement, while the NLO prediction tends to overestimate \RThreeTwo.
This highlights the importance of the higher-order predictions in describing multijet production.
For low cuts on \ptthree and at larger values of \HTTwo, the NNLO prediction slightly overestimates the data.
This is the region where effects from resummation play a more important role, and higher cuts on \ptthree reduce these differences.
Some statistical fluctuations in the NNLO prediction are observed, due to the significant computational requirements of these predictions.
The statistical error of the theory predictions is treated independently between bins.

For each \ptthree cut, the value of $\chi^2$ per degree of freedom (\enquote{\textit{d.o.f.}}) is shown in Table~\ref{tab:chi2}.
The individual experimental and theoretical uncertainties are considered to be uncorrelated with each other, and fully correlated across the \HTTwo bins.
For all \ptthree cuts, the highest \HTTwo bin is excluded from the $\chi^2$ calculation, due to the large statistical fluctuation in the NNLO prediction in this bin: this results in 19 degrees of freedom (bins).
The $\chi^2 / \textit{d.o.f.}$ values for the NLO and NNLO predictions agree well with the data in all bins except the highest \ptthree bin, which has poor agreement due to a fluctuation in the second-highest \HTTwo bin.
The $\chi^2 / \textit{d.o.f.}$ values are often less than 1 for the smallest \ptthree cuts (the two lowest cuts for the NLO comparison, the four lowest for the NNLO comparison).
For the NLO prediction, this is primarily due to the large theory scale uncertainties.
Despite their smaller uncertainty due to scale variations, the NNLO predictions have smaller $\chi^2 / \textit{d.o.f.}$ values than the NLO predictions and are more often less than 1: this is partially due to both an overall improved description of the measured data's shape, and the presence of a non-negligible statistical uncertainty on the prediction.

\begin{table}[t!]
\centering
\caption{Summary of the $\chi^2 / \textit{d.o.f.}$ values from the comparison of the measurement of \RThreeTwo and the NLO and NNLO predictions.}
\begin{tabular}{ l | c | c }
\hline
& \multicolumn{2}{|c}{$\chi^2 / \textit{d.o.f.}$} \\
\hline
& NLO & NNLO \\
\hline
$\ptthree > 60~\GeV$    & 0.48 & 0.36 \\
$\ptthree > 0.05\times\HTTwo$ & 0.55 & 0.32 \\
$\ptthree > 0.10\times\HTTwo$  & 1.05 & 0.24 \\
$\ptthree > 0.20\times\HTTwo$  & 1.11 & 0.30 \\
$\ptthree > 0.30\times\HTTwo$  & 9.24 & 5.49 \\
\hline
\end{tabular}
\label{tab:chi2}
\end{table}

\begin{figure}[tbh]
\centering
\subfigure[]{\includegraphics[width=0.49\textwidth]{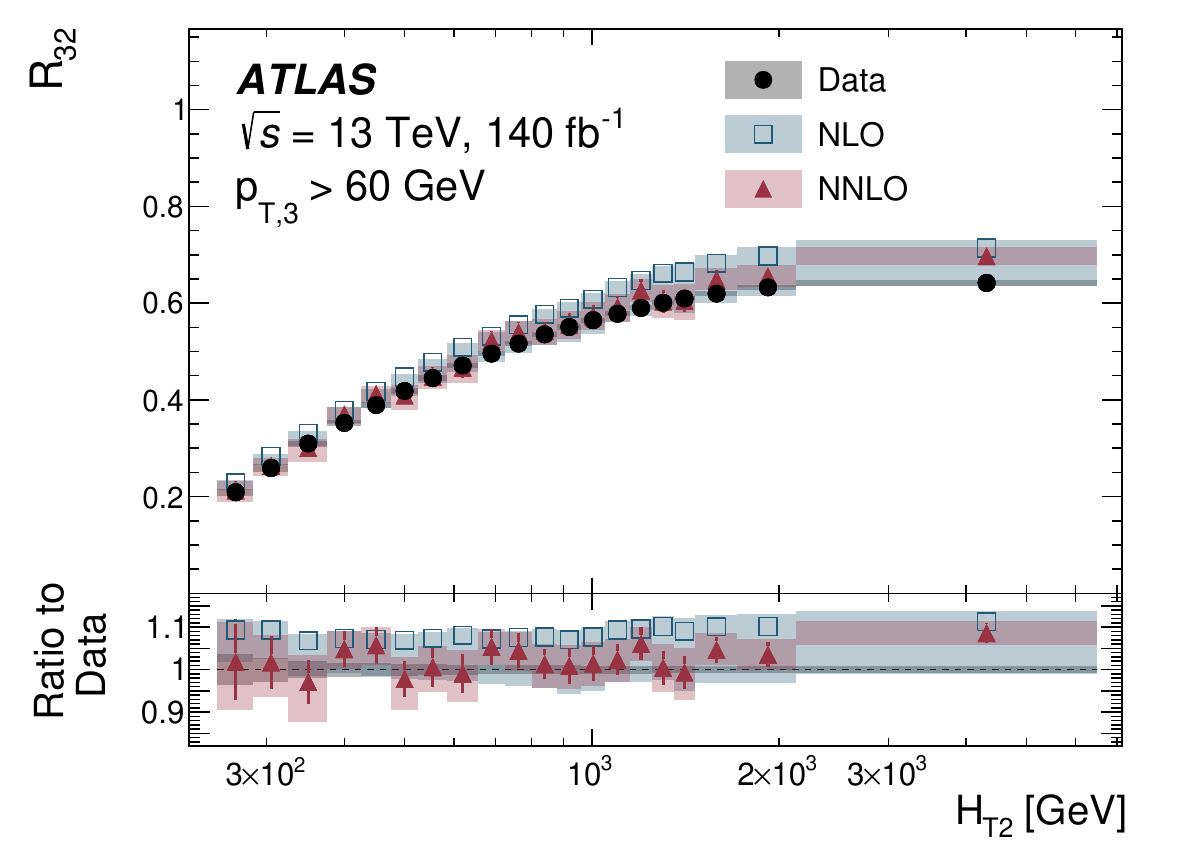}}
\subfigure[]{\includegraphics[width=0.49\textwidth]{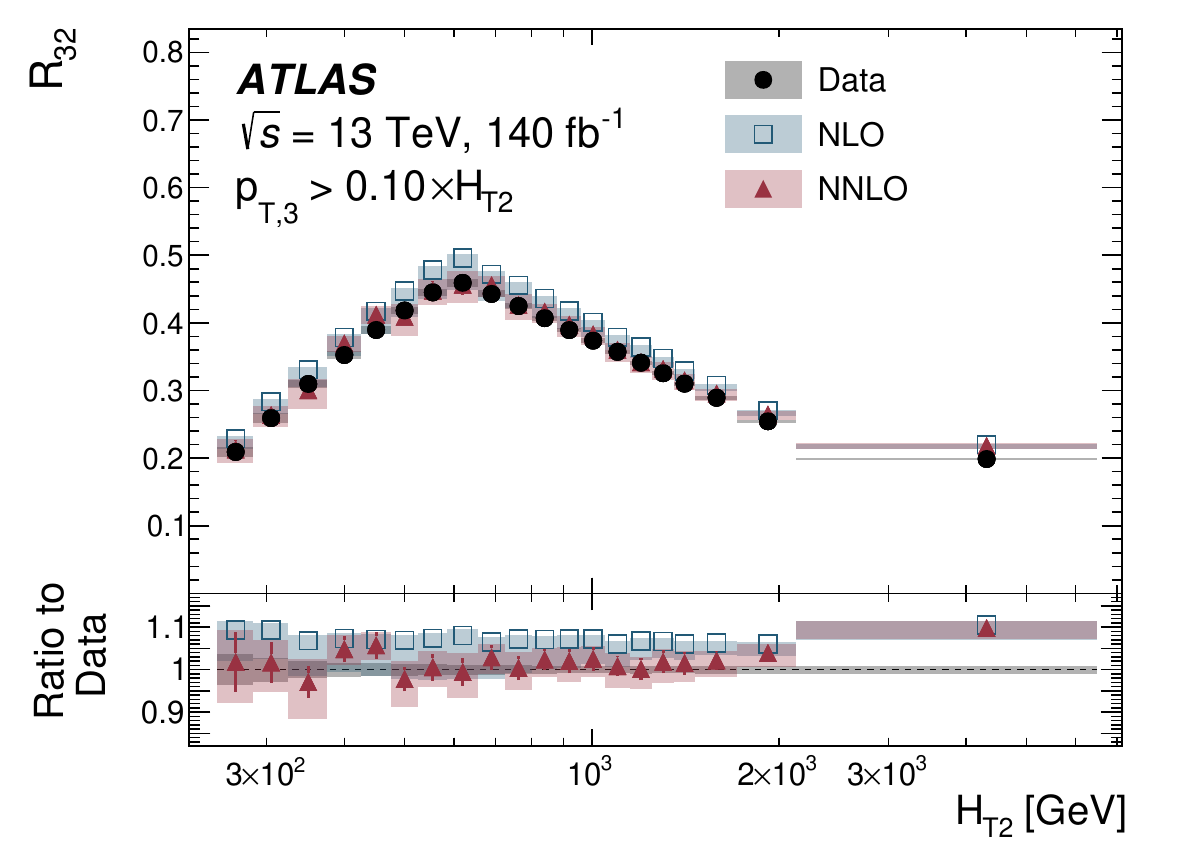}}
\subfigure[]{\includegraphics[width=0.49\textwidth]{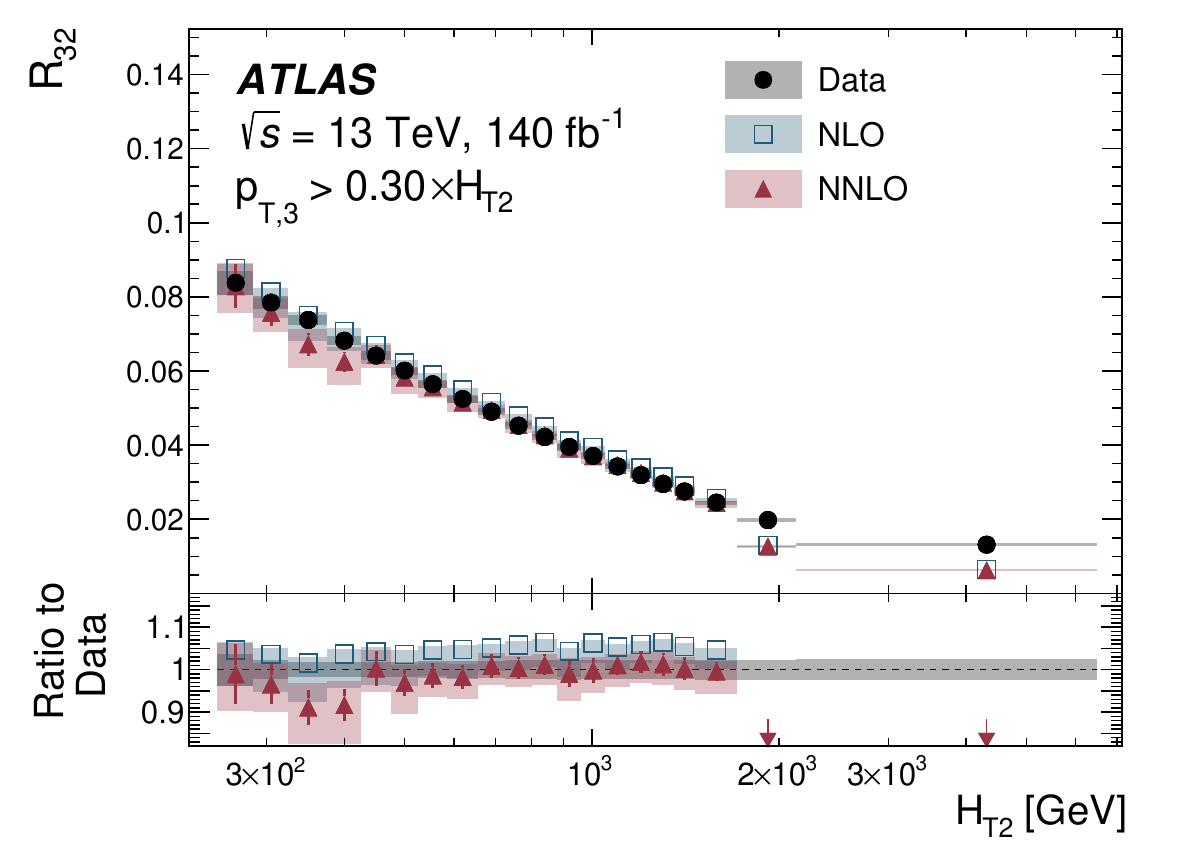}}
\caption{\RThreeTwo vs.~\HTTwo~with~(a) $\ptthree > 60~\GeV$, (b) $\ptthree > 0.10\times\HTTwo$, and (c) $\ptthree > 0.30\times\HTTwo$. The data error bands show the statistical and systematic components summed in quadrature. The theory error bands include contributions from the statistical, PDF, and scale variations, where the scale variations are determined from a seven-point variation of the renormalization and factorization scales used in the prediction. The statistical uncertainty on the theory predictions is illustrated with a vertical line.
The lower figure panels provide ratios of the predictions to the unfolded data. Arrows are used to indicated cases where the ratio falls outside of the plotted ratio range.}
\label{fig:ratios:httwo:pt3:theory}
\end{figure}

The ratios of three-jet to two-jet cross-sections measured as a function of \dyjj, \dymax, \mjj and \mjjmax are compared with HEJ predictions in Figure~\ref{fig:ratios:rapidity:hej}.
The HEJ predictions underestimate the multiplicity at low values of \dyjj for \RThreeTwo, while providing a better prediction of the region with large rapidity differences.
For \dymax, the HEJ predictions significantly overestimate the multiplicity distribution for intermediate rapidity differences while providing a good description of the largest-rapidity-difference region.
The HEJ predictions provide good modeling of the \RThreeTwo distribution for \mjj and \mjjmax, where the results agree with the data within uncertainties.

\begin{figure}[tbh]
\centering
\subfigure[]{\includegraphics[width=0.49\textwidth]{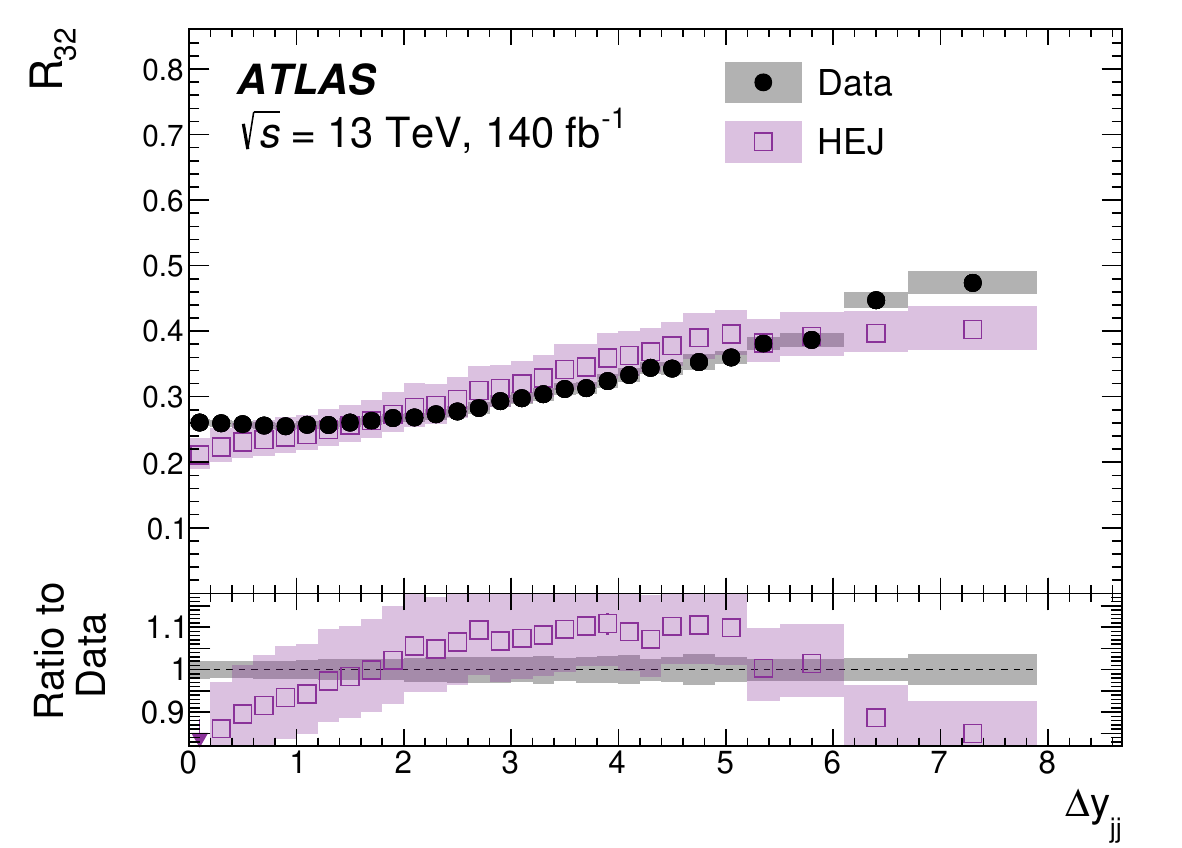}} %
\subfigure[]{\includegraphics[width=0.49\textwidth]{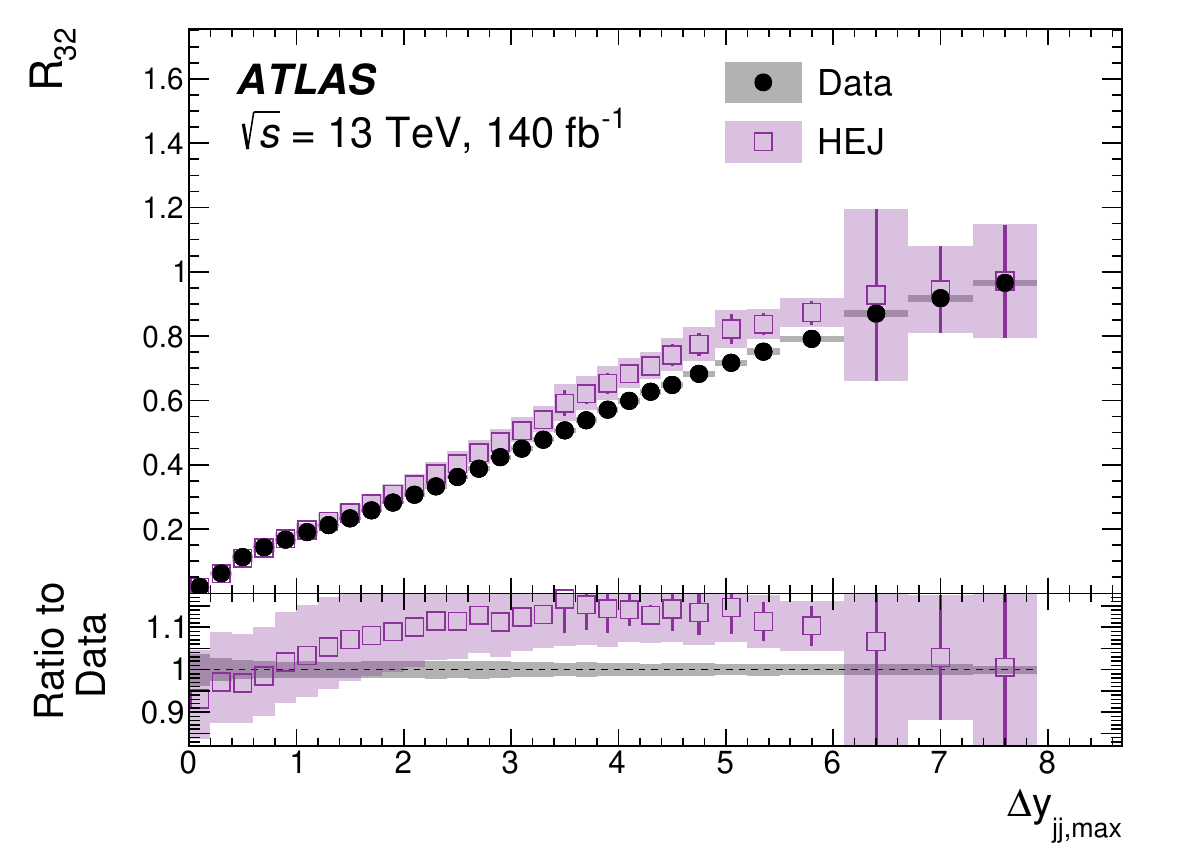}} %
\subfigure[]{\includegraphics[width=0.49\textwidth]{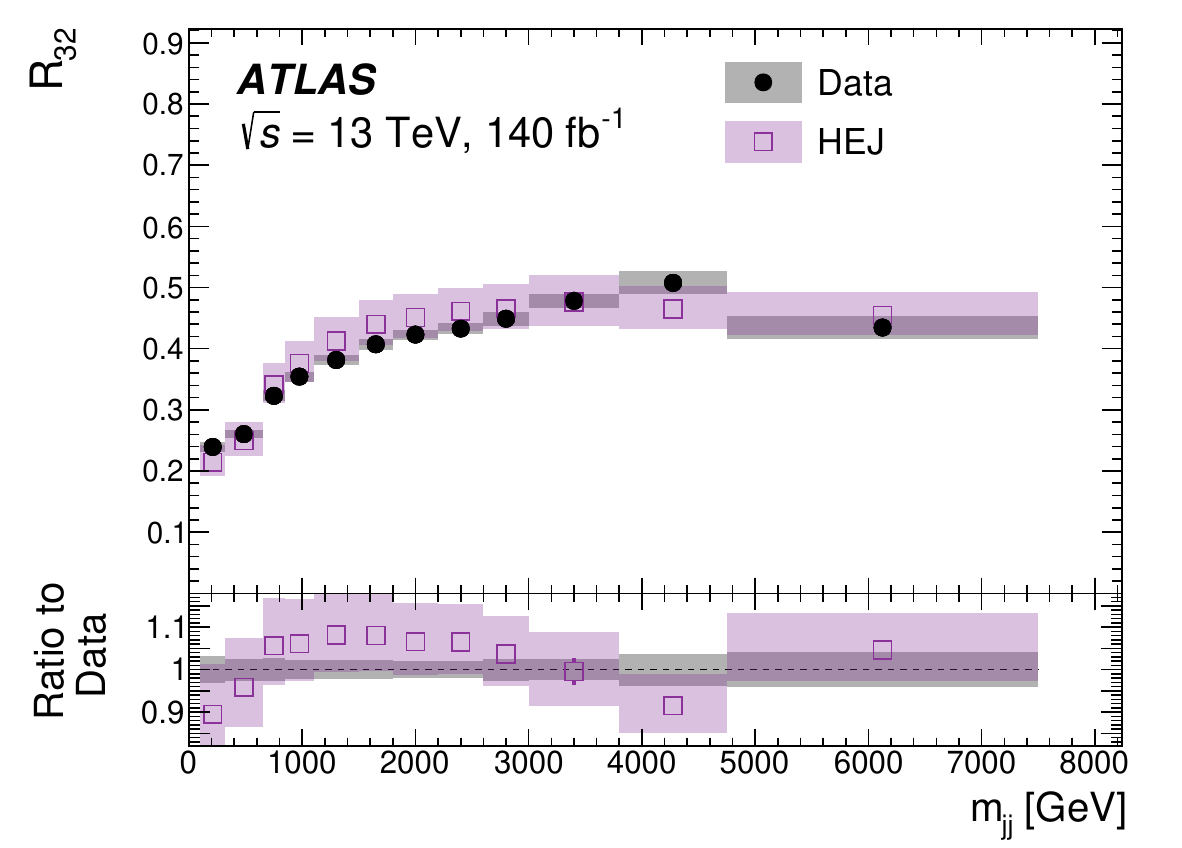}} %
\subfigure[]{\includegraphics[width=0.49\textwidth]{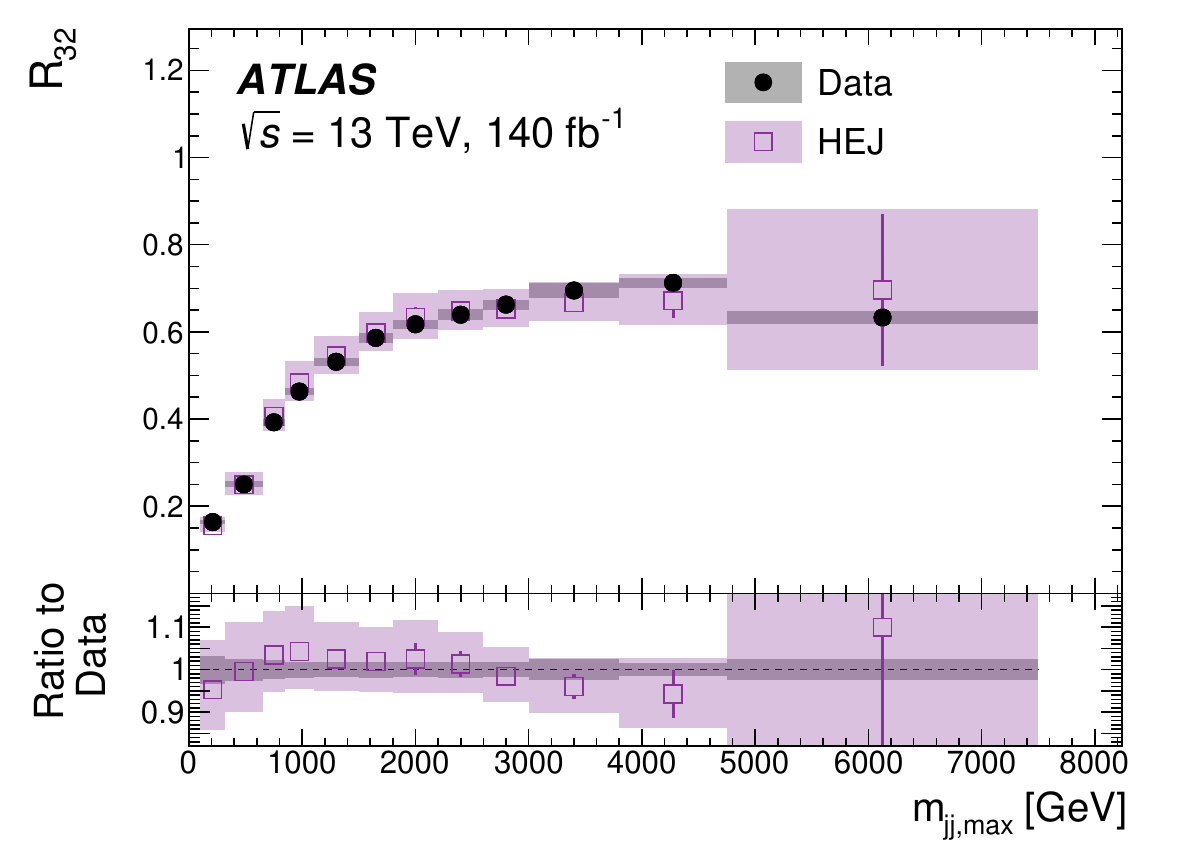}} %
\caption{\RThreeTwo vs. (a)~\dyjj, (b)~\dymax, (c)~\mjj and (d)~\mjjmax~with~$\ptthree > 60~\GeV$. The data error bands show the statistical and systematic components summed in quadrature. The theory error bands are determined from a seven-point variation of the renormalization and factorization scales used in the prediction.
The statistical uncertainty on the theory predictions is illustrated with a vertical line.
The lower figure panels provide ratios of the predictions to the unfolded data. Arrows are used to indicated cases where the ratio falls outside of the plotted ratio range.}
\label{fig:ratios:rapidity:hej}
\end{figure}


%

\FloatBarrier
\section{Concluding remarks}
\label{sec:conclusions}

This paper reports a measurement of jet cross-section ratios between inclusive bins of jet multiplicity, performed in 140~\ifb\ of proton--proton collisions with $\sqrt{s}=13~\TeV$ center-of-mass energy that were recorded with the ATLAS detector at CERN's Large Hadron Collider.
Observables that are sensitive to either the energy-scale (\HTTwo, \ptnincl) or angular distribution (\mjj, \mjjmax, \dyjj, \dymax) of hadronic energy flow in the final state are measured double-differentially, in bins of inclusive jet multiplicity, and the scalar sum of the two leading jets' transverse momenta is measured triple-differentially, in bins of the third jet's transverse momentum as well as bins of jet multiplicity.
Several improvements to the modeling of jet energy scale uncertainties are described, and these result in a significant reduction of the overall ATLAS jet energy scale uncertainty.
In particular, improvements in the Monte Carlo models used to define the \enquote{jet flavor response} uncertainty have reduced that source of uncertainty by up to a factor of two for jets with $\pt=100~\GeV$.
An updated procedure for the jet energy scale uncertainties derived from a single-particle deconvolution method at high jet \pt has reduced that source of uncertainty by roughly a factor of three for jets with $\pt=2~\TeV$, leading to increased precision of the differential cross-sections reported in this work.

The measured distributions are used to construct ratios of the inclusive three-jet to two-jet, four-jet to three-jet, four-jet to two-jet and five-jet to four-jet multiplicity bins, reducing sensitivity to systematic uncertainties and parton distribution functions.
Because uncertainties partially cancel out in the ratios, good precision is achieved for all observables, with the remaining uncertainties typically being less than a few percent for the measurements of the ratios of \HTTwo cross-sections, and less than 10\% for the dijet mass and $\Delta y$ observables.
The \HTTwo distribution is compared with NNLO fixed-order QCD predictions, which are found to model the behavior well across most \ptthree bins.
For the dijet mass and $\Delta y$ observables, significant differences between data and Monte Carlo predictions are observed at large values.
A prediction with additional resummation for logarithmic contributions that arise in topologies characteristic of VBS/VBF events, where high-energy jets are present in the forward region, provides a good description of the measured ratios in regions where the logarithmic terms make significant contributions.

A \textsc{Rivet} routine is available for this measurement~\cite{Buckley:2010ar}, and the measured data points have been made publicly available~\cite{hepdata.105630} for use in future Monte Carlo tuning campaigns and other studies of QCD at the electroweak scale.


%

%
\section*{Acknowledgements}
%

%
%

%
%

We thank CERN for the very successful operation of the LHC and its injectors, as well as the support staff at
CERN and at our institutions worldwide without whom ATLAS could not be operated efficiently.

The crucial computing support from all WLCG partners is acknowledged gratefully, in particular from CERN, the ATLAS Tier-1 facilities at TRIUMF/SFU (Canada), NDGF (Denmark, Norway, Sweden), CC-IN2P3 (France), KIT/GridKA (Germany), INFN-CNAF (Italy), NL-T1 (Netherlands), PIC (Spain), RAL (UK) and BNL (USA), the Tier-2 facilities worldwide and large non-WLCG resource providers. Major contributors of computing resources are listed in Ref.~\cite{ATL-SOFT-PUB-2023-001}.

We gratefully acknowledge the support of ANPCyT, Argentina; YerPhI, Armenia; ARC, Australia; BMWFW and FWF, Austria; ANAS, Azerbaijan; CNPq and FAPESP, Brazil; NSERC, NRC and CFI, Canada; CERN; ANID, Chile; CAS, MOST and NSFC, China; Minciencias, Colombia; MEYS CR, Czech Republic; DNRF and DNSRC, Denmark; IN2P3-CNRS and CEA-DRF/IRFU, France; SRNSFG, Georgia; BMBF, HGF and MPG, Germany; GSRI, Greece; RGC and Hong Kong SAR, China; ISF and Benoziyo Center, Israel; INFN, Italy; MEXT and JSPS, Japan; CNRST, Morocco; NWO, Netherlands; RCN, Norway; MNiSW, Poland; FCT, Portugal; MNE/IFA, Romania; MESTD, Serbia; MSSR, Slovakia; ARRS and MIZ\v{S}, Slovenia; DSI/NRF, South Africa; MICINN, Spain; SRC and Wallenberg Foundation, Sweden; SERI, SNSF and Cantons of Bern and Geneva, Switzerland; MOST, Taipei; TENMAK, T\"urkiye; STFC, United Kingdom; DOE and NSF, United States of America.

Individual groups and members have received support from BCKDF, CANARIE, CRC and DRAC, Canada; CERN-CZ, PRIMUS 21/SCI/017 and UNCE SCI/013, Czech Republic; COST, ERC, ERDF, Horizon 2020, ICSC-NextGenerationEU and Marie Sk{\l}odowska-Curie Actions, European Union; Investissements d'Avenir Labex, Investissements d'Avenir Idex and ANR, France; DFG and AvH Foundation, Germany; Herakleitos, Thales and Aristeia programmes co-financed by EU-ESF and the Greek NSRF, Greece; BSF-NSF and MINERVA, Israel; Norwegian Financial Mechanism 2014-2021, Norway; NCN and NAWA, Poland; La Caixa Banking Foundation, CERCA Programme Generalitat de Catalunya and PROMETEO and GenT Programmes Generalitat Valenciana, Spain; G\"{o}ran Gustafssons Stiftelse, Sweden; The Royal Society and Leverhulme Trust, United Kingdom.

In addition, individual members wish to acknowledge support from CERN: European Organization for Nuclear Research (CERN PJAS); Chile: Agencia Nacional de Investigaci\'on y Desarrollo (FONDECYT 1190886, FONDECYT 1210400, FONDECYT 1230812, FONDECYT 1230987); China: National Natural Science Foundation of China (NSFC - 12175119, NSFC 12275265, NSFC-12075060); Czech Republic: PRIMUS Research Programme (PRIMUS/21/SCI/017); European Union: European Research Council (ERC - 948254, ERC 101089007), Horizon 2020 Framework Programme (MUCCA - CHIST-ERA-19-XAI-00), European Union, Future Artificial Intelligence Research (FAIR-NextGenerationEU PE00000013), Italian Center for High Performance Computing, Big Data and Quantum Computing (ICSC, NextGenerationEU); France: Agence Nationale de la Recherche (ANR-20-CE31-0013, ANR-21-CE31-0013, ANR-21-CE31-0022), Investissements d'Avenir Labex (ANR-11-LABX-0012); Germany: Baden-Württemberg Stiftung (BW Stiftung-Postdoc Eliteprogramme), Deutsche Forschungsgemeinschaft (DFG - 469666862, DFG - CR 312/5-2); Italy: Istituto Nazionale di Fisica Nucleare (ICSC, NextGenerationEU); Japan: Japan Society for the Promotion of Science (JSPS KAKENHI JP21H05085, JSPS KAKENHI JP22H01227, JSPS KAKENHI JP22H04944, JSPS KAKENHI JP22KK0227); Netherlands: Netherlands Organisation for Scientific Research (NWO Veni 2020 - VI.Veni.202.179); Norway: Research Council of Norway (RCN-314472); Poland: Polish National Agency for Academic Exchange (PPN/PPO/2020/1/00002/U/00001), Polish National Science Centre (NCN 2021/42/E/ST2/00350, NCN OPUS nr 2022/47/B/ST2/03059, NCN UMO-2019/34/E/ST2/00393, UMO-2020/37/B/ST2/01043, UMO-2021/40/C/ST2/00187, UMO-2022/47/O/ST2/00148); Slovenia: Slovenian Research Agency (ARIS grant J1-3010); Spain: BBVA Foundation (LEO22-1-603), Generalitat Valenciana (Artemisa, FEDER, IDIFEDER/2018/048), Ministry of Science and Innovation (MCIN \& NextGenEU PCI2022-135018-2, MICIN \& FEDER PID2021-125273NB, RYC2019-028510-I, RYC2020-030254-I, RYC2021-031273-I, RYC2022-038164-I), PROMETEO and GenT Programmes Generalitat Valenciana (CIDEGENT/2019/023, CIDEGENT/2019/027); Sweden: Swedish Research Council (VR 2018-00482, VR 2022-03845, VR 2022-04683, VR grant 2021-03651), Knut and Alice Wallenberg Foundation (KAW 2017.0100, KAW 2018.0157, KAW 2018.0458, KAW 2019.0447, KAW 2022.0358); Switzerland: Swiss National Science Foundation (SNSF - PCEFP2\_194658); United Kingdom: Leverhulme Trust (Leverhulme Trust RPG-2020-004), Royal Society (NIF-R1-231091); United States of America: U.S. Department of Energy (ECA DE-AC02-76SF00515), Neubauer Family Foundation.

%
%


%
%
%
%
%
%

%
%
%
\printbibliography

@article{Sjostrand:2004ef,
    author = "Sj{\"o}strand, T. and Skands, Peter Z.",
    title = "{Transverse-momentum-ordered showers and interleaved multiple interactions}",
    eprint = "hep-ph/0408302",
    archivePrefix = "arXiv",
    reportNumber = "LU-TP-04-29",
    doi = "10.1140/epjc/s2004-02084-y",
    journal = "Eur. Phys. J. C",
    volume = "39",
    pages = "129--154",
    year = "2005"
}

@article{Czakon:2021mjy,
    author = "Czakon, Michal and Mitov, Alexander and Poncelet, Rene",
    title = "{Next-to-Next-to-Leading Order Study of Three-Jet Production at the LHC}",
    eprint = "2106.05331",
    archivePrefix = "arXiv",
    primaryClass = "hep-ph",
    reportNumber = "Cavendish-HEP-21/09, P3H-21-043, TTK-21-20",
    doi = "10.1103/PhysRevLett.127.152001",
    journal = "Phys. Rev. Lett.",
    volume = "127",
    number = "15",
    pages = "152001",
    year = "2021"
}

@article{Andersen:2011hs,
    author = "Andersen, Jeppe R. and Smillie, Jennifer M.",
    title = "{Multiple Jets at the LHC with High Energy Jets}",
    eprint = "1101.5394",
    archivePrefix = "arXiv",
    primaryClass = "hep-ph",
    reportNumber = "CP3-ORIGINS-2011-02, EDINBURGH-2011-03",
    doi = "10.1007/JHEP06(2011)010",
    journal = "JHEP",
    volume = "06",
    pages = "010",
    year = "2011"
}

@article{Andersen:2020yax,
    author = "Andersen, Jeppe R. and Black, James A. and Brooks, Helen M. and Byrne, Emmet P. and Maier, Andreas and Smillie, Jennifer M.",
    title = "{Combined subleading high-energy logarithms and NLO accuracy for W production in association with multiple jets}",
    eprint = "2012.10310",
    archivePrefix = "arXiv",
    primaryClass = "hep-ph",
    reportNumber = "DESY 20-233, DESY-20-233, DCPT/20/134, IPPP/20/67, MCnet-20-27, SAGEX-20-28",
    doi = "10.1007/JHEP04(2021)105",
    journal = "JHEP",
    volume = "04",
    pages = "105",
    year = "2021"
}

@article{Lendermann:2009ah,
    author = "Lendermann, Victor and Haller, Johannes and Herbst, Michael and Kruger, Katja and Schultz-Coulon, Hans-Christian and Stamen, Rainer",
    title = "{Combining Triggers in HEP Data Analysis}",
    eprint = "0901.4118",
    archivePrefix = "arXiv",
    primaryClass = "hep-ex",
    reportNumber = "CERN-OPEN-2009-001, HD-KIP-09-02",
    doi = "10.1016/j.nima.2009.03.173",
    journal = "Nucl. Instrum. Meth. A",
    volume = "604",
    pages = "707--718",
    year = "2009"
}

@article{LUCID2,
  author={G. Avoni and others},
  title={The new LUCID-2 detector for luminosity measurement and monitoring in ATLAS},
  journal={JINST},
  volume={13},
  number={07},
  pages={P07017},
  doi="10.1088/1748-0221/13/07/P07017",
  year={2018},
}

@article{Alioli:2010xa,
      author         = "Alioli, Simone and Hamilton, Keith and Nason, Paolo and
                        Oleari, Carlo and Re, Emanuele",
      title          = "{Jet pair production in POWHEG}",
      journal        = "JHEP",
      volume         = "04",
      year           = "2011",
      pages          = "081",
      doi            = "10.1007/JHEP04(2011)081",
      eprint         = "1012.3380",
      archivePrefix  = "arXiv",
      primaryClass   = "hep-ph",
      reportNumber   = "DESY-10-233, SFB-CPP-10-128, IPPP-10-100, DCPT-10-200,
                        MCNET-10-23",
      SLACcitation   = "%%CITATION = ARXIV:1012.3380;%%"
}

@article{Webber:1983if,
      author         = "Webber, B. R.",
      title          = "{A QCD model for jet fragmentation including soft gluon
                        interference}",
      journal        = "Nucl. Phys. B",
      volume         = "238",
      year           = "1984",
      pages          = "492-528",
      doi            = "10.1016/0550-3213(84)90333-X",
      reportNumber   = "CERN-TH-3713",
      SLACcitation   = "%%CITATION = NUPHA,B238,492;%%"
}

@article{Bellm:2017bvx,
      author         = "Bellm, Johannes and others",
      title          = "{Herwig 7.1 Release Note}",
      year           = "2017",
      eprint         = "1705.06919",
      archivePrefix  = "arXiv",
      primaryClass   = "hep-ph",
      reportNumber   = "CERN-PH-TH-2017-109, CERN-TH-2017-109, MAN-HEP-2017-08,
                        UWTHPH-2017-10, IFJPAN-IV-2017-7, NIKHEF-2017-026,
                        HERWIG-2017-02, KA-TP-19-2017, MCNET-17-08, IPPP-17-40",
      SLACcitation   = "%%CITATION = ARXIV:1705.06919;%%"
}

@misc{unfolding,
     howpublished     = "\url{https://twiki.cern.ch/twiki/bin/viewauth/AtlasProtected/StandardModelUnfolding}",
}

@article{nlojet1,
     author = "{Z. Nagy}",
     title = "{Three-Jet Cross Sections in Hadron-Hadron Collisions at Next-To-Leading-Order}",
     journal = "{Phys. Rev. Lett.}",
     volume = "88",
     pages = "122003",
     year = "2002",
     doi="10.1103/PhysRevLett.88.122003",
     archivePrefix = "arXiv",
     eprint = "hep-ph/0110315",
     primaryClass = "hep-ph"
}

@article{nlojet2,
    author = "{Z. Nagy}",
    title = "{Next-to-leading order calculation of three-jet observables in hadron-hadron collisions}",
    journal = "{Phys. Rev. D}",
    volume = "68",
    pages = "094002",
    year = "2003",
    doi = "10.1103/PhysRevD.68.094002",
    archivePrefix = "arXiv",
    eprint = "hep-ph/0307268",
    primaryClass = "hep-ph"
}

@article{lhapdf,
   author = "{A. Buckley et al.}",
   title = "{LHAPDF6: parton density access in the LHC precision era}",
   journal = "{Eur. Phys. J. C}",
   volume = "75",
   pages = "132",
   year = "2015",
   doi = "10.1140/epjc/s10052-015-3318-8",
   archivePrefix = "arXiv",
   eprint = "1412.7420",
   primaryClass = "hep-ph"
}

@article{D0:2000tld,
    author = "{D0 Collaboration}",
    title = "{Ratios of Multijet Cross Sections in \ppbar collisions at $\sqrt{s} = 1.8$ TeV}",
    eprint = "hep-ex/0009012",
    archivePrefix = "arXiv",
    reportNumber = "FERMILAB-PUB-00-218-E",
    doi = "10.1103/PhysRevLett.86.1955",
    journal = "Phys. Rev. Lett.",
    volume = "86",
    pages = "1955--1960",
    year = "2001"
}

@article{Hou:2019efy,
    author = "Hou, Tie-Jiun and others",
    title = "{New CTEQ global analysis of quantum chromodynamics with high-precision data from the LHC}",
    eprint = "1912.10053",
    archivePrefix = "arXiv",
    primaryClass = "hep-ph",
    reportNumber = "MSUHEP-19-025, PITT-PACC-1911, SMU-HEP-19-03",
    doi = "10.1103/PhysRevD.103.014013",
    journal = "Phys. Rev. D",
    volume = "103",
    number = "1",
    pages = "014013",
    year = "2021"
}

@article{Malaescu:2009dm,
    author = "Malaescu, Bogdan",
    title = "{An Iterative, dynamically stabilized method of data unfolding}",
    eprint = "0907.3791",
    archivePrefix = "arXiv",
    primaryClass = "physics.data-an",
    reportNumber = "LAL-09-107",
    month = "7",
    year = "2009"
}

@article{nnpdf40,
  author = "{NNPDF Collaboration, R.D. Ball et al.}",
  title = "{The path to proton structure at 1\% accuracy}",
  journal = "{Eur. Phys. J. C}",
  volume = "82",
  pages = "428",
  year = "2022",
  doi = "10.1140/epjc/s10052-022-10328-7",
  archivePrefix = "arXiv",
  eprint = "2109.02653",
  primaryClass = "hep-ph"
}

@article{msht20,
    author = "Bailey, S. and Cridge, T. and Harland-Lang, L. A. and Martin, A. D. and Thorne, R. S.",
    title = "{Parton distributions from LHC, HERA, Tevatron and fixed target data: MSHT20 PDFs}",
    eprint = "2012.04684",
    archivePrefix = "arXiv",
    primaryClass = "hep-ph",
    reportNumber = "IPPP/20/58",
    doi = "10.1140/epjc/s10052-021-09057-0",
    journal = "Eur. Phys. J. C",
    volume = "81",
    number = "4",
    pages = "341",
    year = "2021"
}

@article{Adragna:2010zz,
    author = "Adragna, P. and others",
    title = "{Measurement of pion and proton response and longitudinal shower profiles up to 20 nuclear interaction lengths with the ATLAS tile calorimeter}",
    doi = "10.1016/j.nima.2010.01.037",
    journal = "Nucl. Instrum. Meth. A",
    volume = "615",
    pages = "158--181",
    year = "2010"
}

@article{Abdallah:2021gai,
    author = "Abdallah, Jalal and others",
    title = "{Study of energy response and resolution of the ATLAS Tile Calorimeter to hadrons of energies from 16 to 30 GeV}",
    eprint = "2102.04088",
    archivePrefix = "arXiv",
    primaryClass = "physics.ins-det",
    doi = "10.1140/epjc/s10052-021-09292-5",
    journal = "Eur. Phys. J. C",
    volume = "81",
    number = "6",
    pages = "549",
    year = "2021"
}

@article{Cacciari:2007fd,
    author = "Cacciari, Matteo and Salam, Gavin P.",
    title = "{Pileup subtraction using jet areas}",
    eprint = "0707.1378",
    archivePrefix = "arXiv",
    primaryClass = "hep-ph",
    reportNumber = "LPTHE-07-01",
    doi = "10.1016/j.physletb.2007.09.077",
    journal = "Phys. Lett. B",
    volume = "659",
    pages = "119--126",
    year = "2008"
}

@article{Gras:2017jty,
    author = {Gras, Philippe and H\"oche, Stefan and Kar, Deepak and Larkoski, Andrew and L\"onnblad, Leif and Pl\"atzer, Simon and Si\'odmok, Andrzej and Skands, Peter and Soyez, Gregory and Thaler, Jesse},
    title = "{Systematics of quark/gluon tagging}",
    eprint = "1704.03878",
    archivePrefix = "arXiv",
    primaryClass = "hep-ph",
    reportNumber = "MIT-CTP-4885, COEPP-MN-17-2, MCNET-17-04",
    doi = "10.1007/JHEP07(2017)091",
    journal = "JHEP",
    volume = "07",
    pages = "091",
    year = "2017"
}

@article{Komiske:2018vkc,
    author = "Komiske, Patrick T. and Metodiev, Eric M. and Thaler, Jesse",
    title = "{An operational definition of quark and gluon jets}",
    eprint = "1809.01140",
    archivePrefix = "arXiv",
    primaryClass = "hep-ph",
    reportNumber = "MIT-CTP 5042",
    doi = "10.1007/JHEP11(2018)059",
    journal = "JHEP",
    volume = "11",
    pages = "059",
    year = "2018"
}

@article{Begel:2022kwp,
    author = "Begel, M. and others",
    title = "{Precision QCD, Hadronic Structure \& Forward QCD, Heavy Ions: Report of Energy Frontier Topical Groups 5, 6, 7 submitted to Snowmass 2021}",
    eprint = "2209.14872",
    archivePrefix = "arXiv",
    primaryClass = "hep-ph",
    reportNumber = "FERMILAB-CONF-22-733-SCD-T, SMU-HEP-22-06",
    month = "9",
    year = "2022"
}

@article{Nachman:2022emq,
    author = "Bonilla, Johan and others",
    title = "{Jets and Jet Substructure at Future Colliders}",
    eprint = "2203.07462",
    archivePrefix = "arXiv",
    primaryClass = "hep-ph",
    reportNumber = "FERMILAB-PUB-22-186-SCD-T",
    doi = "10.3389/fphy.2022.897719",
    journal = "Front. in Phys.",
    volume = "10",
    pages = "897719",
    year = "2022"
}

@article{Affolder:2001jx,
 author = "{CDF Collaboration}", 
      title          = "{Charged-Particle Multiplicity in $p\bar{p}$
                        Collisions at $\sqrt{s} = 1.8$ TeV}",
      collaboration  = "CDF",
      journal        = "Phys. Rev. Lett.",
      volume         = "87",
      year           = "2001",
      pages          = "211804",
      doi            = "10.1103/PhysRevLett.87.211804",
}

@article{PhysRevLett.72.3145,
  title = {Measurement of the charged multiplicity of $Z^0\rightarrow b\bar{b}$ events},
 author = "{SLD Collaboration}", 
  journal = {Phys. Rev. Lett.},
  volume = {72},
  pages = {3145},
  year = {1994},
  doi = {10.1103/PhysRevLett.72.3145},
  eprint         = "hep-ex/9405004",
      archivePrefix  = "arXiv",
}

@article{ua22,
year={1983},
journal={Z. Phys. C},
volume={20},
doi={10.1007/BF01573214},
title={Measurement of production and properties of jets at the CERN $\bar{p}p$ collider},
author = "{UA2 Collaboration}",
pages={117},
}

@article{BAGNAIA1984291,
title = "Measurement of jet fragmentation properties at the CERN $p\bar{p}$ Collider",
journal = "Phys. Lett. B",
volume = "144",
pages = "291",
year = "1984",
doi = "10.1016/0370-2693(84)91823-9",
author = "{UA2 Collaboration}",

}

@article{Arnison:167874,
      author = "{UA1 Collaboration}",
      title         = "{Analysis of the fragmentation properties of quark and
                       gluon jets at the CERN SPS $p\bar{p}$ collider}",
      journal       = "Nucl. Phys. B",
      volume        = "276",
      pages         = "253",
      year          = "1986",
      doi            = "10.1016/0550-3213(86)90296-8",
}

@article{opallep2,
year={2004},
journal={Eur. Phys. J. C},
volume={37},
doi={10.1140/epjc/s2004-01964-4},
title={Scaling violations of quark and gluon jet fragmentation functions in $ e^ +$$ e^-$ annihilations at $\sqrt s = 91.2$ and $183-209$ GeV},
pages={25},
eprint         = "hep-ex/0404026",
      archivePrefix  = "arXiv",
author = "{OPAL Collaboration}"
}

@article{Akrawy1990617,
title = "A study of coherence of soft gluons in hadron jets ",
journal = "Phys. Lett. B",
volume = "247",
pages = "617",
year = "1990",
doi = "10.1016/0370-2693(90)91911-T",
author = "{OPAL Collaboration}"
}

@article{PhysRevD.56.17,
  title = {Study of gluon versus quark fragmentation in $\ensuremath{\Upsilon}\ensuremath{\rightarrow}\mathrm{gg}\ensuremath{\gamma}$ and ${e}^{+}{e}^{-}\ensuremath{\rightarrow}q\overline{q}\ensuremath{\gamma}$ events at $\sqrt{s}$=10 GeV},
  author = "{CLEO Collaboration}",
  journal = {Phys. Rev. D},
  volume = {56},
  pages = {17},
  year = {1997},
      eprint         = "hep-ex/9701006",
      archivePrefix  = "arXiv",
  doi = {10.1103/PhysRevD.56.17},}

@article{Abreu1999383,
title = "The scale dependence of the hadron multiplicity in quark and gluon jets and a precise determination of $C_A/C_F$ ",
journal = "Phys. Lett. B",
volume = "449",
pages = "383",
year = "1999",
doi = "10.1016/S0370-2693(99)00112-4",
  eprint         = "hep-ex/9903073",
      archivePrefix  = "arXiv",
author = "{DELPHI Collaboration}"
}

@article{ refId1,
author = "{OPAL Collaboration}",
      title = {Experimental properties of gluon 
and quark jets from a point source
},
      DOI= "10.1007/s100529900181",
      journal = {Eur. Phys. J. C},
      year = "1999",
      volume = "11",
      pages = "217",
       eprint         = "hep-ex/9903027",
      archivePrefix  = "arXiv",
}

@article{ refId0,
author = "{OPAL Collaboration}",
      title = {Multiplicity distributions of
gluon and quark jets and tests 
of QCD analytic predictions
},
      DOI= "10.1007/s100520050097",
      journal = {Eur. Phys. J. C},
      year = "1998",
      volume = "1",
      pages = "479",
           eprint         = "hep-ex/9708029",
      archivePrefix  = "arXiv",
}

@article{delphintrack,
year={1996},
journal={Z. Phys. C},
volume={70},
doi={10.1007/s002880050095},
title={Energy dependence of the differences between the quark and gluon jet fragmentation},
pages={179},
author = "{DELPHI Collaboration}"
}

@article{Buskulic1996353,
title = "Quark and gluon jet properties in symmetric three-jet events ",
journal = "Phys. Lett. B",
volume = "384",
pages = "353",
year = "1996",
doi = "10.1016/0370-2693(96)00849-0",
author = "{ALEPH Collaboration}"
}

@article{Alexander1996659,
title = "Test of QCD analytic predictions for the multiplicity ratio between gluon and quark jets ",
journal = "Phys. Lett. B",
volume = "388",
pages = "659",
year = "1996",
doi = "10.1016/S0370-2693(96)01319-6",
author = "{OPAL Collaboration}"
}

@article{Buskulic1995389,
title = "Study of the subjet structure of quark and gluon jets ",
journal = "Phys. Lett. B",
volume = "346",
pages = "389",
year = "1995",
doi = "10.1016/0370-2693(95)00037-L",
author = "{ALEPH Collaboration}"
}

@article{xyz,
year={1995},
journal={Z. Phys. C},
volume={68},
doi={10.1007/BF01566667},
title={A model independent measurement of quark and gluon jet properties and differences},
author="{OPAL Collaboration}",
pages={179},
}

@article{Acton:1993jm,
      author         = "{OPAL Collaboration}",
      title          = "{A study of differences between quark and gluon jets
                        using vertex tagging of quark jets}",
      journal        = "Z. Phys. C",
      volume         = "58",
      year           = "1993",
      pages          = "387",
      doi            = "10.1007/BF01557696",
}

@article{Alexander:1991ce,
      author         = "{OPAL Collaboration}",
      title          = "{A direct observation of quark-gluon jet differences at
                        LEP}",
      journal        = "Phys. Lett. B",
      volume         = "265",
      year           = "1991",
      pages          = "462-474",
      doi            = "10.1016/0370-2693(91)90082-2",
}

@article{PhysRevLett.63.1772,
  title = {Comparison of Quark and Gluon Jets Produced in High-Energy ${e}^{+}$${e}^{\mathrm{-}}$ Annihilations},
  author = "{AMY Collaboration}",
  journal = {Phys. Rev. Lett.},
  volume = {63},
  pages = {1772},
  year = {1989},
  doi = {10.1103/PhysRevLett.63.1772},
}

@article{zphys,
year="1989",
journal="Z. Phys. C",
volume="45",
doi="10.1007/BF01674450",
title="{Charged multiplicity distributions and correlations in $e^+$$e^-$ annihilation at PETRA energies}",
author="{TASSO Collaboration}",
pages="193-208",
}

@article{Bartel:1982ub,
      author         = "{JADE Collabortion.}",
      title          = "{Experimental evidence for differences in $<p_\perp>$ between
                        quark jets and gluon jets}",
      collaboration  = "JADE",
      journal        = "Phys. Lett. B",
      volume         = "123",
      year           = "1983",
      pages          = "460",
      doi            = "10.1016/0370-2693(83)90994-2",
}

@article{PhysRevLett.55.1954,
  title = {Inclusive Charged-Particle Distribution in Nearly Threefold-Symmetric Three-Jet Events at ${E}_{\mathrm{c}.\mathrm{m}.}=29$ GeV},
  author = {{MARK II Collaboration}},
  journal = {Phys. Rev. Lett.},
  volume = {55},
  pages = {1954},
  year = {1985},
  month = {11},
  publisher = {American Physical Society},
  doi = {10.1103/PhysRevLett.55.1954},
}

@article{Derrick:163229,
      author        = "{HRS Collaboration}",
      title         = "{Comparison of charged particle multiplicities in quark
                       and gluon jets produced in $e^+e^-$ annihilation at 29 GeV}",
      journal       = "Phys. Lett. B",
      volume        = "165",
      pages         = "449",
      month         = "9",
      year          = "1985",
      doi = {10.1016/0370-2693(85)91263-8}
}

@article{Derrick:1995ca,
      author         = "{ZEUS Collaboration}",
      title          = "{Measurement of multiplicity and momentum spectra in the
                        current fragmentation region of the Breit frame at HERA}",
      collaboration  = "ZEUS",
      journal        = "Z. Phys. C",
      volume         = "67",
      year           = "1995",
      pages          = "93-108",
      doi            = "10.1007/BF01564824",
      eprint         = "hep-ex/9501012",
      archivePrefix  = "arXiv",
      reportNumber   = "DESY-95-007",
      SLACcitation   = "%%CITATION = HEP-EX/9501012;%%"
}

@article{Aid:1995up,
      author         = "{H1 Collaboration}",
      title          = "{A Study of the fragmentation of quarks in e$^-$p
                        collisions at HERA}",
      collaboration  = "H1",
      journal        = "Nucl. Phys. B",
      volume         = "445",
      year           = "1995",
      pages          = "3-21",
      doi            = "10.1016/0550-3213(95)91599-H",
      eprint         = "hep-ex/9505003",
      archivePrefix  = "arXiv",
      reportNumber   = "DESY-95-072",
      SLACcitation   = "%%CITATION = HEP-EX/9505003;%%"
}

@article{H1:2023fzk,
    author = "{H1 Collaboration}",
    title = "{Unbinned deep learning jet substructure measurement in high $Q^2$ $ep$ collisions at HERA}",
    eprint = "2303.13620",
    archivePrefix = "arXiv",
    primaryClass = "hep-ex",
    reportNumber = "DESY-23-034",
    doi = "10.1016/j.physletb.2023.138101",
    journal = "Phys. Lett. B",
    volume = "844",
    pages = "138101",
    year = "2023"
}

@article{PhysRevD.68.012003,
  title = {Momentum distribution of charged particles in jets in dijet events in $p\overline{p}$ collisions at $\sqrt{s}=1.8~\mathrm{TeV}$ and comparisons to perturbative QCD predictions},
  author = "{CDF Collaboration}",
  journal = {Phys. Rev. D},
  volume = {68},
  pages = {012003},
  numpages = {19},
  year = {2003},
  month = {7},
  publisher = {American Physical Society},
  doi = {10.1103/PhysRevD.68.012003},
}

@article{Aaltonen:2008ae,
  author = "{CDF Collaboration}",
      title          = "{Two-particle momentum correlations in jets produced in
                        $p\bar{p}$ collisions at $\sqrt{s}$ = 1.96~TeV}",
      journal        = "Phys. Rev. D",
      volume         = "77",
      year           = "2008",
      pages          = "092001",
      doi            = "10.1103/PhysRevD.77.092001",
      eprint         = "0802.3182",
      archivePrefix  = "arXiv",
      primaryClass   = "hep-ex",
      reportNumber   = "FERMILAB-PUB-08-037-E",
      SLACcitation   = "%%CITATION = ARXIV:0802.3182;%%"
}

@article{PhysRevLett.102.232002,
  title = {Measurement of the ${k}_{T}$ Distribution of Particles in Jets Produced in $p\overline{p}$ Collisions at $\sqrt{s}=1.96\text{ }\text{ }\mathrm{TeV}$},
  author = "{CDF Collaboration}",
  journal = {Phys. Rev. Lett.},
  volume = {102},
  pages = {232002},
  numpages = {8},
  year = {2009},
  month = {6},
  publisher = {American Physical Society},
  doi = {10.1103/PhysRevLett.102.232002},
  eprint = "0811.2820",
  archivePrefix = "arXiv",
  primaryClass = "hep-ex"
}

@article{Abe:1996zi,
      title          = "{Measurement of the charged multiplicities in b, c and
                        light quark events from $Z^0$ decays}",
       author = "{SLD Collaboration}", 
      journal        = "Phys. Lett. B",
      volume         = "386",
      year           = "1996",
      pages          = "475",
      doi            = "10.1016/0370-2693(96)01025-8",
      eprint         = "hep-ex/9608008",
      archivePrefix  = "arXiv",
}

@article{Czakon:2014oma,
    author = "Czakon, M. and Heymes, D.",
    title = "{Four-dimensional formulation of the sector-improved residue subtraction scheme}",
    eprint = "1408.2500",
    archivePrefix = "arXiv",
    primaryClass = "hep-ph",
    reportNumber = "TTK-14-16",
    doi = "10.1016/j.nuclphysb.2014.11.006",
    journal = "Nucl. Phys. B",
    volume = "890",
    pages = "152--227",
    year = "2014"
}

@article{Czakon:2019tmo,
    author = "Czakon, Micha\l{} and van Hameren, Andreas and Mitov, Alexander and Poncelet, Rene",
    title = "{Single-jet inclusive rates with exact color at $ \mathcal{O} $ ($ {\alpha}_s^4 $)}",
    eprint = "1907.12911",
    archivePrefix = "arXiv",
    primaryClass = "hep-ph",
    doi = "10.1007/JHEP10(2019)262",
    journal = "JHEP",
    volume = "10",
    pages = "262",
    year = "2019"
}

@article{Alvarez:2023fhi,
    author = "Alvarez, Manuel and Cantero, Josu and Czakon, Michal and Llorente, Javier and Mitov, Alexander and Poncelet, Rene",
    title = "{NNLO QCD corrections to event shapes at the LHC}",
    eprint = "2301.01086",
    archivePrefix = "arXiv",
    primaryClass = "hep-ph",
    reportNumber = "Cavendish-HEP-22/11, P3H-22-129, TTK-22-49",
    doi = "10.1007/JHEP03(2023)129",
    journal = "JHEP",
    volume = "03",
    pages = "129",
    year = "2023"
}

@article{Bury:2015dla,
    author = "Bury, M. and van Hameren, A.",
    title = "{Numerical evaluation of multi-gluon amplitudes for High Energy Factorization}",
    eprint = "1503.08612",
    archivePrefix = "arXiv",
    primaryClass = "hep-ph",
    reportNumber = "IFJPAN-IV-2015-4",
    doi = "10.1016/j.cpc.2015.06.023",
    journal = "Comput. Phys. Commun.",
    volume = "196",
    pages = "592--598",
    year = "2015"
}

@article{Abreu:2021oya,
    author = "Abreu, S. and Febres Cordero, F. and Ita, H. and Page, B. and Sotnikov, V.",
    title = "{Leading-color two-loop QCD corrections for three-jet production at hadron colliders}",
    eprint = "2102.13609",
    archivePrefix = "arXiv",
    primaryClass = "hep-ph",
    reportNumber = "CERN-TH-2021-023, FR-PHENO-2021-06, MPP-2021-20",
    doi = "10.1007/JHEP07(2021)095",
    journal = "JHEP",
    volume = "07",
    pages = "095",
    year = "2021"
}

@article{Chicherin:2020oor,
    author = "Chicherin, Dmitry and Sotnikov, Vasily",
    title = "{Pentagon Functions for Scattering of Five Massless Particles}",
    eprint = "2009.07803",
    archivePrefix = "arXiv",
    primaryClass = "hep-ph",
    reportNumber = "MPP-2020-171",
    doi = "10.1007/JHEP12(2020)167",
    journal = "JHEP",
    volume = "20",
    pages = "167",
    year = "2020"
}

@article{CMS-SMP-22-007,
    author = "{CMS Collaboration}",
    title = "{Measurement of the primary Lund jet plane density in proton-proton collisions at $\sqrt{s}$ = 13 TeV}",
    eprint = "2312.16343",
    archivePrefix = "arXiv",
    primaryClass = "hep-ex",
    reportNumber = "CMS-SMP-22-007, CERN-EP-2023-282",
    month = "12",
    year = "2023"
}

@article{STDM-2023-07,
    author = "{ATLAS Collaboration}",
    title = "{Measurements of Lund subjet multiplicities in 13 TeV proton-proton collisions with the ATLAS detector}",
    eprint = "2402.13052",
    archivePrefix = "arXiv",
    primaryClass = "hep-ex",
    reportNumber = "ATLAS-STDM-2023-07, CERN-EP-2024-029",
    month = "2",
    year = "2024"
}

@article{Chahal:2022rid,
    author = "Chahal, Gurpreet Singh and Krauss, Frank",
    title = "{Cluster Hadronisation in Sherpa}",
    eprint = "2203.11385",
    archivePrefix = "arXiv",
    primaryClass = "hep-ph",
    reportNumber = "IPPP/22/14",
    doi = "10.21468/SciPostPhys.13.2.019",
    journal = "SciPost Phys.",
    volume = "13",
    number = "2",
    pages = "019",
    year = "2022"
}

@article{Frixione:1997ks,
    author = "Frixione, Stefano and Ridolfi, Giovanni",
    title = "{Jet photoproduction at HERA}",
    eprint = "hep-ph/9707345",
    archivePrefix = "arXiv",
    reportNumber = "ETH-TH-97-21, GEF-TH-8-1997",
    doi = "10.1016/S0550-3213(97)00575-0",
    journal = "Nucl. Phys. B",
    volume = "507",
    pages = "315--333",
    year = "1997"
}

@article{ALICE:2019cbr,
    author = "{ALICE Collaboration}",
    title = "{Measurement of the production of charm jets tagged with D$^{0}$ mesons in pp collisions at $ \sqrt{\mathrm{s}}=7 $ TeV}",
    eprint = "1905.02510",
    archivePrefix = "arXiv",
    primaryClass = "nucl-ex",
    reportNumber = "CERN-EP-2019-081",
    doi = "10.1007/JHEP08(2019)133",
    journal = "JHEP",
    volume = "08",
    pages = "133",
    year = "2019"
}

@article{ALICE:2021aqk,
    author = "{ALICE Collaboration}",
    title = "{Direct observation of the dead-cone effect in quantum chromodynamics}",
    eprint = "2106.05713",
    archivePrefix = "arXiv",
    primaryClass = "nucl-ex",
    reportNumber = "CERN-EP-2021-107",
    doi = "10.1038/s41586-022-04572-w",
    journal = "Nature",
    volume = "605",
    number = "7910",
    pages = "440--446",
    year = "2022",
    related        = "ALICE:2021aqk-err",
    relatedstring  = "Erratum:"
}

@article{ALICE:2022phr,
    author = "{ALICE Collaboration}",
    title = "{Measurements of Groomed-Jet Substructure of Charm Jets Tagged by $D^0$ Mesons in Proton-Proton Collisions at s=13\,\,TeV}",
    eprint = "2208.04857",
    archivePrefix = "arXiv",
    primaryClass = "nucl-ex",
    reportNumber = "CERN-EP-2022-171",
    doi = "10.1103/PhysRevLett.131.192301",
    journal = "Phys. Rev. Lett.",
    volume = "131",
    number = "19",
    pages = "192301",
    year = "2023"
}

@article{Sherpa:2019gpd,
    author = "Bothmann, Enrico and others",
    collaboration = "Sherpa",
    title = "{Event Generation with Sherpa 2.2}",
    eprint = "1905.09127",
    archivePrefix = "arXiv",
    primaryClass = "hep-ph",
    reportNumber = "FERMILAB-PUB-19-218-T, SLAC-PUB-17433, IPPP/19/42, MCNET-19-11",
    doi = "10.21468/SciPostPhys.7.3.034",
    journal = "SciPost Phys.",
    volume = "7",
    number = "3",
    pages = "034",
    year = "2019"
}

@article{Currie:2015,
      author         = "Currie, James and Gehrmann-De Ridder, Aude and Gehrmann, Thomas and Glover, E. W. Nigel and Huss, Alexander and Pires, Jo\~ao",
      title          = "{Infrared sensitivity of single jet inclusive production at hadron colliders}",
      journal        = "JHEP",
      volume         = "10",
      year           = "2018",
      pages          = "155",
      doi            = "10.1007/JHEP10(2018)155",
      eprint         = "1807.03692",
      archivePrefix  = "arXiv",
      primaryClass   = "hep-ph",
      %reportNumber   = "DESY-10-233, SFB-CPP-10-128, IPPP-10-100, DCPT-10-200,and Jo\~ao Pires
      %                  MCNET-10-23",
      %SLACcitation   = "%%CITATION = ARXIV:1012.3380;%%"
}

@article{Andersen:2009nu,
    author = "Andersen, Jeppe R. and Smillie, Jennifer M.",
    title = "{Constructing All-Order Corrections to Multi-Jet Rates}",
    eprint = "0908.2786",
    archivePrefix = "arXiv",
    primaryClass = "hep-ph",
    reportNumber = "CERN-PH-TH-2009-153, DCPT-09-122, IPPP-09-61",
    doi = "10.1007/JHEP01(2010)039",
    journal = "JHEP",
    volume = "01",
    pages = "039",
    year = "2010"
}

@article{Gieseke:2017clv,
    author = {Gieseke, Stefan and Kirchgae\ss{}er, Patrick and Pl\"atzer, Simon},
    title = "{Baryon production from cluster hadronisation}",
    eprint = "1710.10906",
    archivePrefix = "arXiv",
    primaryClass = "hep-ph",
    reportNumber = "KA-TP-32-2017, UWTHPH-2017-34, HERWIG-2017-03, MCNET-17-25",
    doi = "10.1140/epjc/s10052-018-5585-7",
    journal = "Eur. Phys. J. C",
    volume = "78",
    number = "2",
    pages = "99",
    year = "2018"
}

@article{Christiansen:2015yqa,
    author = "Christiansen, Jesper R. and Skands, Peter Z.",
    title = "{String Formation Beyond Leading Colour}",
    eprint = "1505.01681",
    archivePrefix = "arXiv",
    primaryClass = "hep-ph",
    reportNumber = "COEPP-MN-15-1, LU-TP-15-16, MCNET-15-09, COEPP-MN-15-1, LU-TP-15-16, MCNET-15-09",
    doi = "10.1007/JHEP08(2015)003",
    journal = "JHEP",
    volume = "08",
    pages = "003",
    year = "2015"
}

@article{Gieseke:2018gff,
    author = {Gieseke, Stefan and Kirchgae\ss{}er, Patrick and Pl\"atzer, Simon and Siodmok, Andrzej},
    title = "{Colour Reconnection from Soft Gluon Evolution}",
    eprint = "1808.06770",
    archivePrefix = "arXiv",
    primaryClass = "hep-ph",
    reportNumber = "KA-TP-23-2018, UWTHPH-2018-23, HERWIG-2018-02, MCnet-18-22",
    doi = "10.1007/JHEP11(2018)149",
    journal = "JHEP",
    volume = "11",
    pages = "149",
    year = "2018"
}

@article{Salam:2021tbm,
    author = "Salam, Gavin P. and Slade, Emma",
    title = "{Cuts for two-body decays at colliders}",
    eprint = "2106.08329",
    archivePrefix = "arXiv",
    primaryClass = "hep-ph",
    reportNumber = "OUTP-21-16P",
    doi = "10.1007/JHEP11(2021)220",
    journal = "JHEP",
    volume = "11",
    pages = "220",
    year = "2021"
}

@misc{hepdata.105630,
    author = "{ATLAS Collaboration}",
    title = "{Measurements of jet cross-section ratios in 13 TeV proton--proton collisions with ATLAS}",
    howpublished = "{HEPData (collection)}",
    year = 2024,
    note = "\url{https://doi.org/10.17182/hepdata.105630}"
}

@article{Buckley:2010ar,
    author = "Buckley, Andy and Butterworth, Jonathan and Grellscheid, David and Hoeth, Hendrik and Lonnblad, Leif and Monk, James and Schulz, Holger and Siegert, Frank",
    title = "Rivet user manual",
    eprint = "1003.0694",
    archivePrefix = "arXiv",
    primaryClass = "hep-ph",
    reportNumber = "MCNET-10-03, MCnet/10/03",
    doi = "10.1016/j.cpc.2013.05.021",
    journal = "Comput. Phys. Commun.",
    volume = "184",
    pages = "2803--2819",
    year = "2013"
}

@article{UA1:1985abw,
    author = "{UA1 Collaboration}",
    title = "{Comparison of three-jet and two-jet cross-sections in \ppbar collisions at the CERN SPS \ppbar collider}",
    reportNumber = "CERN-EP/85-98",
    doi = "10.1016/0370-2693(85)90801-9",
    journal = "Phys. Lett. B",
    volume = "158",
    pages = "494",
    year = "1985"
}

@article{UA2:1986bnn,
    author = "{UA2 Collaboration}",
    title = "{A Study of three-jet events at the {CERN} $\bar{p}p$ collider}",
    reportNumber = "CERN-EP/85-189",
    doi = "10.1007/BF01557598",
    journal = "Z. Phys. C",
    volume = "30",
    pages = "341",
    year = "1986"
}

@article{CDF:1995odg,
    author = "{CDF Collaboration}",
    title = "{Properties of High-Mass Multijet Events at the Fermilab Proton-Antiproton Collider}",
    reportNumber = "FERMILAB-PUB-95-038-E, CDF-PUB-JET-PUBLIC-2957",
    doi = "10.1103/PhysRevLett.75.608",
    journal = "Phys. Rev. Lett.",
    volume = "75",
    pages = "608--612",
    year = "1995"
}

@article{D0:1995bph,
    author = "{D0 Collaboration}",
    title = "{Studies of topological distributions of inclusive three- and four-jet events in $\bar{p}p$ collisions at $\sqrt{s}=1800$ GeV with the D0 detector}",
    eprint = "hep-ex/9509005",
    archivePrefix = "arXiv",
    reportNumber = "FERMILAB-PUB-95-296-E, D0-2735, D0-PUB-95-11",
    doi = "10.1103/PhysRevD.53.6000",
    journal = "Phys. Rev. D",
    volume = "53",
    pages = "6000--6016",
    year = "1996"
}

@Article{Cacciari:2008gp,
     author    = "Cacciari, Matteo and Salam, Gavin P. and Soyez, Gregory",
     title     = "{The anti-\(k_{t}\) jet clustering algorithm}",
     journal   = "JHEP",
     volume    = "04",
     year      = "2008",
     pages     = "063",
     eprint    = "0802.1189",
     archivePrefix = "arXiv",
     primaryClass  =  "hep-ph",
     doi       = "10.1088/1126-6708/2008/04/063",
     SLACcitation  = "%%CITATION = 0802.1189;%%"
}

@Article{Fastjet,
      author         = "Cacciari, Matteo and Salam, Gavin P. and Soyez, Gregory",
      title          = "{FastJet user manual}",
      journal        = "Eur. Phys. J. C",
      volume         = "72",
      year           = "2012",
      pages          = "1896",
      doi            = "10.1140/epjc/s10052-012-1896-2",
      eprint         = "1111.6097",
      archivePrefix  = "arXiv",
      primaryClass   = "hep-ph",
      reportNumber   = "CERN-PH-TH-2011-297",
      SLACcitation   = "%%CITATION = ARXIV:1111.6097;%%"
}

@article{Lai:2010vv,
      author         = "Lai, H.-L. and others",
      title          = "{New parton distributions for collider physics}",
      journal        = "Phys. Rev. D",
      volume         = "82",
      pages          = "074024",
      doi            = "10.1103/PhysRevD.82.074024",
      year           = "2010",
      eprint         = "1007.2241",
      archivePrefix  = "arXiv",
      primaryClass   = "hep-ph",
      reportNumber   = "MSUHEP-100707, SMU-HEP-10-10",
      SLACcitation   = "%%CITATION = ARXIV:1007.2241;%%",
}

@article{Dulat:2015mca,
      author         = "Dulat, Sayipjamal and Hou, Tie-Jiun and Gao, Jun and
                        Guzzi, Marco and Huston, Joey and Nadolsky, Pavel and
                        Pumplin, Jon and Schmidt, Carl and Stump, Daniel and Yuan,
                        C. P.",
      title          = "{New parton distribution functions from a global analysis
                        of quantum chromodynamics}",
      journal        = "Phys. Rev. D",
      volume         = "93",
      year           = "2016",
      number         = "3",
      pages          = "033006",
      doi            = "10.1103/PhysRevD.93.033006",
      eprint         = "1506.07443",
      archivePrefix  = "arXiv",
      primaryClass   = "hep-ph",
      SLACcitation   = "%%CITATION = ARXIV:1506.07443;%%"
}

@article{Ball:2012cx,
      author         = "{NNPDF Collaboration} and Ball, Richard D. and others",
      title          = "{Parton distributions with LHC data}",
      journal        = "Nucl. Phys. B",
      volume         = "867",
      year           = "2013",
      pages          = "244",
      doi            = "10.1016/j.nuclphysb.2012.10.003",
      eprint         = "1207.1303",
      archivePrefix  = "arXiv",
      primaryClass   = "hep-ph",
      reportNumber   = "EDINBURGH-2012-08, IFUM-FT-997, FR-PHENO-2012-014,
                        RWTH-TTK-12-25, CERN-PH-TH-2012-037, SFB-CPP-12-47\,
                        --CERN-PH-TH-2012-037",
      SLACcitation   = "%%CITATION = ARXIV:1207.1303;%%"
}

@article{Ball:2014uwa,
      author         = "{NNPDF Collaboration} and Ball, Richard D. and others",
      title          = "{Parton distributions for the LHC run II}",
      journal        = "JHEP",
      volume         = "04",
      year           = "2015",
      pages          = "040",
      doi            = "10.1007/JHEP04(2015)040",
      eprint         = "1410.8849",
      archivePrefix  = "arXiv",
      primaryClass   = "hep-ph",
      reportNumber   = "EDINBURGH-2014-15, IFUM-1034-FT, CERN-PH-TH-2013-253,
                        OUTP-14-11P, CAVENDISH-HEP-14-11",
      SLACcitation   = "%%CITATION = ARXIV:1410.8849;%%"
}

@article{Skands:2014pea,
      author         = "Skands, Peter and Carrazza, Stefano and Rojo, Juan",
      title          = "{Tuning PYTHIA 8.1: the Monash 2013 Tune}",
      journal        = "Eur. Phys. J. C",
      volume         = "74",
      year           = "2014",
      number         = "8",
      pages          = "3024",
      doi            = "10.1140/epjc/s10052-014-3024-y",
      eprint         = "1404.5630",
      archivePrefix  = "arXiv",
      primaryClass   = "hep-ph",
      reportNumber   = "CERN-PH-TH-2014-069, MCNET-14-08, OUTP-14-05P",
      SLACcitation   = "%%CITATION = ARXIV:1404.5630;%%"
}

@article{Sjostrand:2014zea,
      author         = "Sj{\"o}strand, Torbj{\"o}rn and Ask, Stefan and Christiansen,
                        Jesper R. and Corke, Richard and Desai, Nishita and Ilten,
                        Philip and Mrenna, Stephen and Prestel, Stefan and
                        Rasmussen, Christine O. and Skands, Peter Z.",
      title          = "{An introduction to PYTHIA 8.2}",
      journal        = "Comput. Phys. Commun.",
      volume         = "191",
      year           = "2015",
      pages          = "159",
      doi            = "10.1016/j.cpc.2015.01.024",
      eprint         = "1410.3012",
      archivePrefix  = "arXiv",
      primaryClass   = "hep-ph",
      reportNumber   = "LU-TP-14-36, MCNET-14-22, CERN-PH-TH-2014-190,
                        FERMILAB-PUB-14-316-CD, DESY-14-178, SLAC-PUB-16122,
                        --FERMILAB-PUB-14-316-CD",
      SLACcitation   = "%%CITATION = ARXIV:1410.3012;%%"
}

@article{Andersson:1983ia,
      author         = "Andersson, Bo and Gustafson, G. and Ingelman, G. and
                        Sj{\"o}strand, T.",
      title          = "{Parton fragmentation and string dynamics}",
      journal        = "Phys. Rept.",
      volume         = "97",
      year           = "1983",
      pages          = "31-145",
      doi            = "10.1016/0370-1573(83)90080-7",
      reportNumber   = "LU-TP-83-10",
      SLACcitation   = "%%CITATION = PRPLC,97,31;%%"
}

@Article{Lange:2001uf,
      author         = "Lange, D. J.",
      title          = "{The EvtGen particle decay simulation package}",
      booktitle      = "{Proceedings, 7th International Conference on B physics
                        at hadron machines (BEAUTY 2000)}",
      journal        = "Nucl. Instrum. Meth. A",
      volume         = "462",
      year           = "2001",
      pages          = "152",
      doi            = "10.1016/S0168-9002(01)00089-4",
      SLACcitation   = "%%CITATION = NUIMA,A462,152;%%"
}

@Article{Bahr:2008pv,
      author         = "B{\"a}hr, M. and others",
      title          = "{Herwig++ physics and manual}",
      journal        = "Eur. Phys. J. C",
      volume         = "58",
      year           = "2008",
      pages          = "639",
      doi            = "10.1140/epjc/s10052-008-0798-9",
      eprint         = "0803.0883",
      archivePrefix  = "arXiv",
      primaryClass   = "hep-ph",
      reportNumber   = "CERN-PH-TH-2008-038, CAVENDISH-HEP-08-03, KA-TP-05-2008,
                        DCPT-08-22, IPPP-08-11, CP3-08-05",
      SLACcitation   = "%%CITATION = ARXIV:0803.0883;%%"
}

@Article{Bellm:2015jjp,
      author         = "Bellm, Johannes and others",
      title          = "{Herwig 7.0/Herwig++ 3.0 release note}",
      journal        = "Eur. Phys. J. C",
      volume         = "76",
      year           = "2016",
      number         = "4",
      pages          = "196",
      doi            = "10.1140/epjc/s10052-016-4018-8",
      eprint         = "1512.01178",
      archivePrefix  = "arXiv",
      primaryClass   = "hep-ph",
      reportNumber   = "CERN-PH-TH-2015-289, MAN-HEP-2015-15, IFJPAN-IV-2015-13,
                        HERWIG-2015-01, KA-TP-18-2015, DCPT-15-142, MCNET-15-28,
                        IPPP-15-71, --HERWIG-2015-01",
      SLACcitation   = "%%CITATION = ARXIV:1512.01178;%%"
}

@Article{Nason:2004rx,
      author         = "Nason, Paolo",
      title          = "{A new method for combining NLO QCD with shower Monte Carlo algorithms}",
      journal        = "JHEP",
      volume         = "11",
      pages          = "040",
      doi            = "10.1088/1126-6708/2004/11/040",
      year           = "2004",
      eprint         = "hep-ph/0409146",
      archivePrefix  = "arXiv",
}

@Article{Frixione:2007vw,
      author         = "Frixione, Stefano and Nason, Paolo and Oleari, Carlo",
      title          = "{Matching NLO QCD computations with parton shower
                        simulations: the POWHEG method}",
      journal        = "JHEP",
      volume         = "11",
      pages          = "070",
      doi            = "10.1088/1126-6708/2007/11/070",
      year           = "2007",
      eprint         = "0709.2092",
      archivePrefix  = "arXiv",
      primaryClass   = "hep-ph",
}

@Article{Alioli:2010xd,
      author         = "Alioli, Simone and Nason, Paolo and Oleari, Carlo and Re,
                        Emanuele",
      title          = "{A general framework for implementing NLO calculations in
                        shower Monte Carlo programs: the POWHEG BOX}",
      journal        = "JHEP",
      volume         = "06",
      pages          = "043",
      doi            = "10.1007/JHEP06(2010)043",
      year           = "2010",
      eprint         = "1002.2581",
      archivePrefix  = "arXiv",
      primaryClass   = "hep-ph",
}

@Article{Schumann:2007mg,
      author         = "Schumann, Steffen and Krauss, Frank",
      title          = "{A parton shower algorithm based on Catani--Seymour dipole factorisation}",
      journal        = "JHEP",
      volume         = "03",
      pages          = "038",
      doi            = "10.1088/1126-6708/2008/03/038",
      year           = "2008",
      eprint         = "0709.1027",
      archivePrefix  = "arXiv",
      primaryClass   = "hep-ph",
}

@article{Buccioni:2019sur,
    author = "Buccioni, Federico and Lang, Jean-Nicolas and Lindert, Jonas M. and Maierh{\"o}fer, Philipp and Pozzorini, Stefano and Zhang, Hantian and Zoller, Max F.",
    title = "{OpenLoops 2}",
    eprint = "1907.13071",
    archivePrefix = "arXiv",
    primaryClass = "hep-ph",
    reportNumber = "IPPP/19/62, FR-PHENO-2019-12, PSI-PR-19-15, ZU-TH 37/19",
    doi = "10.1140/epjc/s10052-019-7306-2",
    journal = "Eur. Phys. J. C",
    volume = "79",
    number = "10",
    pages = "866",
    year = "2019"
}

@article{Agostinelli:2002hh,
      author         = "Agostinelli, S. and others",
      title          = "{\textsc{Geant4} -- a simulation toolkit}",
      journal        = "Nucl. Instrum. Meth. A",
      volume         = "506",
      year           = "2003",
      pages          = "250",
      doi            = "10.1016/S0168-9002(03)01368-8",
      reportNumber   = "SLAC-PUB-9350, FERMILAB-PUB-03-339",
      SLACcitation   = "%%CITATION = NUIMA,A506,250;%%"
}

@Inproceedings{Adye:2011gm,
    author = "Adye, Tim",
    title = "{Unfolding algorithms and tests using RooUnfold}",
    booktitle = "{Proceedings, 2011 Workshop on Statistical Issues Related to Discovery Claims in Search Experiments and Unfolding (PHYSTAT 2011)}",
    venue = "CERN, Geneva, Switzerland",
    eventdate = "2011-01-17/2011-01-20",
    pages = "313-318",
    eprint = "1105.1160",
    archivePrefix = "arXiv",
    primaryClass = "physics.data-an",
    doi = "10.5170/CERN-2011-006.313",
}

@article{DAgostini:1994fjx,
      author         = "D'Agostini, G.",
      title          = "{A multidimensional unfolding method based on Bayes' theorem}",
      journal        = "Nucl. Instrum. Meth. A",
      volume         = "362",
      year           = "1995",
      pages          = "487-498",
      doi            = "10.1016/0168-9002(95)00274-X",
      reportNumber   = "DESY-94-099",
      SLACcitation   = "%%CITATION = NUIMA,A362,487;%%"
}

@Article{PERF-2007-01,
    author         = "{ATLAS Collaboration}",
    title          = "{The ATLAS Experiment at the CERN Large Hadron Collider}",
    journal        = "JINST",
    volume         = "3",
    year           = "2008",
    pages          = "S08003",
    doi            = "10.1088/1748-0221/3/08/S08003",
    primaryClass   = "hep-ex",
}

@Article{SOFT-2010-01,
    author         = "{ATLAS Collaboration}",
    title          = "{The ATLAS Simulation Infrastructure}",
    journal        = "Eur. Phys. J. C",
    volume         = "70",
    year           = "2010",
    pages          = "823",
    doi            = "10.1140/epjc/s10052-010-1429-9",
    eprint         = "1005.4568",
    archivePrefix  = "arXiv",
    primaryClass   = "physics.ins-det",
}

@Article{STDM-2010-10,
    author         = "{ATLAS Collaboration}",
    title          = "{Study of jet shapes in inclusive jet production in \(pp\) collisions at \(\sqrt{s} = 7\,\text{TeV}\) using the ATLAS detector}",
    journal        = "Phys. Rev. D",
    volume         = "83",
    year           = "2011",
    pages          = "052003",
    doi            = "10.1103/PhysRevD.83.052003",
    reportNumber   = "CERN-PH-EP-2010-077",
    eprint         = "1101.0070",
    archivePrefix  = "arXiv",
    primaryClass   = "hep-ex",
}

@Article{STDM-2010-12,
    author         = "{ATLAS Collaboration}",
    title          = "{Measurement of multi-jet cross sections in proton--proton collisions at a \(7\,\text{TeV}\) center-of-mass energy}",
    journal        = "Eur. Phys. J. C",
    volume         = "71",
    year           = "2011",
    pages          = "1763",
    doi            = "10.1140/epjc/s10052-011-1763-6",
    reportNumber   = "CERN-PH-EP-2011-098",
    eprint         = "1107.2092",
    archivePrefix  = "arXiv",
    primaryClass   = "hep-ex",
}

@Article{PERF-2011-05,
    author         = "{ATLAS Collaboration}",
    title          = "{Single hadron response measurement and calorimeter jet energy scale uncertainty with the ATLAS detector at the LHC}",
    journal        = "Eur. Phys. J. C",
    volume         = "73",
    year           = "2013",
    pages          = "2305",
    doi            = "10.1140/epjc/s10052-013-2305-1",
    reportNumber   = "CERN-PH-EP-2012-005",
    eprint         = "1203.1302",
    archivePrefix  = "arXiv",
    primaryClass   = "hep-ex",
}

@Article{STDM-2011-38,
    author         = "{ATLAS Collaboration}",
    title          = "{ATLAS measurements of the properties of jets for boosted particle searches}",
    journal        = "Phys. Rev. D",
    volume         = "86",
    year           = "2012",
    pages          = "072006",
    doi            = "10.1103/PhysRevD.86.072006",
    reportNumber   = "CERN-PH-EP-2012-149",
    eprint         = "1206.5369",
    archivePrefix  = "arXiv",
    primaryClass   = "hep-ex",
}

@Article{PERF-2012-01,
    author         = "{ATLAS Collaboration}",
    title          = "{Jet energy measurement and its systematic uncertainty in proton--proton collisions at \(\sqrt{s} = 7\,\text{TeV}\) with the ATLAS detector}",
    journal        = "Eur. Phys. J. C",
    volume         = "75",
    year           = "2015",
    pages          = "17",
    doi            = "10.1140/epjc/s10052-014-3190-y",
    reportNumber   = "CERN-PH-EP-2013-222",
    eprint         = "1406.0076",
    archivePrefix  = "arXiv",
    primaryClass   = "hep-ex",
}

@Article{TRIG-2012-01,
    author         = "{ATLAS Collaboration}",
    title          = "{The performance of the jet trigger for the ATLAS detector during 2011 data taking}",
    journal        = "Eur. Phys. J. C",
    volume         = "76",
    year           = "2016",
    pages          = "526",
    doi            = "10.1140/epjc/s10052-016-4325-0",
    reportNumber   = "CERN-EP-2016-031",
    eprint         = "1606.07759",
    archivePrefix  = "arXiv",
    primaryClass   = "hep-ex",
}

@Article{PERF-2013-02,
    author         = "{ATLAS Collaboration}",
    title          = "{Light-quark and gluon jet discrimination in \(pp\) collisions at \(\sqrt{s} = 7\,\text{TeV}\) with the ATLAS detector}",
    journal        = "Eur. Phys. J. C",
    volume         = "74",
    year           = "2014",
    pages          = "3023",
    doi            = "10.1140/epjc/s10052-014-3023-z",
    reportNumber   = "CERN-PH-EP-2014-058",
    eprint         = "1405.6583",
    archivePrefix  = "arXiv",
    primaryClass   = "hep-ex",
}

@Article{PERF-2014-03,
    author         = "{ATLAS Collaboration}",
    title          = "{Performance of pile-up mitigation techniques for jets in \(pp\) collisions at \(\sqrt{s} = 8\,\text{TeV}\) using the ATLAS detector}",
    journal        = "Eur. Phys. J. C",
    volume         = "76",
    year           = "2016",
    pages          = "581",
    doi            = "10.1140/epjc/s10052-016-4395-z",
    reportNumber   = "CERN-PH-EP-2015-206",
    eprint         = "1510.03823",
    archivePrefix  = "arXiv",
    primaryClass   = "hep-ex",
}

@Article{PERF-2014-07,
    author         = "{ATLAS Collaboration}",
    title          = "{Topological cell clustering in the ATLAS calorimeters and its performance in LHC Run~1}",
    journal        = "Eur. Phys. J. C",
    volume         = "77",
    year           = "2017",
    pages          = "490",
    doi            = "10.1140/epjc/s10052-017-5004-5",
    reportNumber   = "CERN-PH-EP-2015-304",
    eprint         = "1603.02934",
    archivePrefix  = "arXiv",
    primaryClass   = "hep-ex",
}

@Article{PERF-2015-05,
    author         = "{ATLAS Collaboration}",
    title          = "{A measurement of the calorimeter response to single hadrons and determination of the jet energy scale uncertainty using LHC Run-1 \(pp\)-collision data with the ATLAS detector}",
    journal        = "Eur. Phys. J. C",
    volume         = "77",
    year           = "2017",
    pages          = "26",
    doi            = "10.1140/epjc/s10052-016-4580-0",
    reportNumber   = "CERN-EP-2016-149",
    eprint         = "1607.08842",
    archivePrefix  = "arXiv",
    primaryClass   = "hep-ex",
}

@Article{PERF-2015-09,
    author         = "{ATLAS Collaboration}",
    title          = "{Jet reconstruction and performance using particle flow with the ATLAS Detector}",
    journal        = "Eur. Phys. J. C",
    volume         = "77",
    year           = "2017",
    pages          = "466",
    doi            = "10.1140/epjc/s10052-017-5031-2",
    reportNumber   = "CERN-EP-2017-024",
    eprint         = "1703.10485",
    archivePrefix  = "arXiv",
    primaryClass   = "hep-ex",
}

@Article{PERF-2016-06,
    author         = "{ATLAS Collaboration}",
    title          = "{Identification and rejection of pile-up jets at high pseudorapidity with the ATLAS detector}",
    journal        = "Eur. Phys. J. C",
    volume         = "77",
    year           = "2017",
    pages          = "580",
    doi            = "10.1140/epjc/s10052-017-5081-5",
    reportNumber   = "CERN-EP-2017-055",
    eprint         = "1705.02211",
    archivePrefix  = "arXiv",
    primaryClass   = "hep-ex",
    related        = "PERF-2016-06-err",
    relatedstring  = "Erratum:",
}

@Article{TRIG-2016-01,
    author         = "{ATLAS Collaboration}",
    title          = "{Performance of the ATLAS trigger system in 2015}",
    journal        = "Eur. Phys. J. C",
    volume         = "77",
    year           = "2017",
    pages          = "317",
    doi            = "10.1140/epjc/s10052-017-4852-3",
    reportNumber   = "CERN-EP-2016-241",
    eprint         = "1611.09661",
    archivePrefix  = "arXiv",
    primaryClass   = "hep-ex",
}

@Article{STDM-2017-04,
    author         = "{ATLAS Collaboration}",
    title          = "{Measurement of the Soft-Drop Jet Mass in \(pp\) Collisions at \(\sqrt{s} = 13\,\text{TeV}\) with the ATLAS detector}",
    journal        = "Phys. Rev. Lett.",
    volume         = "121",
    year           = "2018",
    pages          = "092001",
    doi            = "10.1103/PhysRevLett.121.092001",
    reportNumber   = "CERN-EP-2017-231",
    eprint         = "1711.08341",
    archivePrefix  = "arXiv",
    primaryClass   = "hep-ex",
}

@Article{STDM-2017-16,
    author         = "{ATLAS Collaboration}",
    title          = "{Properties of jet fragmentation using charged particles measured with the ATLAS detector in \(pp\) collisions at \(\sqrt{s} = 13\,\text{TeV}\)}",
    journal        = "Phys. Rev. D",
    volume         = "100",
    year           = "2019",
    pages          = "052011",
    doi            = "10.1103/PhysRevD.100.052011",
    reportNumber   = "CERN-EP-2019-090",
    eprint         = "1906.09254",
    archivePrefix  = "arXiv",
    primaryClass   = "hep-ex",
}

@Article{DAPR-2018-01,
    author         = "{ATLAS Collaboration}",
    title          = "{ATLAS data quality operations and performance for 2015--2018 data-taking}",
    journal        = "JINST",
    volume         = "15",
    year           = "2020",
    pages          = "P04003",
    doi            = "10.1088/1748-0221/15/04/P04003",
    reportNumber   = "CERN-EP-2019-207",
    eprint         = "1911.04632",
    archivePrefix  = "arXiv",
    primaryClass   = "physics.ins-det",
}

@Article{EGAM-2018-01,
    author         = "{ATLAS Collaboration}",
    title          = "{Electron and photon performance measurements with the ATLAS detector using the 2015--2017 LHC proton--proton collision data}",
    journal        = "JINST",
    volume         = "14",
    year           = "2019",
    pages          = "P12006",
    doi            = "10.1088/1748-0221/14/12/P12006",
    reportNumber   = "CERN-EP-2019-145",
    eprint         = "1908.00005",
    archivePrefix  = "arXiv",
    primaryClass   = "hep-ex",
}

@Article{JETM-2018-05,
    author         = "{ATLAS Collaboration}",
    title          = "{Jet energy scale and resolution measured in proton--proton collisions at \(\sqrt{s} = 13\,\text{TeV}\) with the ATLAS detector}",
    journal        = "Eur. Phys. J. C",
    volume         = "81",
    year           = "2021",
    pages          = "689",
    doi            = "10.1140/epjc/s10052-021-09402-3",
    reportNumber   = "CERN-EP-2020-083",
    eprint         = "2007.02645",
    archivePrefix  = "arXiv",
    primaryClass   = "hep-ex",
}

@Article{JETM-2018-08,
    author         = "{ATLAS Collaboration}",
    title          = "{Measurement of the energy response of the ATLAS calorimeter to charged pions from \(W^{\pm}\to\tau^{\pm}(\to\pi^{\pm}\nu_{\tau})\nu_{\tau}\) events in Run~2 data}",
    journal        = "Eur. Phys. J. C",
    volume         = "82",
    year           = "2022",
    pages          = "223",
    doi            = "10.1140/epjc/s10052-022-10117-2",
    reportNumber   = "CERN-EP-2021-147",
    eprint         = "2108.09043",
    archivePrefix  = "arXiv",
    primaryClass   = "hep-ex",
}

@Article{PIX-2018-001,
    author         = "Abbott, B. and others",
    title          = "{Production and integration of the ATLAS Insertable B-Layer}",
    journal        = "JINST",
    volume         = "13",
    year           = "2018",
    pages          = "T05008",
    doi            = "10.1088/1748-0221/13/05/T05008",
    eprint         = "1803.00844",
    archivePrefix  = "arXiv",
    primaryClass   = "physics.ins-det",
}

@Article{STDM-2018-57,
    author         = "{ATLAS Collaboration}",
    title          = "{Measurement of the Lund Jet Plane Using Charged Particles in \(13\,\text{TeV}\) Proton--Proton Collisions with the ATLAS Detector}",
    journal        = "Phys. Rev. Lett.",
    volume         = "124",
    year           = "2020",
    pages          = "222002",
    doi            = "10.1103/PhysRevLett.124.222002",
    reportNumber   = "CERN-EP-2020-030",
    eprint         = "2004.03540",
    archivePrefix  = "arXiv",
    primaryClass   = "hep-ex",
}

@Article{STDM-2019-02,
    author         = "{ATLAS Collaboration}",
    title          = "{Measurement of hadronic event shapes in high-\(p_{\text{T}}\) multijet final states at \( \sqrt{s} = 13\,\text{TeV}\) with the ATLAS detector}",
    journal        = "JHEP",
    volume         = "01",
    year           = "2021",
    pages          = "188",
    doi            = "10.1007/JHEP01(2021)188",
    reportNumber   = "CERN-EP-2020-077",
    eprint         = "2007.12600",
    archivePrefix  = "arXiv",
    primaryClass   = "hep-ex",
    related        = "STDM-2019-02-err",
    relatedstring  = "Erratum:",
}

@Article{STDM-2020-04,
    author         = "{ATLAS Collaboration}",
    title          = "{Measurements of jet cross-section ratios in \(13\,\text{TeV}\) proton--proton collisions with ATLAS}",
    year           = "2024",
    reportNumber   = "CERN-EP-2024-119",
    eprint         = "2405.20206",
    archivePrefix  = "arXiv",
    primaryClass   = "hep-ex",
}

@Article{STDM-2020-20,
    author         = "{ATLAS Collaboration}",
    title          = "{Measurements of multijet event isotropies using optimal transport with the ATLAS detector}",
    journal        = "JHEP",
    volume         = "10",
    year           = "2023",
    pages          = "060",
    doi            = "10.1007/JHEP10(2023)060",
    reportNumber   = "CERN-EP-2023-079",
    eprint         = "2305.16930",
    archivePrefix  = "arXiv",
    primaryClass   = "hep-ex",
}

@Article{STDM-2020-32,
    author         = "{ATLAS Collaboration}",
    title          = "{Determination of the parton distribution functions of the proton using diverse ATLAS data from \(pp\) collisions at \(\sqrt{s} = 7, 8\) and \(13\,\text{TeV}\)}",
    journal        = "Eur. Phys. J. C",
    volume         = "82",
    year           = "2022",
    pages          = "438",
    doi            = "10.1140/epjc/s10052-022-10217-z",
    reportNumber   = "CERN-EP-2021-239",
    eprint         = "2112.11266",
    archivePrefix  = "arXiv",
    primaryClass   = "hep-ex",
}

@Article{DAPR-2021-01,
    author         = "{ATLAS Collaboration}",
    title          = "{Luminosity determination in \(pp\) collisions at \(\sqrt{s} = 13\,\text{TeV}\) using the ATLAS detector at the LHC}",
    journal        = "Eur. Phys. J. C",
    volume         = "83",
    year           = "2023",
    pages          = "982",
    doi            = "10.1140/epjc/s10052-023-11747-w",
    reportNumber   = "CERN-EP-2022-281",
    eprint         = "2212.09379",
    archivePrefix  = "arXiv",
    primaryClass   = "hep-ex",
}

@Booklet{ATL-SOFT-PUB-2021-001,
    author         = "{ATLAS Collaboration}",
    title          = "{The ATLAS Collaboration Software and Firmware}",
    howpublished   = "{ATL-SOFT-PUB-2021-001}",
    url            = "https://cds.cern.ch/record/2767187",
    year           = "2021",
}

@Booklet{ATL-SOFT-PUB-2023-001,
    author         = "{ATLAS Collaboration}",
    title          = "{ATLAS Computing Acknowledgements}",
    howpublished   = "{ATL-SOFT-PUB-2023-001}",
    url            = "https://cds.cern.ch/record/2869272",
    year           = "2023",
}

@Report{ATLAS-TDR-19,
    author         = "{ATLAS Collaboration}",
    title          = "{ATLAS Insertable B-Layer: Technical Design Report}",
    type           = "ATLAS-TDR-19; CERN-LHCC-2010-013",
    year           = "2010",
    url            = "https://cds.cern.ch/record/1291633",
    related        = "ATLAS-TDR-19-addm",
    relatedstring  = "Addendum:",
}

@Article{CMS-QCD-10-012,
    author         = "{CMS Collaboration}",
    title          = "{Measurement of the Ratio of the 3-jet to 2-jet Cross Sections in \(pp\) Collisions at \(\sqrt{s} = 7\,\text{TeV}\)}",
    journal        = "Phys. Lett. B",
    volume         = "702",
    year           = "2011",
    pages          = "336",
    doi            = "10.1016/j.physletb.2011.07.067",
    reportNumber   = "CERN-PH-EP-2011-044",
    eprint         = "1106.0647",
    archivePrefix  = "arXiv",
    primaryClass   = "hep-ex",
}

@Article{CMS-QCD-11-003,
    author         = "{CMS Collaboration}",
    title          = "{Measurement of the ratio of the inclusive 3-jet cross section to the inclusive 2-jet cross section in \(pp\) collisions at \(\sqrt{s} = 7\,\text{TeV}\) and first determination of the strong coupling constant in the TeV range}",
    journal        = "Eur. Phys. J. C",
    volume         = "73",
    year           = "2013",
    pages          = "2604",
    doi            = "10.1140/epjc/s10052-013-2604-6",
    reportNumber   = "CERN-PH-EP-2013-057",
    eprint         = "1304.7498",
    archivePrefix  = "arXiv",
    primaryClass   = "hep-ex",
}

@Article{CMS-SMP-16-010,
    author         = "{CMS Collaboration}",
    title          = "{Measurements of the differential jet cross section as a function of the jet mass in dijet events from proton--proton collisions at \(\sqrt{s} = 13\,\text{TeV}\)}",
    journal        = "JHEP",
    volume         = "11",
    year           = "2018",
    pages          = "113",
    doi            = "10.1007/JHEP11(2018)113",
    reportNumber   = "CERN-EP-2018-180",
    eprint         = "1807.05974",
    archivePrefix  = "arXiv",
    primaryClass   = "hep-ex",
}

@Article{CMS-SMP-16-011,
    author         = "{CMS Collaboration}",
    title          = "{Measurement of the triple-differential dijet cross section in proton--proton collisions at \(\sqrt{s} = 8\,\text{TeV}\) and constraints on parton distribution functions}",
    journal        = "Eur. Phys. J. C",
    volume         = "77",
    year           = "2017",
    pages          = "746",
    doi            = "10.1140/epjc/s10052-017-5286-7",
    reportNumber   = "CERN-EP-2017-061",
    eprint         = "1705.02628",
    archivePrefix  = "arXiv",
    primaryClass   = "hep-ex",
}

@Article{CMS-SMP-20-010,
    author         = "{CMS Collaboration}",
    title          = "{Study of quark and gluon jet substructure in \(Z\)+jet and dijet events from \(pp\) collisions}",
    journal        = "JHEP",
    volume         = "01",
    year           = "2022",
    pages          = "188",
    doi            = "10.1007/JHEP01(2022)188",
    reportNumber   = "CERN-EP-2021-161",
    eprint         = "2109.03340",
    archivePrefix  = "arXiv",
    primaryClass   = "hep-ex",
}

@Booklet{ATLAS-CONF-2015-029,
    author         = "{ATLAS Collaboration}",
    title          = "{Selection of jets produced in \(13~\text{TeV}\) proton--proton collisions with the ATLAS detector}",
    howpublished   = "{ATLAS-CONF-2015-029}",
    url            = "https://cds.cern.ch/record/2037702",
    year           = "2015",
}

@Booklet{ATL-PHYS-PUB-2014-021,
    author         = "{ATLAS Collaboration}",
    title          = "{ATLAS Pythia~8 tunes to \(7~\text{TeV}\) data}",
    howpublished   = "{ATL-PHYS-PUB-2014-021}",
    url            = "https://cds.cern.ch/record/1966419",
    year           = "2014",
}

@Booklet{ATL-PHYS-PUB-2016-017,
    author         = "{ATLAS Collaboration}",
    title          = "{The Pythia~8 A3 tune description of ATLAS minimum bias and inelastic measurements incorporating the Donnachie--Landshoff diffractive model}",
    howpublished   = "{ATL-PHYS-PUB-2016-017}",
    url            = "https://cds.cern.ch/record/2206965",
    year           = "2016",
}

@Booklet{ATL-PHYS-PUB-2019-017,
    author         = "{ATLAS Collaboration}",
    title          = "{Multijet simulation for \(13\,\text{TeV}\) ATLAS Analyses}",
    howpublished   = "{ATL-PHYS-PUB-2019-017}",
    url            = "https://cds.cern.ch/record/2672252",
    year           = "2019",
}

@Booklet{ATL-PHYS-PUB-2021-011,
    author         = "{ATLAS Collaboration}",
    title          = "{Evaluating statistical uncertainties and correlations using the bootstrap method}",
    howpublished   = "{ATL-PHYS-PUB-2021-011}",
    url            = "https://cds.cern.ch/record/2759945",
    year           = "2021",
}

@Booklet{ATL-PHYS-PUB-2022-021,
    author         = "{ATLAS Collaboration}",
    title          = "{Dependence of the Jet Energy Scale on the Particle Content of Hadronic Jets in the ATLAS Detector Simulation}",
    howpublished   = "{ATL-PHYS-PUB-2022-021}",
    url            = "https://cds.cern.ch/record/2808016",
    year           = "2022",
}
\clearpage
 
\begin{flushleft}
\hypersetup{urlcolor=black}
{\Large The ATLAS Collaboration}

\bigskip

\AtlasOrcid[0000-0002-6665-4934]{G.~Aad}$^\textrm{\scriptsize 103}$,
\AtlasOrcid[0000-0001-7616-1554]{E.~Aakvaag}$^\textrm{\scriptsize 16}$,
\AtlasOrcid[0000-0002-5888-2734]{B.~Abbott}$^\textrm{\scriptsize 121}$,
\AtlasOrcid[0000-0002-1002-1652]{K.~Abeling}$^\textrm{\scriptsize 55}$,
\AtlasOrcid[0000-0001-5763-2760]{N.J.~Abicht}$^\textrm{\scriptsize 49}$,
\AtlasOrcid[0000-0002-8496-9294]{S.H.~Abidi}$^\textrm{\scriptsize 29}$,
\AtlasOrcid[0009-0003-6578-220X]{M.~Aboelela}$^\textrm{\scriptsize 44}$,
\AtlasOrcid[0000-0002-9987-2292]{A.~Aboulhorma}$^\textrm{\scriptsize 35e}$,
\AtlasOrcid[0000-0001-5329-6640]{H.~Abramowicz}$^\textrm{\scriptsize 152}$,
\AtlasOrcid[0000-0002-1599-2896]{H.~Abreu}$^\textrm{\scriptsize 151}$
% ...
\bigskip
\\

% ...
$^{16}$Department for Physics and Technology, University of Bergen, Bergen; Norway.\\
% ...
$^{29}$Physics Department, Brookhaven National Laboratory, Upton NY; United States of America.\\
% ...
$^{35}$$^{(a)}$Facult\'e des Sciences Ain Chock, R\'eseau Universitaire de Physique des Hautes Energies - Universit\'e Hassan II, Casablanca;$^{(b)}$Facult\'{e} des Sciences, Universit\'{e} Ibn-Tofail, K\'{e}nitra;$^{(c)}$Facult\'e des Sciences Semlalia, Universit\'e Cadi Ayyad, LPHEA-Marrakech;$^{(d)}$LPMR, Facult\'e des Sciences, Universit\'e Mohamed Premier, Oujda;$^{(e)}$Facult\'e des sciences, Universit\'e Mohammed V, Rabat;$^{(f)}$Institute of Applied Physics, Mohammed VI Polytechnic University, Ben Guerir; Morocco.\\
% ...
$^{44}$Physics Department, Southern Methodist University, Dallas TX; United States of America.\\
% ...
$^{49}$Fakult\"{a}t Physik , Technische Universit{\"a}t Dortmund, Dortmund; Germany.\\
% ...
$^{55}$II. Physikalisches Institut, Georg-August-Universit\"{a}t G\"ottingen, G\"ottingen; Germany.\\
% ...
$^{103}$CPPM, Aix-Marseille Universit\'e, CNRS/IN2P3, Marseille; France.\\
% ...
$^{121}$Homer L. Dodge Department of Physics and Astronomy, University of Oklahoma, Norman OK; United States of America.\\
% ...
$^{151}$Department of Physics, Technion, Israel Institute of Technology, Haifa; Israel.\\
$^{152}$Raymond and Beverly Sackler School of Physics and Astronomy, Tel Aviv University, Tel Aviv; Israel.\\
% ...
$^{a}$ Also Affiliated with an institute covered by a cooperation agreement with CERN.\\
$^{b}$ Also at An-Najah National University, Nablus; Palestine.\\
$^{c}$ Also at Borough of Manhattan Community College, City University of New York, New York NY; United States of America.\\
$^{d}$ Also at Center for High Energy Physics, Peking University; China.\\
$^{e}$ Also at Center for Interdisciplinary Research and Innovation (CIRI-AUTH), Thessaloniki; Greece.\\
$^{f}$ Also at Centro Studi e Ricerche Enrico Fermi; Italy.\\
% ...

\end{flushleft}


%

%
%

%

%
%
%
%

\end{document}